\documentclass{aa}
\usepackage{natbib}
\bibpunct{(}{)}{;}{a}{}{,} 
\usepackage{graphicx}
\usepackage{txfonts}

\newcommand{\textscown}[1]{{\fontfamily{ptm}\fontseries{m}\fontshape{sc}\selectfont{#1}}}

\begin{document}
\title{Role of emission angular directionality in spin determination of accreting black holes with a broad iron line}
\author{J.~Svoboda, M.~Dov\v{c}iak, R.~Goosmann,$\!$\thanks{Present address: Observatoire Astronomique de Strasbourg, 11~rue de l'Universit\'e, F--67000 Strasbourg, France} \and V.~Karas}
\institute{Astronomical Institute, Academy of Sciences, Bo\v{c}n\'{\i}~II~1401, CZ-14131~Prague, Czech Republic}

\authorrunning{J.~Svoboda et al.}
\titlerunning{Role of angular directionality in spin determination of accreting black holes\ldots}
\date{Received 24 February 2009 / Accepted 4 August 2009}
\offprints{J.~Svoboda, email: svoboda@astro.cas.cz}
\abstract{}{The spin of an accreting black hole can be determined by
spectroscopy of the emission and absorption features produced in the
inner regions of an accretion disc. In this work, we discuss the method
employing the relativistic line profiles of iron in the X-ray domain,
where the emergent spectrum is blurred by general relativistic
effects.}{Precision of the spectra fitting procedure could be compromised
by inappropriate accounting for the angular distribution of the disc
emission. 
Often a unique profile is assumed, 
invariable over the~entire range of radii in the~disc and energy in the spectral band. 
An isotropic distribution or a particular limb-darkening law have been frequently
set, although some radiation transfer computations exhibit an
emission excess towards the grazing angles (i.e., the~limb brightening). 
By assuming a rotating black hole in the centre of an accretion disc, we
perform radiation transfer computations of an X-ray irradiated disc
atmosphere (NOAR code) to determine the~directionality of outgoing
X-rays in the $2$--$10$~keV energy band. Based on these computations, we
produce a new extension to the~\textscown{ky} software  package for X-ray
spectra fitting of relativistic accretion discs.}{We study how
sensitive the spin determination is to the assumptions about 
the~intrinsic angular distribution of the~emitted photons. The~uncertainty
of the~directional emission distribution translates to $\simeq20$\%
uncertainty in the determination of the~marginally stable orbit. 
We implemented the simulation results as a new extension to the {\textscown{ky}} 
software package for X-ray spectra fitting of relativistic accretion disc models. 
Although the parameter space is rather complex, leading to a rich
variety of possible outcomes, we find that on average the isotropic
directionality reproduces our model data to the best precision.
Our results also suggest that an improper use of limb darkening
can partly mimic a~steeper profile of radial emissivity.
We demonstrate these results in
the case of XMM-Newton observation of the Seyfert galaxy 
MCG--6-30-15, for which we
construct confidence levels of $\chi^2$  statistics, and on 
the~simulated data for the~future X-ray IXO mission.
Our simulations, with the~tentative IXO response, show a significant improvement that can
qualitatively enhance the~accuracy of spin determination.
}{}
\maketitle

\section{Introduction}
There is now strong evidence suggesting that the radiation of 
active galactic nuclei as well as some compact binary stars in the Galaxy
originates in part from the gas of an accretion disc orbiting around 
a central black hole \citep{1998bhrs.conf...79R,1998bhad.conf.....K}.
A great deal of information about these objects has been obtained
via X-rays \citep{1997iagn.book.....P,1999agnc.book.....K}. 
Therefore, reliable models are needed to describe
production of high-energy photons emerging from the accretion disc 
surface and subsequent propagation of these photons towards a 
distant observer.

Directional distribution of the outgoing radiation is among the important
aspects that must be addressed. Limb darkening traditionally refers 
to the gradual diminution of intensity in the image of the surface of 
a star as one moves from the centre of the image to the edge. It is a 
consequence of uneven angular distribution of the radiation flux emerging 
from the stellar surface \citep{1960ratr.book.....C,1978stat.book.....M}. 
Limb darkening results as a combination of two effects: 
(i)~the density of the surface layers decreases in the outward
direction, and (ii)~temperature also drops as the distance from the
centre of the star increases. The outgoing radiance is therefore
distributed in a non-uniform manner. The actual form of the angular
distribution depends on the physical mechanism responsible for the
emission and the geometrical proportions of the source. 
The limb darkening law is widely applied also to describe radiation coming 
from an accretion disc around a black hole. In this case,
relativistic effects can significantly enhance the observed anisotropy
by bending the light rays and boosting the photon energy in the
direction of motion of the emitter. As a result of this interplay
between local physics of light emission and the global effects of the
gravitational field, different processes contribute to the final
(observed) directional anisotropy of the emission.

We consider the angular distribution of the emitted X-rays in the
context of accreting black holes. The source is an equatorial accretion
disc in which the energy is drawn by conversion of gravitational energy
of the orbiting matter gradually sinking into the black hole
\citep{1973A&A....24..337S,1981ARA&A..19..137P,2002apa..book.....F},
and electromagnetic process creating coronal flares and illuminating 
the underlying accretion disc \citep[][and further references cited therein]{1979ApJ...229..318G,2000ApJ...540L..37N,2004A&A...420....1C}.
The relevant objects are accreting supermassive black holes in nuclei of
galaxies and stellar-mass black holes in compact binary systems
\citep[see, e.g.,][for broad reviews of the subject]{2006csxs.book..157M,2007ARA&A..45..441M,2007MNRAS.382..194N}.
The form of the X-ray spectrum and, in particular, the discrete features at
$6$--$7$~keV provide a powerful tool to explore the physical properties of the
emitting region, close to the black hole horizon. The line profile
often appears to be smeared and asymmetric, with the red wing extending 
down to $\sim3$~keV or even less. Spectral modelling can help us to determine
the system parameters, namely, the black hole spin 
\citep{2005Ap&SS.300...71R,2009ApJ...697..900M}.

It is important to realise that the emission angular directionality probes 
the physical conditions of the emitting medium.
Apart from the above-mentioned temperature stratification, it depends
also on the opacity profile. The latter is typically a 
frequency-dependent quantity, hence the observed radiation also forms at
a variable physical depth within the disc. Therefore, the limb
darkening/brightening property depends on the wavelength. We
take this into account by maintaining the energy resolution in our
computations.

The resulting radiation is dominated by X-rays in a few kiloelectronvolt (keV)
band, which come from the innermost regions of the accreting system
and bear imprints of a strong gravitational field and rapid orbital motion
at their origin \citep{2000PASP..112.1145F,2008AN....329..155F}. The spectrum is a
combination of a multi-colour thermal component from the accreting medium
of the disc \citep{1974ApJ...191..499P,2005ApJS..157..335L}, the
power-law component originating from the disc corona
\citep{1979ApJ...229..318G,2004A&A...420....1C}, and the reflection and
absorption features arising on the disc surface and in the extended
corona along the line of sight 
\citep{1989MNRAS.238..729F,2003PhR...377..389R}. For active galaxies,
the thermal multi-colour black body component dominates in the UV 
and rarely reaches the sub keV energies. This component, however,
occurs at higher energy of the spectrum of black hole binaries.

This paper is organised as follows. Section~\ref{sec:model} formulates 
the model and presents its basic assumptions.
Section~\ref{sec:anisotropy1} explores how the constraints on the black
hole spin depend on the directionality of the spectrum
emerging around the iron line energy. To this end we construct the
confidence level diagrams of the $\chi^2$ statistics which
demonstrate the level of expected sensitivity of the model. First
we explore simple analytical approximations of the limb-darkening profile,
which do not depend on energy, and then we consider numerical results of
our self-consistent radiative transfer computations of both the continuum and the
spectral features between $\sim2$--$10$~keV. We demonstrate the results
by ``fitting'' the artificially simulated data, for which all parameters
are perfectly under our control, and by reanalysing the XMM-Newton
observation of the Seyfert~1 galaxy MCG--6-30-15.
We discuss our results in section~\ref{sec:discussion} and briefly
summarise them in section~\ref{sec:conclusions}.

\section{Model assumptions and requisites}
\label{sec:model}

\subsection{Gravitational field near a rotating black hole}
According to general relativity, the gravitational field of a rotating black
hole influences the intrinsic spectrum of a nearby source by changing
the energy of photons and bending the shape of light rays
\citep{1973grav.book.....M}. The assumed geometry of the source, i.e.\
the accretion disc with high orbital velocity near its inner rim, is
another factor contributing to significant anisotropy of the emission in
the observer's frame. We remark that geometrical optics provides an
appropriate framework describing these effects near astrophysical black
holes, where the wavelength of X-rays is much shorter than the typical 
length-scale of the system represented by the black hole gravitational 
radius.\footnote{The gravitational radius is
$r_{\rm{g}}=GM/c^2\approx1.48\times10^{13}\,M_8$~cm, where the central
black hole mass is expressed in terms of $M_8{\equiv}M/(10^8M_{\sun})$.
The velocity of the Keplerian orbital motion near a black hole is
$v_{_{\rm{}K}}\approx2.1\times10^{10}
(r/r_{\rm{}g})^{-1/2}{\rm{cm\;s}}^{-1}$, and the corresponding orbital
period is $T_{_{\rm{}K}}\approx3.1\times10^3(r/r_{\rm{}g})^{3/2}M_8$~s
(here we neglect an order-of-unity correction factor due to the
relativistic frame-dragging effect which, however, will be included later
in the paper).} In the innermost regions of accretion discs, bulk motion
of the gas takes place at a significant fraction of the speed of light,
indicating that special relativistic aberration and Doppler boosting
should be important. 

The gravitational field is described by the Kerr metric 
\citep[][chapt.~33]{1973grav.book.....M}:
\begin{equation} \label{metric} 
  {\rm d}s^{2} =
  -\frac{\Delta\Sigma}{A}\,{\rm d}t^{2} 
  +\frac{A\,\sin^{2}\theta}{\Sigma}\; 
   \big({\rm d}\phi-\omega\,{\rm d}t\big)^2
  +\frac{\Sigma}{\Delta}\,{\rm d}r^{2} 
  +\Sigma\,{\rm d}\theta^{2}, 
\end{equation}
where the Boyer-Lindquist spheroidal coordinates ($t,r,\theta,\phi$) and
geometrised units ($c=G=1$) are used. Metric functions are: 
$\Delta(r)=r^{2}-2r+a^2$,
$\Sigma(r,\theta)=r^{2}+a^{2}\cos^{2}\theta$,
$A(r,\theta)=(r^{2}+a^{2})^{2}-\Delta(r)\,a^{2}\sin^{2}\theta$,
and $\omega(r,\theta)=2ar/A(r,\theta)$;
$a$ denotes the specific rotational angular momentum (spin) of the central
body. These functions are assumed not to be perturbed by gravitational
effects of the external matter (accretion disc self-gravity is neglected). 
Such an assumption is well established
if we consider the inner regions of the disc, which within the distance
of several gravitational radii contains only
very low mass compared to the central black hole mass
\citep{2003MNRAS.339..937G,2004CQGra..21R...1K}.

\begin{figure}[tbh!]
\begin{center}
\includegraphics[width=0.5\textwidth]{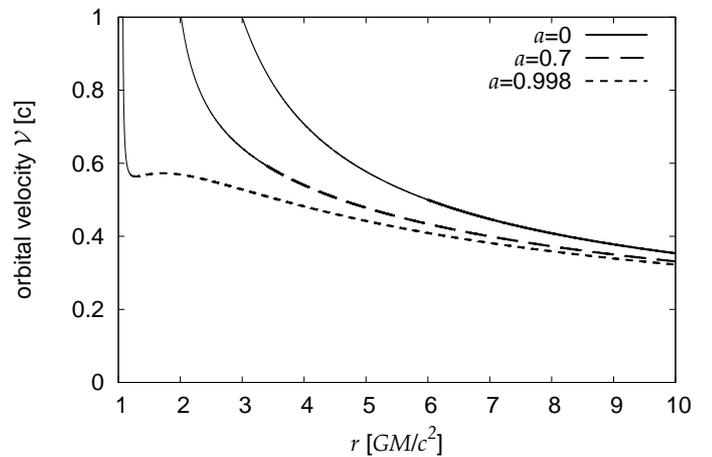}
\caption{Orbital velocity 
${\cal V}(r)$ of co-rotational motion near a rotating black hole, 
as given by formula~(\ref{vr}) for three values of the
black-hole dimensionless angular momentum parameter $a$ (spin). The thick part 
of each curve indicates the range of radii above the marginally stable
orbit, $r\geq r_{\rm ms}(a)$, where the circular motion is stable. The thin curve
indicates an unstable region at small radii.}
\label{fig1}
\end{center}
\end{figure}

Unlike weakly gravitating, non-compact bodies (e.g.\ main-sequence
stars), black holes strongly affect light passing near their horizon.
This influence bears not only the signature of black hole mass but it
also depends on its rotation via frame-dragging. Although the latter
effect is rather weak, it does shape the details of the final spectrum and
hence can be employed to measure the black hole angular momentum. The
conversion factor from the angular momentum $J_{\rm{}phys}$ (in
physical units) to the angular momentum $J$ (in geometrical units)
reads: $J=(G/c^3)J_{\rm{}phys}$. The geometrised dimension of $J$ is 
the square of the length [cm$^2$].

It is convenient to make all geometrised quantities dimensionless by 
scaling them with the appropriate power of mass $M$.
The dimensionless specific angular momentum, $a\equiv J/M^2$, spans the
range $-1\leq a\leq1$, where the positive/negative value refers 
to the motion co/counter-rotating with respect to the black hole. 
We will further assume co-rotational motion only ($a\geq0$).
We can scale all lengths by $M$ to 
reach dimensionless units. 

For the Keplerian angular velocity of the orbital motion, 
we obtain \citep{1972ApJ...178..347B}
\begin{equation}
\Omega_{_{\rm{}K}}(r)=\frac{1}{r^{3/2} + a}.
\end{equation}
For the linear velocity with respect to a locally 
non-rotating observer, we have (see Figure~\ref{fig1})
\begin{equation}
 {\cal V}(r) = \frac{r^2-2ar^{1/2}+{a}^2}{\Delta^{1/2}\left(r^{3/2}
 +{a}\right)}.
\label{vr}
\end{equation}
The velocity at the marginally stable orbit reaches a considerable
fraction of the light speed $c$ and has a similar value 
${\cal V}(r) \approx 0.5-0.6\,c$ for any value of the angular momentum.
For large spin, a small dip develops in the velocity profile 
near the horizon. Although it is an interesting feature \citep[see][]{2005PhRvD..71b4037S},
its magnitude is far too small to be recognised in our analysis 
of the complicated structure of the model $\chi^2$ space.

Notice that the above-mentioned interval of $a$-values follows from the assumption, 
traditionally accepted by astronomers, that we are dealing with a Kerr 
black hole, for which the outer horizon is located at
$r_+=1+(1-a^2)^{1/2}$. Hence, the magnitude of $a$ is thought to be less
than unity in order to have a regular horizon and avoid the case of naked 
singularity, although the latter possibility cannot be straightforwardly 
rejected just on the basis of spectra fitting.

The standard disc scenario assumes that the accretion
disc stops at the innermost stable circular orbit (ISCO, also called
the marginally stable orbit), $r_{\rm in}=r_{\rm ms}(a)$, where
\citep{1972ApJ...178..347B}
\begin{equation}
r_{\rm ms} = 3+Z_2-\big[\left(3-Z_1)(3+Z_1+2Z_2\right)\big]^{1 \over 2},
\label{velocity1} 
\end{equation}
$Z_1 = 1+(1-a^2)^{1 \over 3}[(1+a)^{1 \over 3}+(1-a)^{1 \over 3}]$ and
$Z_2 = (3a^2+Z_1^2)^{1 \over 2}$. Notice that $r_{\rm ms}(a)$ spans the
range of radii from $r_{\rm ms}=1$ for $a=1$ (the case of maximally
co-rotating black hole) to $r_{\rm ms}=6$ for $a=0$ (static black hole).
As discussed further below, Laor's (\citeyear{1991ApJ...376...90L}) model adopts
the standard disc scheme and further assumes that rotation of the black
hole is limited by an equilibrium value, $a\dot{=}0.998$, because of
capture of photons from the disc \citep{1974ApJ...191..507T}. This
implies $r_{\rm{}ms}\dot{=}1.23$.

\subsection{Black-hole signatures in accretion disc radiation}
The physical parameters of the black hole -- namely, its angular momentum
$a$ -- are imprinted in the spacetime geometry and they influence
the velocity of the orbiting material as well as the radiation propagating
through that spacetime. Thus, the observed spectrum is affected -- the 
equivalent width and the centroid energy of spectral lines
differ from the value expected on the basis of a purely Newtonian model
\citep[e.g.][and further references cited
therein]{2000MNRAS.312..817M,2001MNRAS.321..605D,2003MNRAS.344L..22M,2004MNRAS.348.1415V}.
In other words, the relativistic effects offer the possibility to reveal
the presence of a compact body in the centre of the accretion disc, and
to determine its parameters. This assertion is based on an implicit
assumption that we can constrain the local physics forming the intrinsic 
spectrum of the source with sufficient accuracy. 

The effects of strong gravity may be concealed from us if the
intrinsic spectrum of the accretion disc is not known well enough. 
This is particularly an issue with respect to the angular
momentum determination because the geometrical effect of frame-dragging
due to a rotating black hole is a rather subtle one. In order to assess
potential inaccuracies, we investigate how sensitive the
determination of the $a$ parameter is to the angular distribution of the
disc emission, i.e., the limb-darkening (or the limb-brightening) law in
the spectral range of $2\lesssim E\lesssim10$~keV.

Numerical as well as semi-analytical methods were developed and combined to
study the spectral features from relativistic accretion discs,
especially the profiles of the iron-line complex between $6$--$7$ keV.
An important practical attribute from the view point of applicability of any
particular method is its implementation within the XSPEC data analysis
package \citep{1996ASPC..101...17A}. In this context, over almost two decades the most
widely used model of the relativistic disc spectral line is the one by
\citet{1991ApJ...376...90L}, which includes the effects
of a rapidly rotating Kerr black hole. The \textscown{laor}
model sets the dimensionless angular momentum $a$ to the canonical value
of $a=0.998$, so it cannot be the subject of a rigorous data fitting
procedure \citep[although see][]{SvobodaWDS}. 

Although the mathematical properties of the extreme
Kerr spacetime ($a^2=1$) are significantly different from the sub-extreme
case ($a^2<1$), the graphs of the redshift factor as well as the emission
angle are practically identical. Therefore we do not expect any detectable
differences of the observed spectra between these two cases. Secondly,
the main effect on the standard accretion disc spectrum and the
relativistic broad line formation arises from the assumed link between
the inner edge of the disc and the black hole spin parameter
\citep{2008ApJ...675.1048R}. 
\citet{SvobodaWDS} have examined the accuracy of the spin fitting
using the \textscown{laor} model. They demonstrate that, although such an approach is not
self-consistent, the accuracy of the resulting $a$-values is actually quite
good.

\citet{2004ApJS..153..205D} have relaxed the limitation of the \textscown{laor}
model regarding the fixed value of the spin. These authors
allowed for $a$ to be fitted to the data by $\chi^2$ minimisation in a
suite of \textscown{ky} models. Furthermore, \citet{2004MNRAS.352..353B} 
have developed the \textscown{kdline}, and \citet{2006ApJ...652.1028B} have
developed the \textscown{kerrdisk} models which are endowed with similar
functionality. The latter authors performed useful tests
demonstrating that \textscown{ky} and \textscown{kerrdisk} give compatible results
when they are set to equivalent parameter values. Numerical codes have
been developed independently by several other groups
\citep{2000MNRAS.312..817M,1993A&A...272..355V,2004AdSpR..34.2544Z,2004A&A...424..733F,2005MNRAS.363..177C} using different techniques. The \textscown{laor} kernel has been
recently applied by \citet{2007PASJ...59S.315M} who reanalysed the 
prototypical source MCG--6-30-15 using Suzaku observations to determine 
the relativistic blurring and to measure the black hole parameters. 
Their results confirm the high value of the spin.

\citet{2009ApJ...697..900M} applied the modelling of relativistic spectral 
features to estimate the black hole spin from a sample of stellar-mass black 
holes. They used the joint constraints from both the disc reflection and continuum
and found evidence for a broad range of black hole spin parameters.

The angular emissivity law, ${\cal M}(\mu_{\rm e},r_{\rm e},E_{\rm e})$,
defines the distribution of the intrinsic intensity outgoing from each
radius $r_{\rm e}$ of the disc surface with respect to the perpendicular
direction. The emission angle $\theta_{\rm e}=\arccos\mu_{\rm e}$ is
measured from the disc normal direction to the equatorial plane, in
the disc co-moving frame, i.e.\ in the local Keplerian frame orbiting
with the angular velocity $\Omega_{_{\rm{}K}}(r)$. Likewise the
intrinsic energy $E_{\rm e}$ is measured with respect to the local frame. The
total disc emission can be  written in the form of product 
\begin{equation}
I(r_{\rm
e},\mu_{\rm e},E_{\rm e})\equiv {\cal R}(r_{\rm e})\,\mu_{\rm e}\,{\cal M}(\mu_{\rm
e},r_{\rm e},E_{\rm e})\,{\cal E}(E_{\rm e}), 
\label{iloc}
\end{equation}
where the radial part is
well approximated by a power law, 
\begin{equation}
{\cal R}(r_{\rm e})=r_{\rm e}^{-q}\qquad (q=\mbox{const}),
\label{rloc}
\end{equation}
and $\mathcal{E}(E_{\rm e})$ is the energy dependence of the locally emitted radiation.
When needed we will also use the broken power-law profile which is a combination of
two profiles (\ref{rloc}) matched at the transition radius, also
coded in the \textscown{ky} set of models.

The redshift factor $g$ and the emission angle $\theta_{\rm e}$ are
\begin{equation} 
 g  =  \frac{{\cal{C}}}{{\cal{B}}-r^{-3/2}\xi}, 
 \quad 
 \theta_{\rm e}  =  \arccos\frac{g\sqrt{\eta}}{r}, 
\label{gtheta}
\end{equation}
where ${\cal{B}}=1+{a}r^{-3/2}$, ${\cal{C}}=1-3r^{-1}+2{a}r^{-3/2}$;
$\xi$ and $\eta$ are constants of motion connected with symmetries of
the Kerr spacetime. 

The observed radiation flux is then obtained by integrating the
intrinsic emission over the entire disc surface, from the inner edge ($r=r_{\rm in}$) to the outer
edge ($r=r_{\rm out}$), weighted by the transfer function
$T(r_{\rm e},\phi_{\rm e},\theta_{\rm o},a)$ determining the impact of relativistic
energy change (Doppler and gravitational) as well as the lensing effect
for a distant observer directed along the inclination angle $\theta_{\rm o}$
\citep[see][]{1975ApJ...202..788C,1989PASJ...41..763A,1992MNRAS.259..569K,2006AN....327..961K}. 


Given the high
velocity of the orbital motion and the strong-gravity light bending near the 
black hole, the effect of directional anisotropy of the local emission is enhanced.
For this reason it is important to describe the angular
distribution in a correct manner; ad hoc choices of the limb-darkening
law may lead to errors in the determination of the model best fit
parameters, including the inaccuracy in $a$ parameter which are
difficult to control, or they may prevent us from estimating the statistical
confidence of the model. In the case of black hole accretion discs, this
complication becomes important because the aberration, beaming and
light-bending effects grow rapidly towards the inner edge of the disc.
Therefore, even a small discrepancy between the assumed and the correct
angular emissivity profiles becomes greatly enhanced in the observer
frame.

\begin{figure*}[tbh!]
\begin{center}
\includegraphics[width=0.49\textwidth]{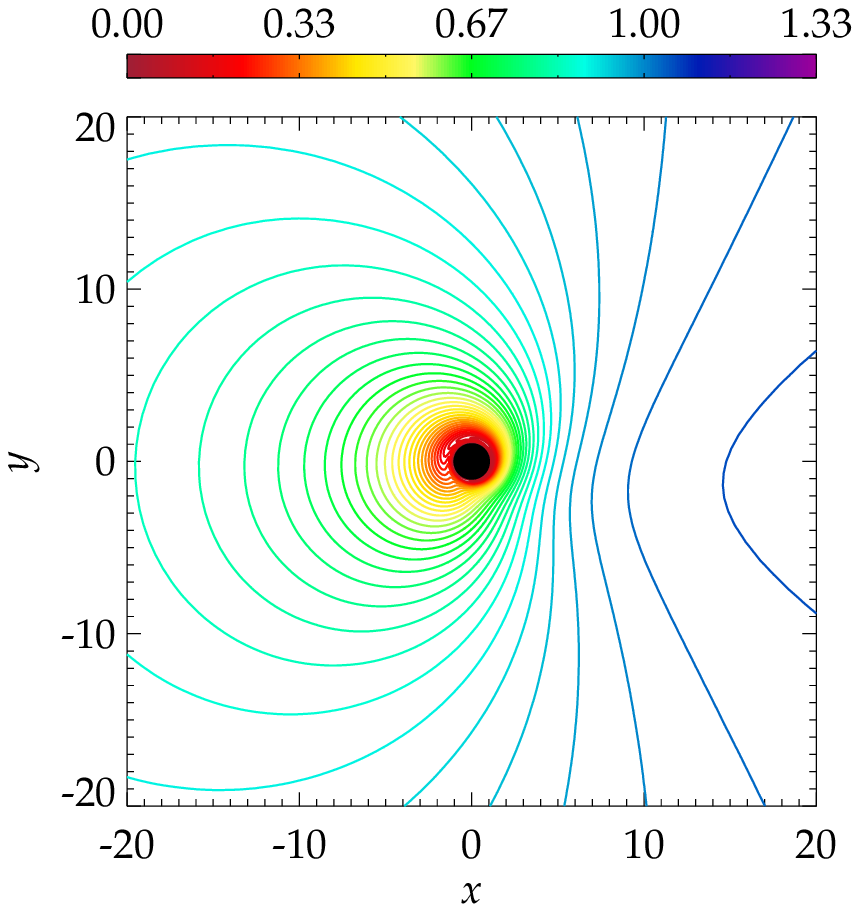}
\includegraphics[width=0.49\textwidth]{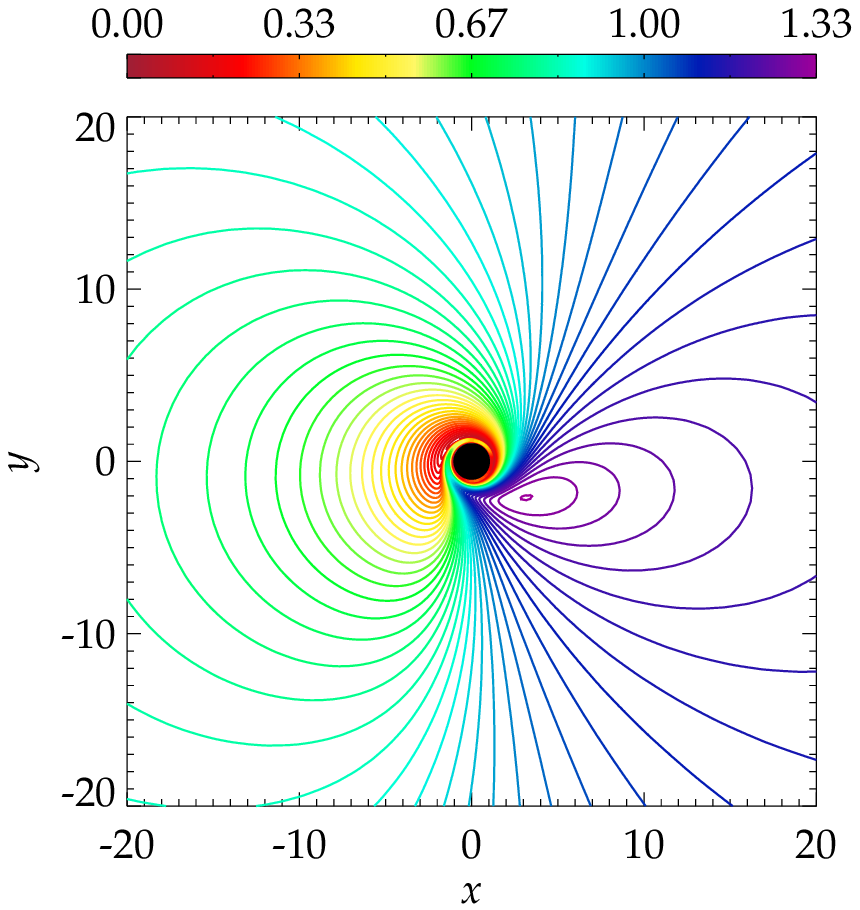}
\caption{Contours of redshift factor $g(r,\phi)$ near a 
maximally rotating black hole, $a=1$, depicted in the equatorial plane $(x,y)$. 
The black hole and the accretion disc rotate counter clock-wise.
The inner region is shown up to $r=20$ gravitational radii
from the black hole (denoted by dark circle of unit radius
around the centre). A distant observer is located towards the top
of the figure. Two cases of different observer inclinations
are shown.
Left: $\theta_{\rm o}=30$~deg. Right: $\theta_{\rm o}=70$~deg.
The colour bar encodes the range acquired by $g(r,\phi)$, 
where $g>1$ corresponds to blueshift (approaching side of the disc),
while  $g<1$ is for redshift.}
\label{fig2}
\end{center}
\end{figure*}

\begin{figure*}[tbh!]
\begin{center}
\includegraphics[width=0.49\textwidth]{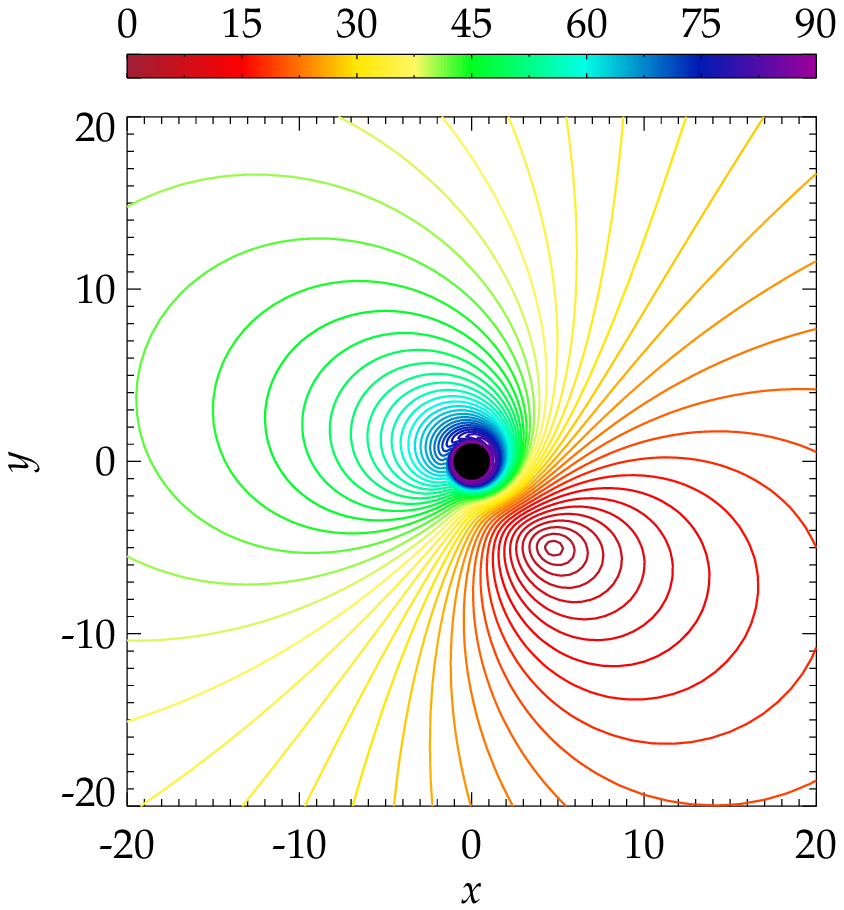}
\includegraphics[width=0.49\textwidth]{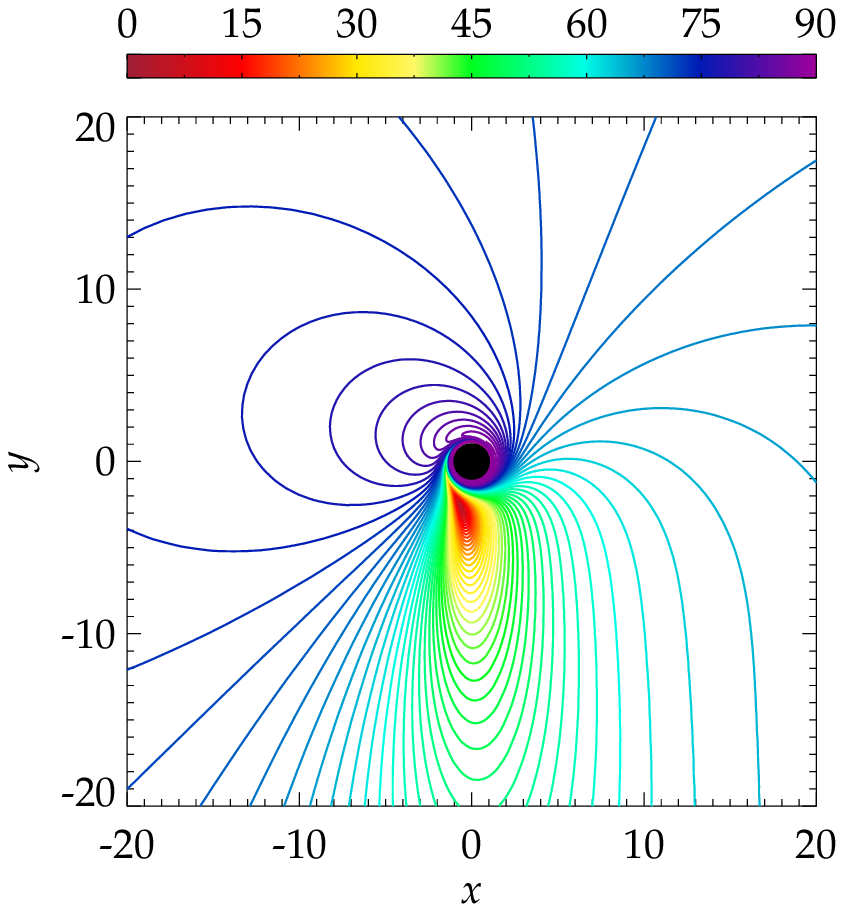}
\caption{Contours 
of the local emission angle, $\theta_{\rm e}(r,\phi)$. 
The same parameters as in the previous figure. The local emission angle spans 
the entire range, from $0$ to $90$ degrees. This is due to the combined effects 
of aberration and light bending which grow greatly near the inner rim 
of the disc. In the inner region the photons are boosted in the direction of 
rotation and they emerge along grazing angles.
Asymptotically, $\theta_{\rm e}(r,\phi)\rightarrow \theta_{\rm o}$
for $r\rightarrow\infty$.}
\label{fig3}
\end{center}
\end{figure*}

\subsection{Effects of the emission angular directionality}
In optics, Lambert's cosine law describes an emitter producing 
a radiation intensity that is directly proportional to $\mu_{\rm e}$.
Lambertian surfaces exhibit the same apparent radiance when viewed from
any angle $\theta_{\rm o}$. Likewise, Lambertian scattering refers to
the situation when the surface radiates as a result of external
irradiation by a primary source and the scattered light is distributed
according to the same cosine law. This is, however, a very special
circumstance; directionality of the emergent light is sensitive to
the details of the radiation mechanism. For example, the classical
result of the Eddington approximation for stellar atmospheres states that
the effective optical depth of the continuum is $\tau=\frac{2}{3}$, and
so the emergent intensity is described by the limb-darkening law,
$I(\mu_{\rm e})\propto\mu_{\rm e}+\frac{2}{3}$. 

In the case of a fluorescence iron line produced by an illuminated plane-parallel 
slab, the angular distribution was investigated by various authors
\citep{1978ApJ...223..268B,1991MNRAS.249..352G,1993ApJ...413..680H,1994MNRAS.267..743G,2000ApJ...540..143R}.
In that case a complicated interplay arises among the angular
distribution of the primary irradiation, reflection and scattering in
the disc atmosphere. Several authors pointed out that it is essential for the reliable
determination of the model parameters to determine the angular directionality of
the broad line emission correctly. \citet{2000MNRAS.312..817M} noted, by
employing the lamp-post model, that ``{\em ...the broadening of the observed
spectral features is particularly evident when strongly anisotropic
emissivity laws, resulting from small $h$ [i.e., the lamp-post elevation
above the equatorial plane], are considered.}`` 

The important role of
the emission angular directionality was clearly spelled out by
\citet{2004MNRAS.352..353B}: ``{\em ...the angular emissivity law (limb
darkening or brightening) can make significant changes to the derived line
profiles where light bending is important``} (see their Fig.~9--13).
Similarly, \citet{2004ragt.meet...33D,2004ApJS..153..205D} and 
\citet{2005ragt.meet...29B} compared the relativistic broad lines produced
under different assumptions about the emission angular directionality. 
However, to verify the real sensitivity of the models to the
mentioned effect of directionality, it is necessary to connect the
radiative transfer computations with the spectral fitting procedure, and to
carry out a systematic analysis of the resulting spectra, taking into
account both the line and the continuum in the full relativistic regime.
Here we report on our results from such computations.

\citet[][sec. 4.3]{2004MNRAS.352..205R} argue that
the combined effect of photoelectric absorption in the disc and Compton 
scattering in the corona more affect the iron line photons emerging
along grazing light rays than continuum photons. They conclude that
the line equivalent width should be diminished for observers viewing the
accretion disc at high inclination angles. Such a trend can be seen also
in the lamp-post model of \citet{1992A&A...257...63M} and \citet[][see their
Figure~11]{2000MNRAS.312..817M}. However, in the latter work this diminution
is less pronounced when we compare it with the case of intrinsically isotropic
emissivity.

More recently,
\citet{2008MNRAS.386..759N} studied the effect of different
limb-darkening laws on the iron-line profiles. They point out that the
role of emission directionality can be quite significant once the radial
emissivity of the line is fixed with sufficient confidence. However,
this is a serious assumption. In reality, the radial emissivity is not
well constrained by current models.

The angular dependence of the outgoing radiation is determined by the
whole interconnection of various effects. We describe them in more
detail below (sec.\ \ref{sec-titan-noar-modelling}).
Briefly, the conclusion is such that realistic models require numerical
computations of the full radiative transfer.
We have developed a complete and consistent approach to such radiative
transfer computations in the context of the broad iron-line modelling {\em
together} with the underlying continuum computations. As described 
in considerable detail below, we performed the extensive
computations which are necessary in order to reliably determine the impact
of the emission angular anisotropy on spectral fitting results (namely, on
the determination of the black hole angular momentum). In particular, we
describe detailed results from the investigation of the model goodness
(by employing an adequate statistical analysis of the complicated $\chi^2$
parameter space). Such the analysis has not been performed so far in previous
papers because detailed stepping through the parameter space and proper
re-fitting of the model parameters was not possible due to the enormous
complexity of the models and extensive computational costs.

\begin{figure*}[tbh!]
\begin{center}
\includegraphics[width=0.49\textwidth]{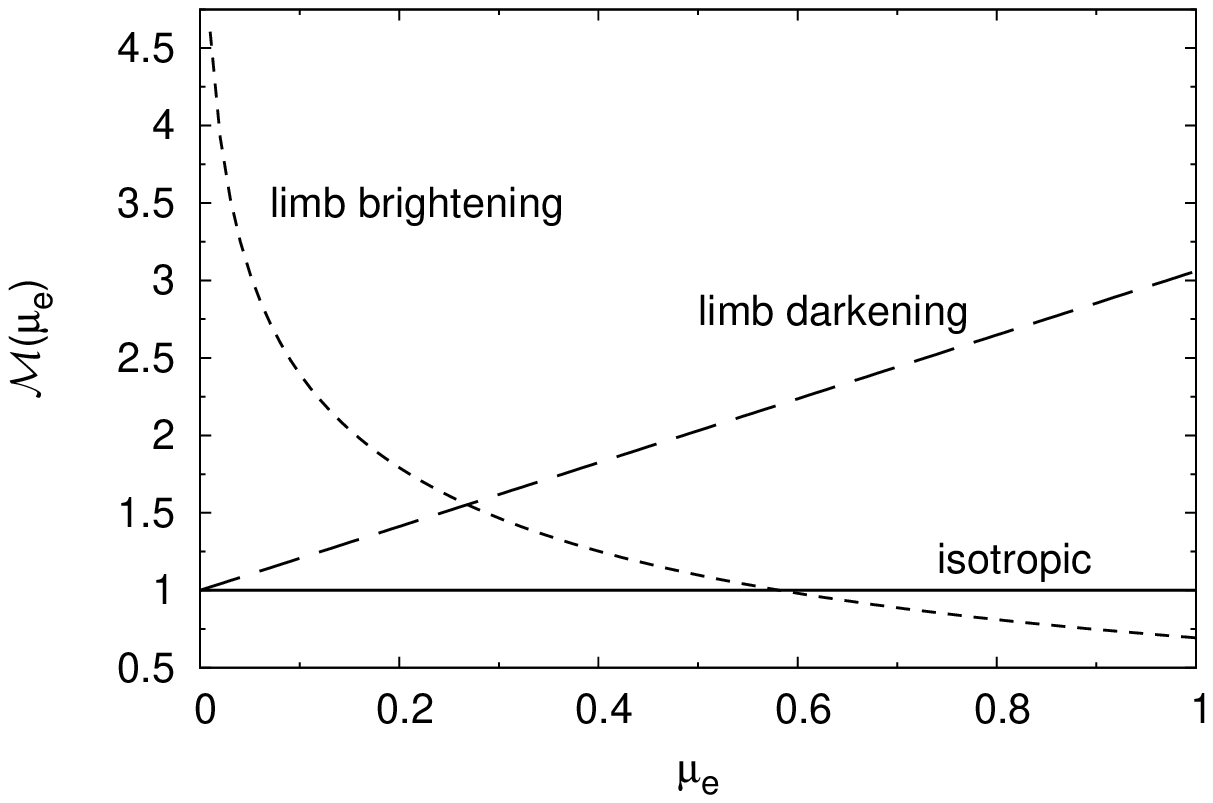}
\hfill~\includegraphics[width=0.49\textwidth]{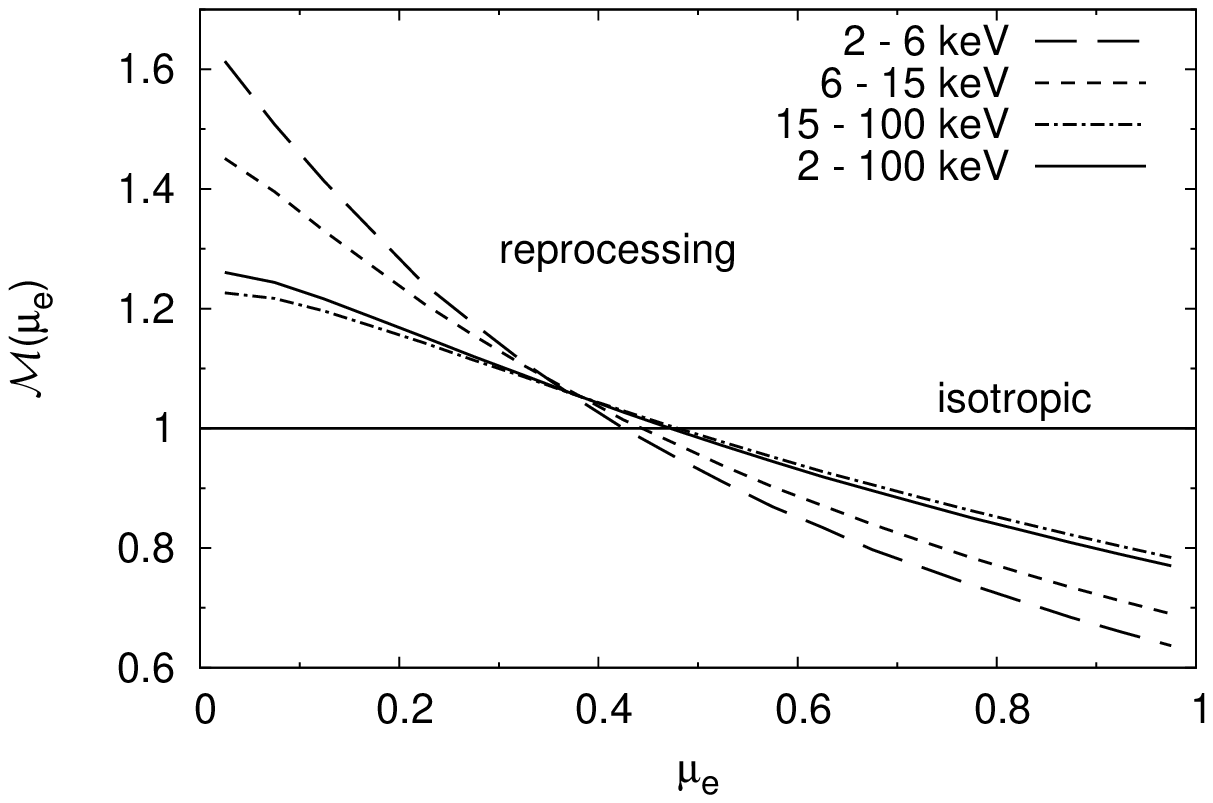}
\caption{Left: directional distribution of the intrinsic emissivity
(see sec.~\ref{sec:anisotropy31} for details). Right: an example of the 
directional distribution from the numerical computations, showing
results from the reprocessing model with the continuum photon index $\Gamma=1.9$,
integrated over $2$--$100$ keV energy range (solid line), and in several
different energy sub-ranges (line styles are indicated in the inset). 
The latter graph demonstrates the presence of 
limb-brightening effect in comparison with respect to the isotropic emission. 
This effect is clearly visible as a function of the 
emission angle $\mu_{\rm e}$, although its magnitude is smaller than the 
limb-brightening approximation
used in the left panel. See sec.~\ref{sec-titan-noar-modelling}
for details.}
\label{fig4}
\end{center}
\end{figure*}

In regular stars and their accretion discs, the relativistic effects 
hardly affect the emerging radiation. The situation is very
different in the inner regions of a black hole accretion disc,
where the energy shift and gravitational lensing are significant.
The observed  signal can be boosted or
diminished by the Doppler effect combined with gravitational redshift:
$I(\mu_{\rm e})/I(\mu_{\rm o})\equiv g^3$, where the $g$ values span 
more than a decade. Figure~\ref{fig2} shows the
redshift factor $g$ near the black hole, proving that general relativistic
effects indeed change the photon energy significantly.

As mentioned above, many authors have adopted the defining choice
\citep{1960ratr.book.....C,1991ApJ...376...90L} of the cosine profile
for the line angular emissivity: ${\cal M}(\mu_{\rm e},r_{\rm e},E_{\rm
e})=  1+2.06\,\mu_{\rm e}$. This relation describes the
energy-independent limb-darkening type of profile. However, the choice
is somewhat arbitrary in the sense that the physical assumptions behind
this law are not satisfied at every radius over the entire surface of 
the accretion disc. It has been argued that the limb-darkening
characteristics need to be modified, or even replaced
by some kind of limb brightening in the case of X-ray irradiated disc
atmospheres with Compton reflection
\citep{1993ApJ...413..680H,1994MNRAS.267..743G,1994MNRAS.266..653Z}. The latter should
include the energy dependence, as the Compton reprocessing of the
reflected component plays a significant role. The angle-dependent 
computations of the Compton reflection demonstrate these effects
convincingly \citep{2004A&A...420....1C}.  Indeed, the same effect is
seen also in our computations, as shown in the right panel of Figure~\ref{fig4}.
The increase of emissivity
with the emission angle strongly depends on the ionisation state of the
reflecting material, so the actual situation can be quite complex 
\citep{2007A&A...475..155G}. 

A question arises of whether the current determinations of the black hole
angular momentum might be affected by the uncertainty in the actual
emissivity angular distribution, and to what degree. In fact, this may
be of critical concern when future high-resolution data become available
from the new generation of detectors. We have therefore carried out a
systematic investigation using the \textscown{ky} code to reveal how sensitive
the constraints on the dimensionless $a$ parameter are with respect to
the possible variations in the angular part ${\cal M}(\mu_{\rm e})$ of
the emissivity.

Figure~\ref{fig3} shows the contours of the local
emission angle $\theta_{\rm e}$ from the accretion disc, taking into account their
distortion by the central black hole and assuming that the emitted photons
reach a distant observer at a given view angle $\theta_{\rm o}$. Although
the contours are most dramatically distorted near the horizon, where
both aberration and light bending effects grow significantly, the emission
angle is visibly different from the observer inclination even quite far
from the horizon, at a distance of several tens $r_{\rm{}g}$. This is
mainly due to special-relativistic aberration which decays slowly with
the distance as the disc obeys Keplerian rotation at all radii.

In summary, figures \ref{fig2}--\ref{fig3} demonstrate the main attributes of the
photon propagation in black hole spacetime relevant to our problem: the
energy shift and the direction of emission depend on the view angle of
the observer as well as on the angular momentum of the black hole.
Notice that, near the inner rim, the local emission angle is indeed
highly inclined towards the equatorial plane where at the same time the
outgoing radiation is boosted. These effects are further enhanced by
gravitational focusing, which we also take into account in our
calculations.

It should be noted that the energy shifts and emission angles
near a black hole, such as those shown in Figs. 2--3, have been
studied by a number of authors in mutually complementary ways. 
The figures shown here have been produced by plotting directly the
content of FITS format files that are encoded in the corresponding
\textscown{ky} XSPEC routines (see \citet{2004astro.ph.11605D}, where an
atlas of contour plots is presented for different inclinations and spins).
Analogous figures were also shown in \cite{2004MNRAS.352..353B}, who
depict the dependence of the cosine of the emission angle on the
energy shift of the received photon and the emission radius.

The above quoted papers concentrated mainly on the discussion of the
energy shifts and the emission directions of the individual photons,
or they isolated the role of relativistic effects on the predicted
shape of the spectral line profile. They clearly demonstrated that
the impact of relativistic effects can be very significant. However,
what is still lacking is a more systematic analysis which would
reveal how these effects, when integrated over the entire source,
influence the results of spectra fitting. To this end one needs to
perform an extensive analysis of the model spectra including the
continuum component, match the predicted spectra to the data by
appropriate spectra fitting procedures, and to investigate the
robustness of the fit by varying the model parameters and exploring
the confidence contours.

One might anticipate the directional effects of the local emission
to be quite unimportant. The argument for such an expectation
suggests that the role of directionality should grow with the source
inclination, whereas the unobscured Seyfert 1 type AGNs (where the
relativistically broadened and skewed iron line is usually expected)
are thought to have only small or moderate inclinations. However,
this qualitative trend cannot be used to quantitatively constrain
the model parameters and perform any kind of precise analysis,
needed to determine the black hole spins from current and future
high-quality data. Such an analysis has not been performed so far,
and we embark on it here for the first time.

\section{Iron K$\alpha$ line band examined with different directionalities 
of the intrinsic emissivity}
\label{sec:anisotropy1}

\begin{figure*}[tbh!]
\begin{center}
\includegraphics[width=0.45\textwidth]{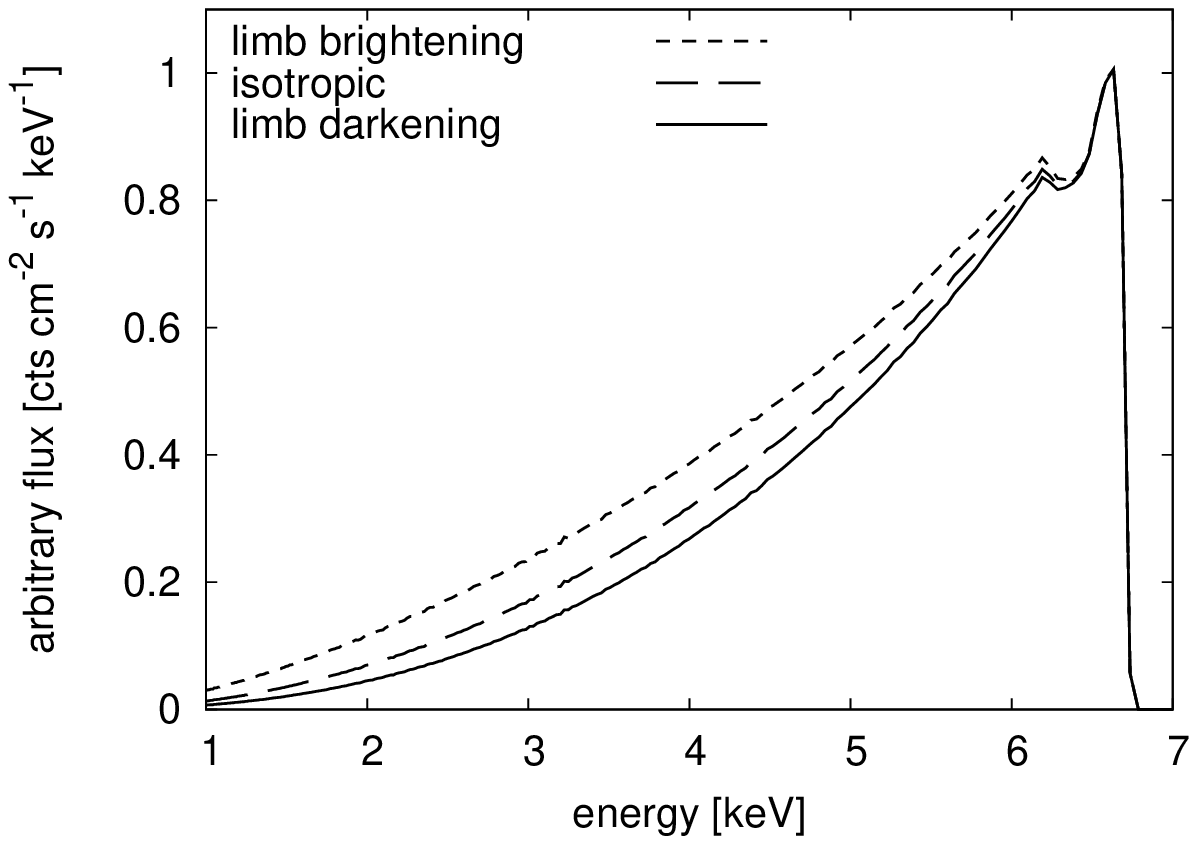}
\hfill
\includegraphics[width=0.45\textwidth]{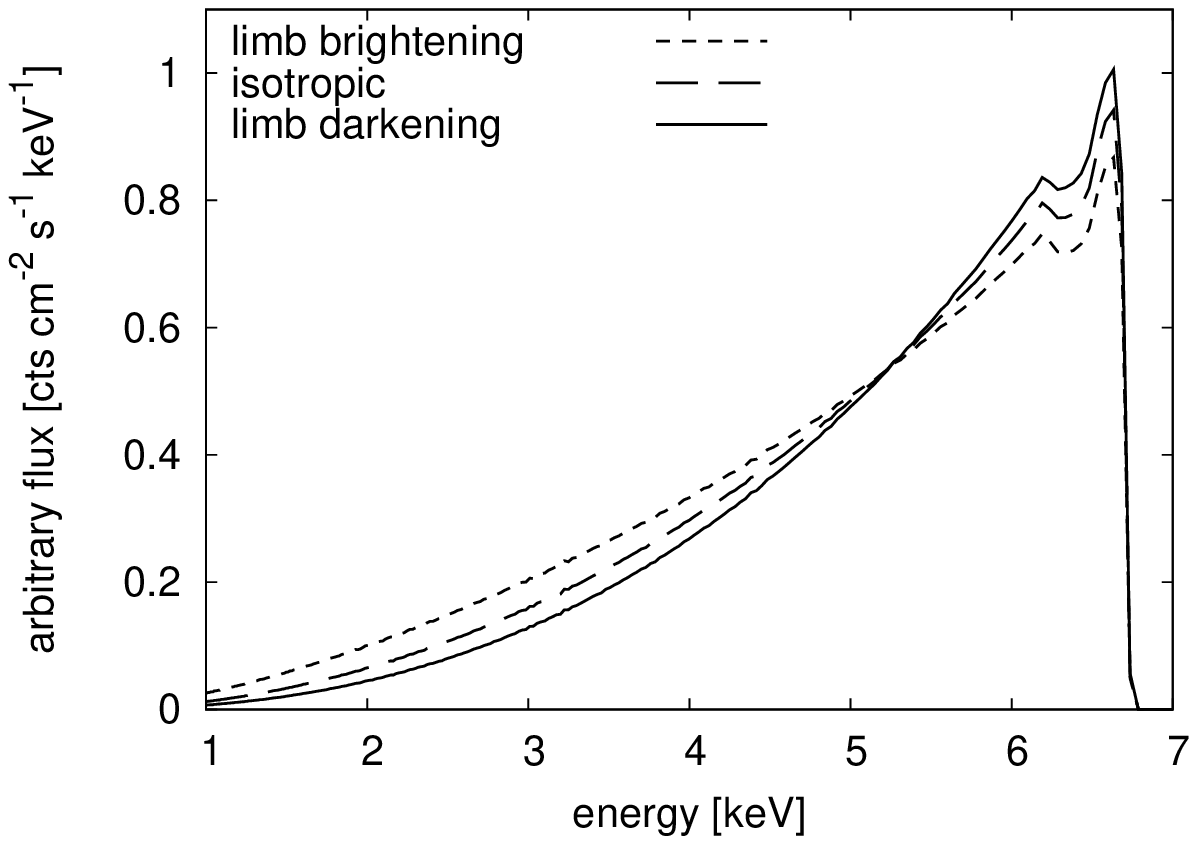}
\caption{Left: theoretical
profiles of the relativistic line (the \textscown{kyrline} model, without continuum),
corresponding to the three cases in the left panel of Fig.~\ref{fig4}.
The lines are normalised with respect to the height of the blue peak.
Model parameters are $a=0.9982$, $q=3$, $r_{\rm in}=r_{\rm ms}(a)=1.23$, $r_{\rm out}=400$,
$\theta_{\rm o}=30\deg$, $E_0=6.4$~keV. 
Right: the same as in the left panel, but with the normalisation set in such a way that
the radiation flux is identical in all three profiles.}
\label{fig4a}
\end{center}
\end{figure*}

\begin{figure*}[tbh!]
\begin{center}
\includegraphics[width=0.49\textwidth]{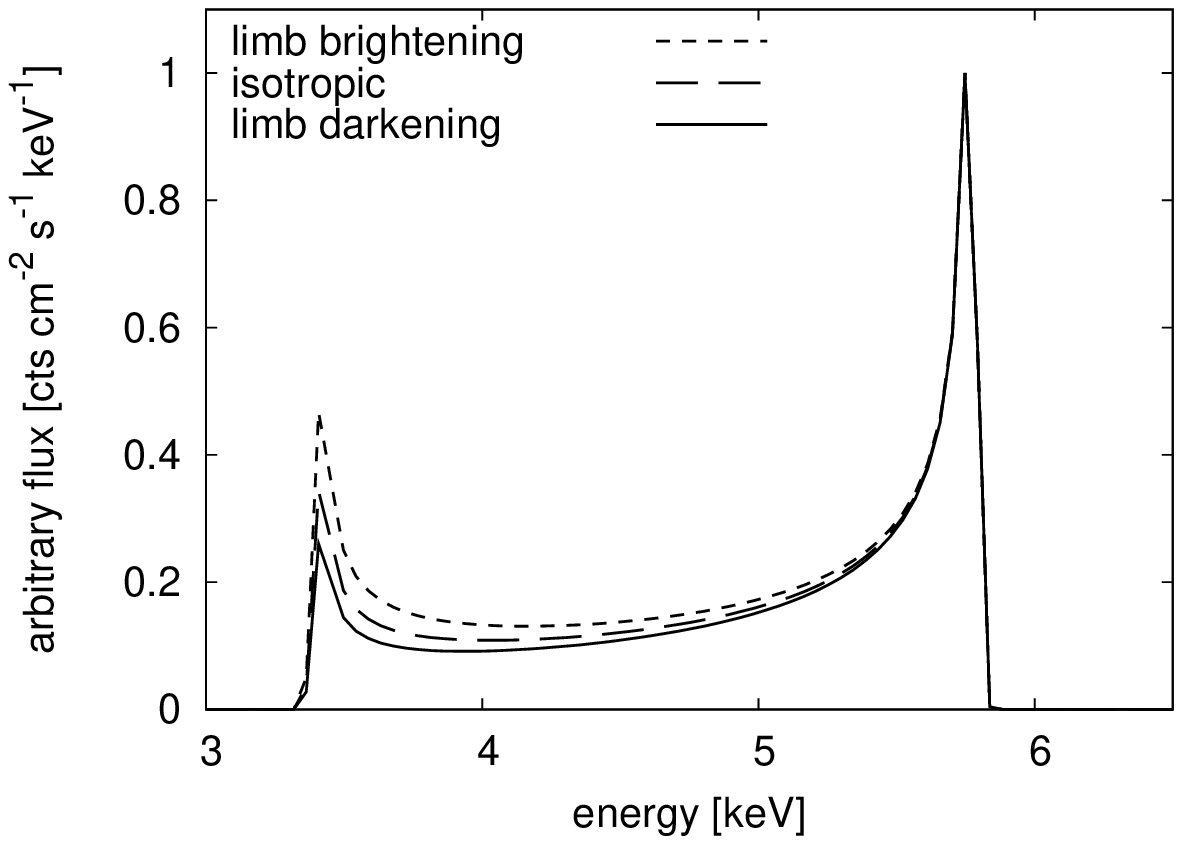}
\hfill
\includegraphics[width=0.47\textwidth]{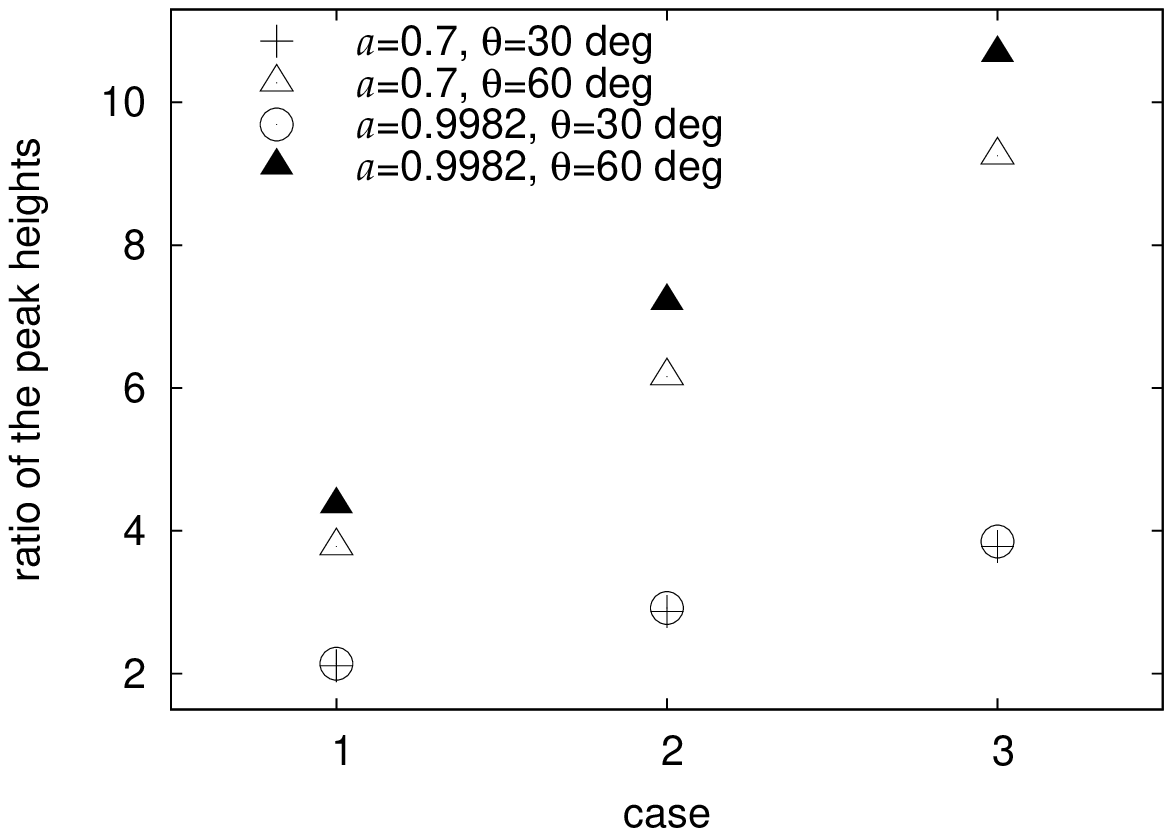}
\caption{Left: The same as in Fig.~\ref{fig4a}, but for a narrow ring 
with $r_{\rm in} \doteq 4.7$, $r_{\rm out} \doteq 4.8$ (other parameters are
$a=0.9982$,  $q=3$, $\theta_{\rm o}=30\deg$, $E_0=6.4$~keV). 
The lines are normalised with respect to the height of
the blue peak. Right: The ratio of the two peak heights for the three
cases of the emission angular directionality according to
eq.~(\ref{cases}) and for the two values of the spin and the inclination angle
(crosses and circles are for $\theta_{\rm o}=30\deg$; empty and filled
triangles are for $\theta_{\rm o}=60\deg$).} 
\label{fig4b}
\end{center}
\end{figure*}

\begin{figure*}[tbh!]
\begin{center}
\includegraphics[width=0.49\textwidth]{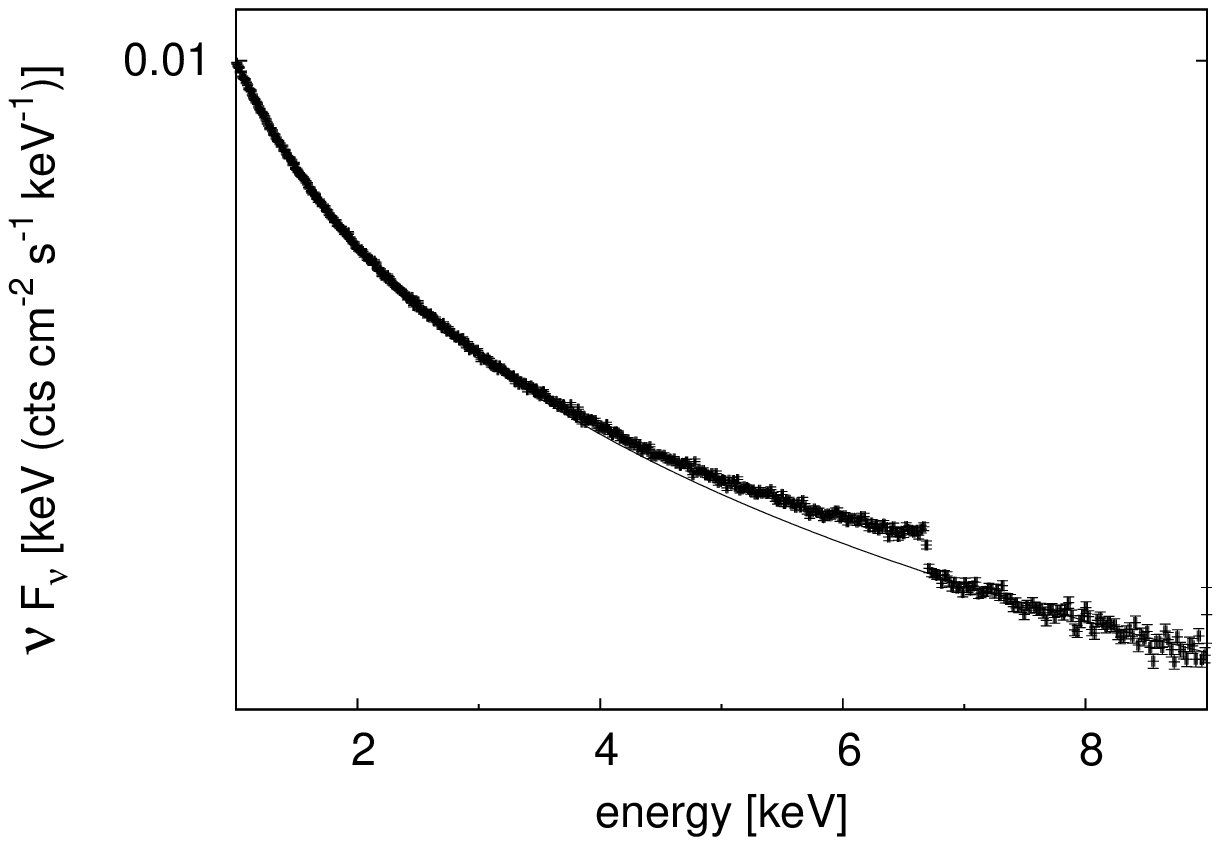}
\hfill
\includegraphics[width=0.49\textwidth]{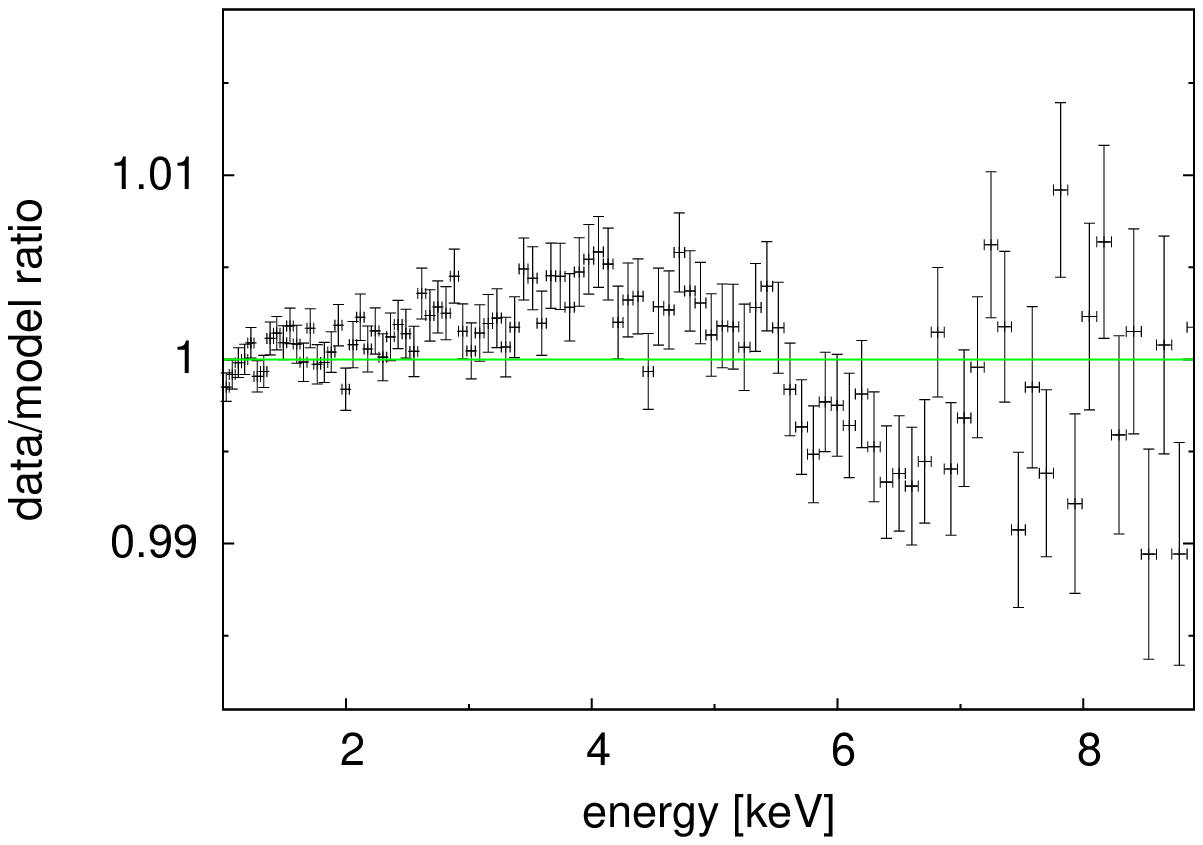}
\caption{Left: unfolded spectrum generated by the 
\textscown{powerlaw + kyrline} model with parameters of the line identical as
in previous figure. We assumed the limb-brightening profile, i.e.\ Case~1
of eq.~(\ref{case123}). Right: the ratio of the model on the left to
the same model with the Case~3 (limb-darkening) profile.
The normalisation of the line was allowed to vary during the fit. The
$\chi^{2}_{\rm red}$ changed from $1.04$ to $1.30$ between the two cases.
The details about the simulation of the data points are the same as described
in the section \ref{simulated}.}
\label{fig5}
\end{center}
\end{figure*}

\subsection{Approximations to the angular emission profile}
\label{sec:anisotropy31}
We describe the methodology which we adopted in order to explore the
effects of the spectral line emission directionality. To this end we
first employ simple approximations, neglecting any dependence on the
photon energy and the emission radius. We set the line intrinsic
emissivity from the planar disc to be described by one of the following
angular profiles,
\begin{equation}
\begin{array}{l}
\mbox{Case 1:}\quad \\ \mbox{Case 2: \rule[-1.3em]{0pt}{3em}}\quad \\ \mbox{Case 3:}\quad
\end{array}
{\cal M}(\mu_{\rm e})= \left\{
\begin{array}{l}
\ln(1+\mu_{\rm e}^{-1}) \quad\mbox{\citep{1993ApJ...413..680H}}\\ 1 \rule[-1.3em]{0pt}{3em} \quad \mbox{(locally isotropic emission)}\\ 1+2.06\,\mu_{\rm e}\quad\mbox{\citep{1991ApJ...376...90L}} 
\end{array}
\right.
\label{cases}
\label{case123}
\end{equation}
with the radial profiles being a power law ${\cal R}(r_{\rm e})=r_{\rm e}^{-q}$, 
and ${\cal E}(E_{\rm
e})=\delta(E_{\rm e}-E_0)$. The three cases correspond, respectively, to
the limb-brightened, isotropic, and limb-darkened angular profiles of the 
line emission. 

The limb-brightening law by \citet{1993ApJ...413..680H} describes the
angular distribution of a fluorescent iron line emerging from an
accretion disc that is irradiated by an extended X-ray source. The
relation was obtained from geometrical considerations and agrees well
with more detailed Monte-Carlo computations \citep{1991MNRAS.249..352G, 1992A&A...257...63M}. 
The physical circumstances relevant for the limb-darkening law are
different, and we include this case mainly because it is implemented
in the {\sc laor} model and frequently used in the data analysis. 
The isotropic case, dividing all limb-brightening and limb-darkening 
emissivity laws, is included in our analysis for comparison.
Notice that an extra geometrical factor $\mu_{\rm e}$ is involved 
due to simple geometrical projection of the disc surface in eq.~(\ref{iloc}). 

The radial profile of the emission is set to a unique 
power law, eq.~(\ref{rloc}), over the entire range of radii across the disc.
The directionality formula (\ref{case123}) of the intrinsic
emissivity and the resulting spectral profiles are illustrated in
Fig.~\ref{fig4} (left panel). Naturally, more elaborate and accurate approximations
have been discussed in the literature for some time. For example, \citet{1994MNRAS.267..743G}
in their eq.~(2) include higher-order terms in $\mu_{\rm e}$ to describe
the X-ray reprocessing in the single-scattering Rayleigh approximation. However,
at this stage the first-order terms are sufficient for us to demonstrate 
the differences between the three cases. Later on we will proceed towards
numerical radiation transfer computations that are necessary to derive
realistic profiles of the emission angular distribution and to keep their 
energy dependence.

The dependence of the spectral profiles on the angular
distribution ${\cal M}(\mu_{\rm e})$ of the 
intrinsic emission is shown in figures \ref{fig4a}--{\ref{fig4b}.
It is apparent from Figure~\ref{fig4a} that the limb brightening case 
makes the line profile broader (in the left panel)
and the height of the blue peak lower (in the right panel) 
than the limb darkening case for the same set of parameters, 
which can consequently lead to discrepant evaluation of the spin.
While Figure~\ref{fig4a} shows the line profile for an extended
disc, Figure~\ref{fig4b} deals with a narrow ring. In this
case, a typical double-horn profile develops. Although the energy of the
peaks is almost entirely insensitive to the emission angular directionality,
the peak heights are influenced by the adopted limb darkening law. 
In the right panel of Figure~\ref{fig4b}, the ratio of the two peaks
is shown for the three cases of the emission angular directionality according to
eq.~(\ref{cases}). The influence of the directionality is apparent 
and it is comparable to the effect of the inclination angle. 
The spin value has only little impact for large inclinations.

In general, we notice that the extension and prominence of the
red wing of the line are indeed related to the intrinsic emission
directionality. This was examined in detail by \citet{2004MNRAS.352..353B},
who plotted the expected broad-line profiles for different angular emissivity
and explored how the red wing of the line changes as a result
of this undetermined angular distribution.
Similarly, the inferred spin of the black hole
must depend on the assumed profile to a certain degree.
This is especially so if the spin is determined by the extremal
redshift of the line's red wing. It has been recently argued
\citep{2008MNRAS.386..759N} that the
directionality effect could be a real issue for the accuracy of line 
fitting if other parameters (such as the radial emissivity profile) 
are well constrained. 

However, simple arguments are
insufficient to assess how inaccurate the spin determination might be
in realistic situations, as
the spectral fitting procedure involves several components extending
over a range of energy above and below the iron-line band.
We illustrate this in Figure~\ref{fig5}, noticing that
the different prescriptions produce very similar results outside the line
energy, but they do differ at the broad line energy range. 
The theoretical (background-subtracted) profiles of the relativistic line
cannot alone be used to make any firm conclusion about the error of the best
fit parameters that could result from the poorly known angular emissivity.
To this end one has to study a consistent model of the full spectrum.
With a real observation, the sensitivity to the
problem of directionality (as well as any other uncertainty inherent in
the theoretical model) will depend also on the achieved resolution, 
energy binning and the error bars of the data.

\subsection{Example: MCG--6-30-15 reanalysis}

\begin{figure*}[tbh!]
\begin{center}
\includegraphics[width=0.49\textwidth]{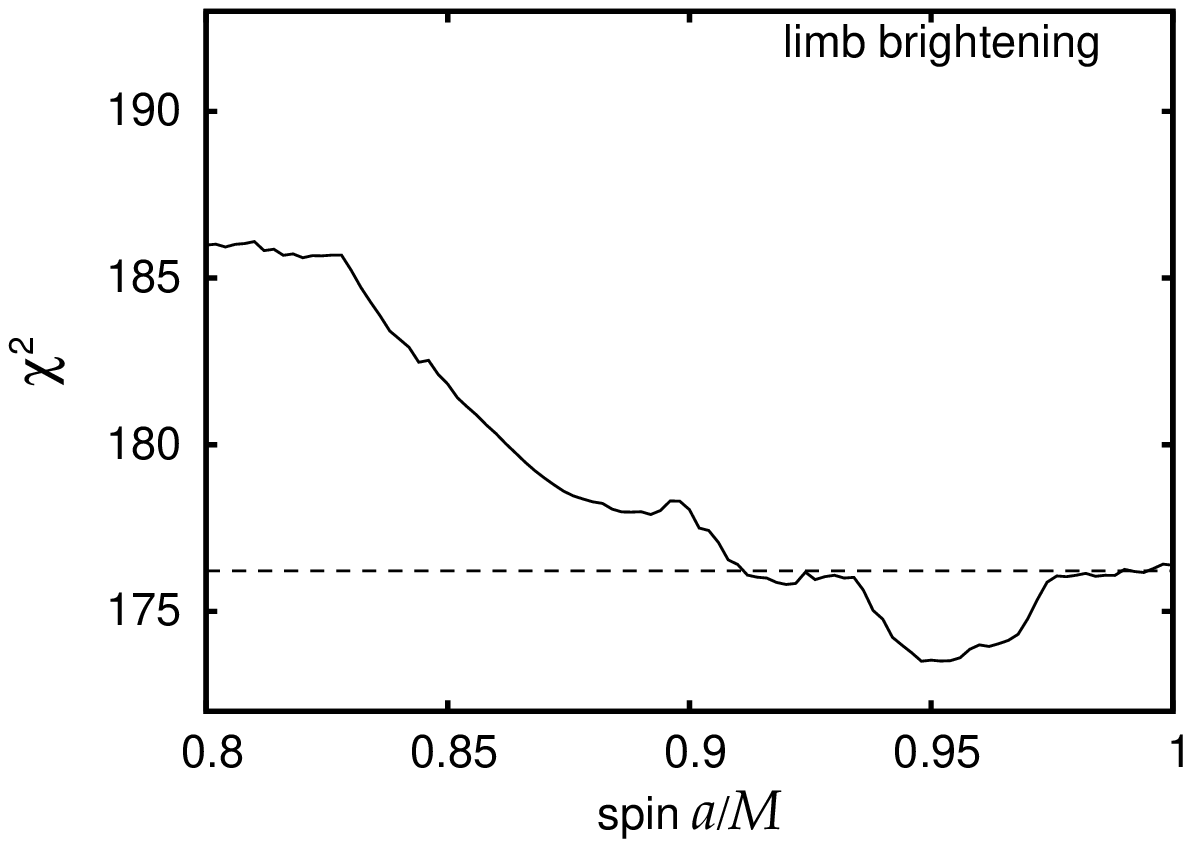}
\hfill
\includegraphics[width=0.49\textwidth]{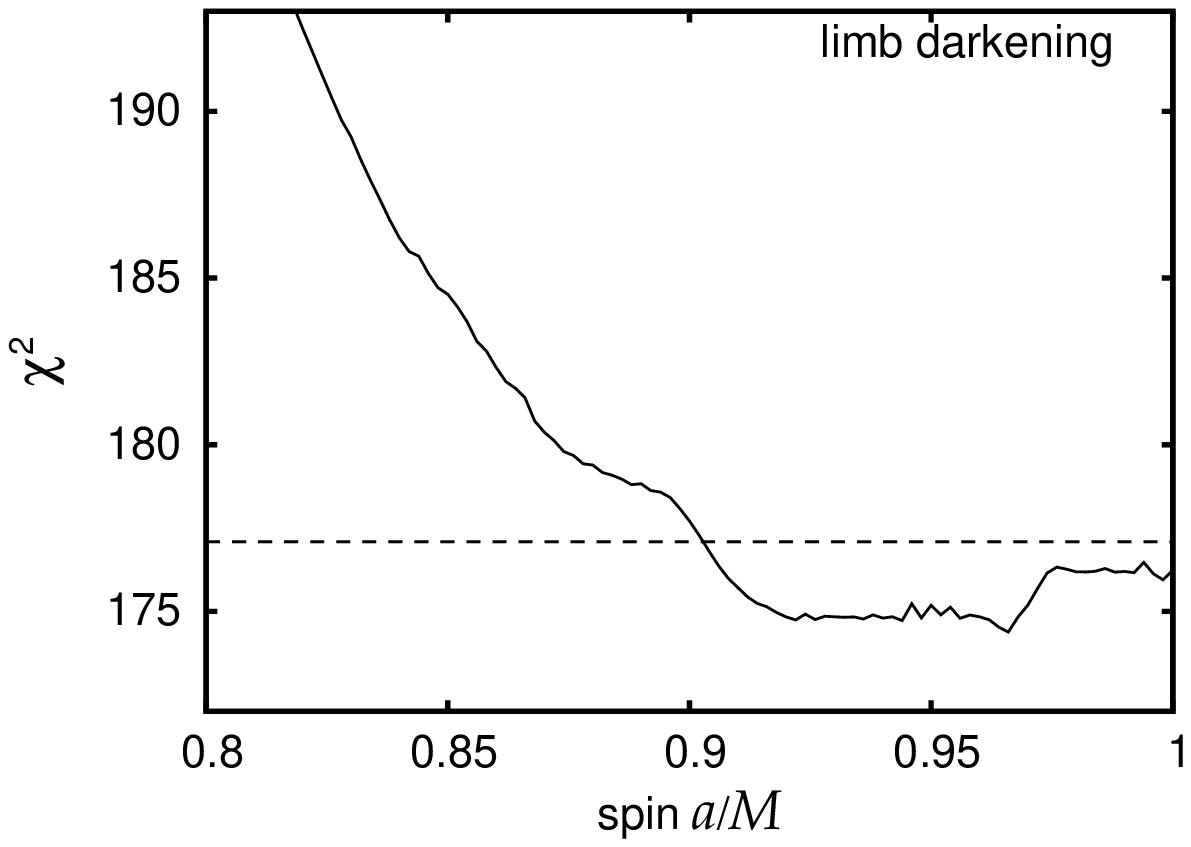}
\caption{The best-fit values of $\chi^2$ statistics for the spin parameter,
which we obtained by gradually stepping $a$ from $0.8$ to $1$.
XMM-Newton data for MCG--6-30-15 were employed 
\citep{2002MNRAS.335L...1F}. Left: the limb-brightening
profile (Case~1). Right: the limb darkening profile (Case~3).
The dashed line is the 90\% confidence level
(2 sigma). See the text for a detailed description of the model.}
\label{fig6}
\end{center}
\end{figure*}

We reanalysed a long XMM-Newton observation of a nearby Seyfert 1 galaxy MCG-6-30-15 
\citep{2002MNRAS.335L...1F}
to test whether the different directionality approximations can be distinguished 
in the current data.
The observation took place in summer 2001 and the acquired exposure time 
was about 350\,ks. The spectra reveal the presence of a broad and skewed
iron line \citep{2002MNRAS.335L...1F, 2003MNRAS.342..239B, 2004MNRAS.348.1415V}.
We reduced the EPN data from three sequential revolutions (301, 302, 303)
using the SAS software version 7.1.2.\footnote{http://xmm.esac.esa.int/sas}
We used standard tools for preparing and fitting the data 
contained in the HEASOFT software package version~6.4.\footnote{http://heasarc.gsfc.nasa.gov}
Using the FTOOL MATHPHA (part of the HEASOFT) we joined the three spectra to 
improve the statistical significance. 
Further, we used the XSPEC version 12.2 for the spectral analysis.
We rebinned all the data channels to oversample 
the instrumental energy resolution maximally by a factor of 3 
and to have at least 20 counts per bin. The first condition is much stronger 
with respect to the total flux of the source -- $4\times10^{-11}$\,erg\,cm$^{-2}$\,s$^{-1}$ 
($1.1\times10^{6}$\,cts) in the 2--10\,keV energy interval. 

We applied the same continuum model
as presented in \cite{2002MNRAS.335L...1F}: the power law component
(photon index $\Gamma =1.9$) absorbed by neutral 
hydrogen (column density $n_{\rm H} = 0.41\times 10^{21}$\,cm$^{-2}$).
This simple model is sufficient to fit the data above $\approx 2.5$\,keV, 
which is also satisfactory for our goal of reproducing the overall shape of
the broad iron line.
Other components need to be added to the model in order to fully understand 
the spectrum formation and to decide between viable alternatives, including
the presence of outflows and a combination of absorption and reflection effects
\citep{2008MNRAS.483..437,2009A&ARv..17...47T}.
However, our goal here is not to give precedence to any of the particular schemes.
Instead, we rely on the model of a broad line and we test
and compare different angular laws for the emission. 

The residuals from the simple power law model can be explained 
by a complex of a broad line and two narrow lines.
Due to the complexity of the model, it is hard to
distinguish between the narrow absorption line 
at $E\approx6.75$\,keV and an emission line at 
$E\approx6.97$\,keV \citep{2002MNRAS.335L...1F}.
Although the level of $\chi^{2}$ values stays almost the same,
the parameters of the broad iron line do depend on the
continuum model and the presence of narrow lines.
By adding an absorption line at 
$E\approx6.75$\,keV, the best fit rest energy of the broad 
component comes out to be $E=6.7$\,keV, and so the broad
line originates in a moderately ionised disc.
This result is consistent with \citet{2003MNRAS.342..239B}.

More details of our reanalysis of the broad
iron line shape are described in \citet{SvobodaWDS}.
Here, we present the results of the one-dimensional {\sl steppar} command in Figure~\ref{fig6}, 
which demonstrates the expected confidence of $a$-parameter best-fit values
for the two extreme cases of directionality, Case~1 and Case~3.
The model used in the XSPEC syntax is: \textscown{phabs}*(\textscown{powerlaw}+\textscown{zgauss}+\textscown{zgauss}+\textscown{kyrline}).
The fixed parameters of the model are the column density 
$n_{\rm H} = 0.41\times 10^{21}$\,cm$^{-2}$, 
the photon index of power law $\Gamma =1.9$, the redshift factor $z = 0.008$, 
the energy of the narrow emission line $E_{\rm em} = 6.4$\,keV,
and the energy of the narrow absorption line $E_{\rm abs} = 6.77$\,keV.
The parameters of the broad iron line were allowed to vary during the fitting procedure. 
Their default values were $E_{\rm broad} = 6.7$\,keV for the energy of the broad iron line,
$\theta_{\rm o} = 30$\,deg for the emission angle, $q_{1} = 4.5$, $q_{2} = 2$ and 
$r_{\rm b} = 10$\,$r_{\rm g}$ for the radial dependence of the emissivity
(the radial part of the intensity needs to be rather complicated
to fit the data and can be expressed as a broken power law:
${\cal R}(r_{\rm e})=r_{\rm e}^{-q_1}$ for $r_{\rm e} < r_{\rm b}$, 
and ${\cal R}(r_{\rm e})=r_{\rm e}^{-q_2}$ for $r_{\rm e} > r_{\rm b}$).

The determined best-fit values for the spin are virtually the same for both cases, 
independent of the details of the limb-brightening/darkening profile. 
However, this result arises on account of the growing complexity of the model.
The differences between the two cases become hidden in different 
values of the other parameters -- especially in $q_1$, $q_2$ and $r_{\rm b}$, 
i.e.\ the parameters characterising the radial dependence of the line
emissivity in \textscown{kyrline} as a broken power law with a break radius $r_{\rm b}$.
We find: (i) $E_{\rm broad} = 6.60(1)$, $\theta_{\rm o} = 31.5(7)$\,deg,
$q_1 = 3.7(1)$, $q_2 = 2.1(1)$, $r_{\rm b} = 18(1)$\,$r_{\rm g}$ for Case~1;
and (ii) $E_{\rm broad} = 6.67(1)$\,keV, $\theta_{\rm o} = 26.7(7)$\,deg,
$q_1 = 5.3(1)$, $q_2 = 2.8(1)$, $r_{\rm b} = 4.9(2)$\,$r_{\rm g}$ for Case~3.
The errors in brackets are evaluated as the 90\% confidence region 
for a single interesting parameter when the values of the other parameters
are fixed.
The combination of three parameters $q_1$, $q_2$, $r_{\rm b}$ thus adjusts 
the best fit in XSPEC. Nonetheless, the clear differences between 
the models occur consistently with theoretical expectations: 
for Case~3 the lower values of the spin,
$a<0.87$, produce larger $\chi^2$  and the best-fit spin can
reach the extreme value within the 90\% confidence threshold.

\subsection{Analysis of simulated data for next generation X-ray missions}
\label{simulated}

\begin{table}
\begin{center}
\begin{tabular}{c|cc|cc}
\rule[-0.8em]{0pt}{1.5em}Case & \multicolumn{2}{c}{$a_{\rm f}=0.7$} & \multicolumn{2}{c}{$a_{\rm f}=0.9982$} \\ 
\rule[-0.8em]{0pt}{1.5em}no. & $\theta_{\rm f}=30^{\circ}$ & $\theta_{\rm f}=60^{\circ}$ & $\theta_{\rm f}=30^{\circ}$ & $\theta_{\rm f}=60^{\circ}$\\ \hline
\rule{0pt}{2em} 1 & $0.56^{+0.04}_{-0.03}$ & $0.69^{+0.03}_{-0.04}$ & $0.92^{+0.03}_{-0.03}$ & $0.981^{+0.013}_{-0.031}$ \\
\rule{0pt}{2em} 2 & $0.66^{+0.05}_{-0.05}$ & $0.70^{+0.02}_{-0.04}$ & $\geq0.966$ & $\geq0.986$\\
\rule[-1em]{0pt}{3em} 3 & $0.74^{+0.05}_{-0.03}$ & $0.70^{+0.03}_{-0.03}$ & $\geq0.991$ & $\geq0.993$\\ 
\end{tabular}
\caption{The best-fit spin values inferred for the three cases of the
limb darkening/brightening law, eq.~(\ref{case123}), for the \textscown{kyrline} model.
The artificial data were generated using the \textscown{kyrline} model with isotropic 
directionality and the fiducial values of parameters (denoted by the subscript ``f'').
See the main text for details.}
\label{tab1}
\end{center}
\end{table}

In order to evaluate the feasibility of determining the spin of a rotating black hole
and to assess the expected constraints from future X-ray data, we produced a set of
artificial spectra. We used a simple model prescription and preliminary
response matrices for the International X-ray Observatory (IXO) mission.\footnote{We used 
the current version of provisional 
response matrices available at {\sf http://ixo.gsfc.nasa.gov/science/responseMatrices.html}
for the glass core calorimeter, dated October 30, 2008.
We used the energy resolution of 5\,eV per bin.} 
Here we limit the energy band in the range $2.5$--$10$~keV. The adopted model
consists of \textscown{powerlaw} for continuum (photon index $\Gamma$ and the
corresponding normalisation $K_\Gamma$), plus \textscown{kyrline} model
\citep{2004ApJS..153..205D} for the broad line. 
The normalisation factors of the model were chosen in such a way that
the model flux matches the flux of MCG--6-30-15.
In this section, the simulated flux is around $3.1\times10^{-11}$\,erg\,cm$^{-2}$\,s$^{-1}$
with about 3\% of the flux linked to the broad iron line component.
The simulated exposure time was 100\,ks.


\begin{figure*}[tbh!]
\begin{center}
\begin{tabular}{ccc}
  \includegraphics[width=0.31\textwidth]{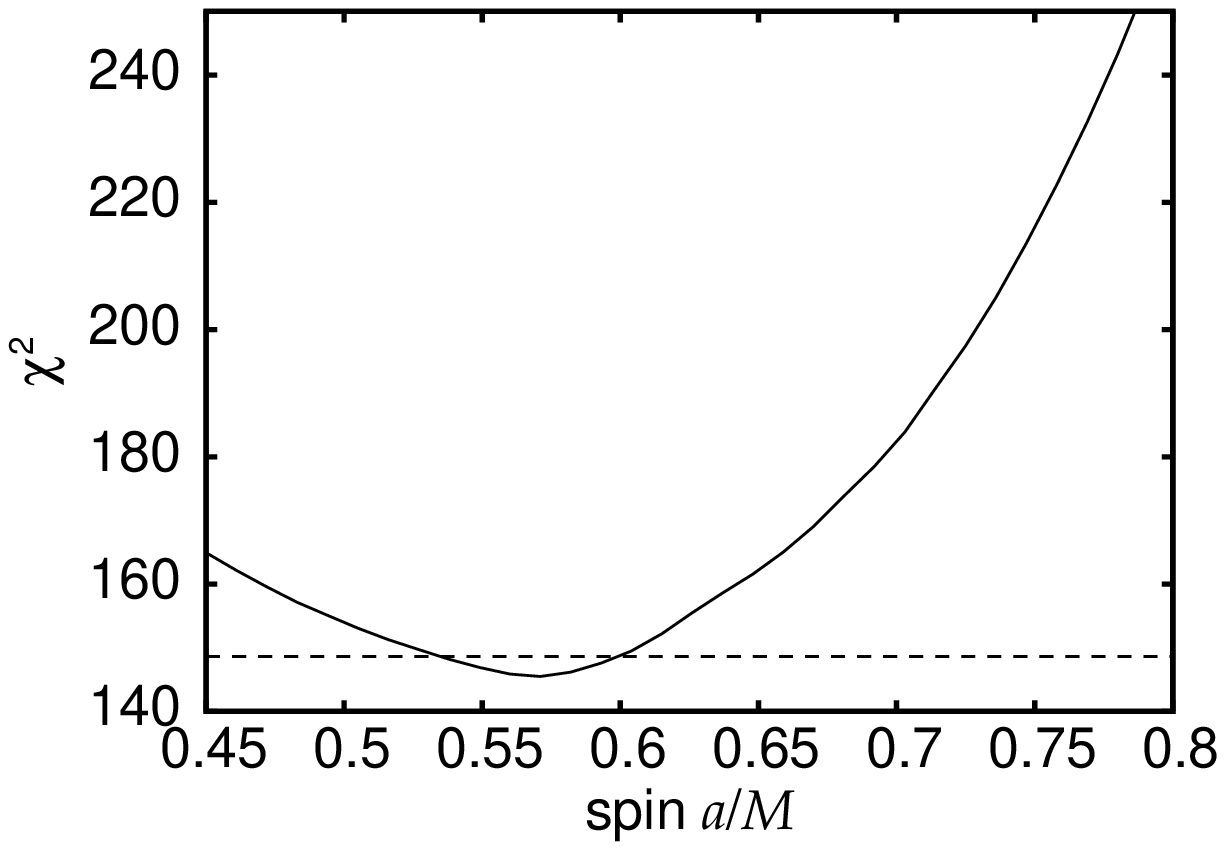} &
  \includegraphics[width=0.31\textwidth]{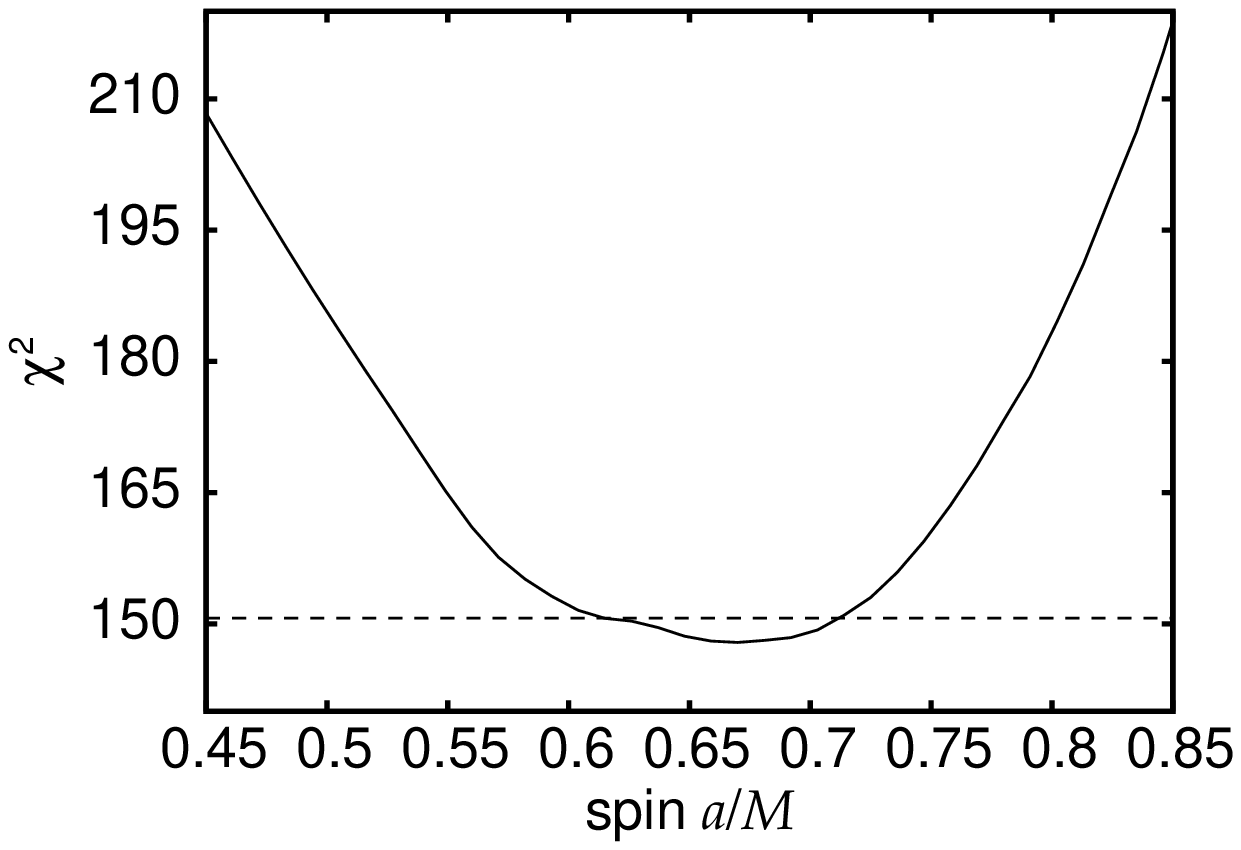} &
  \includegraphics[width=0.31\textwidth]{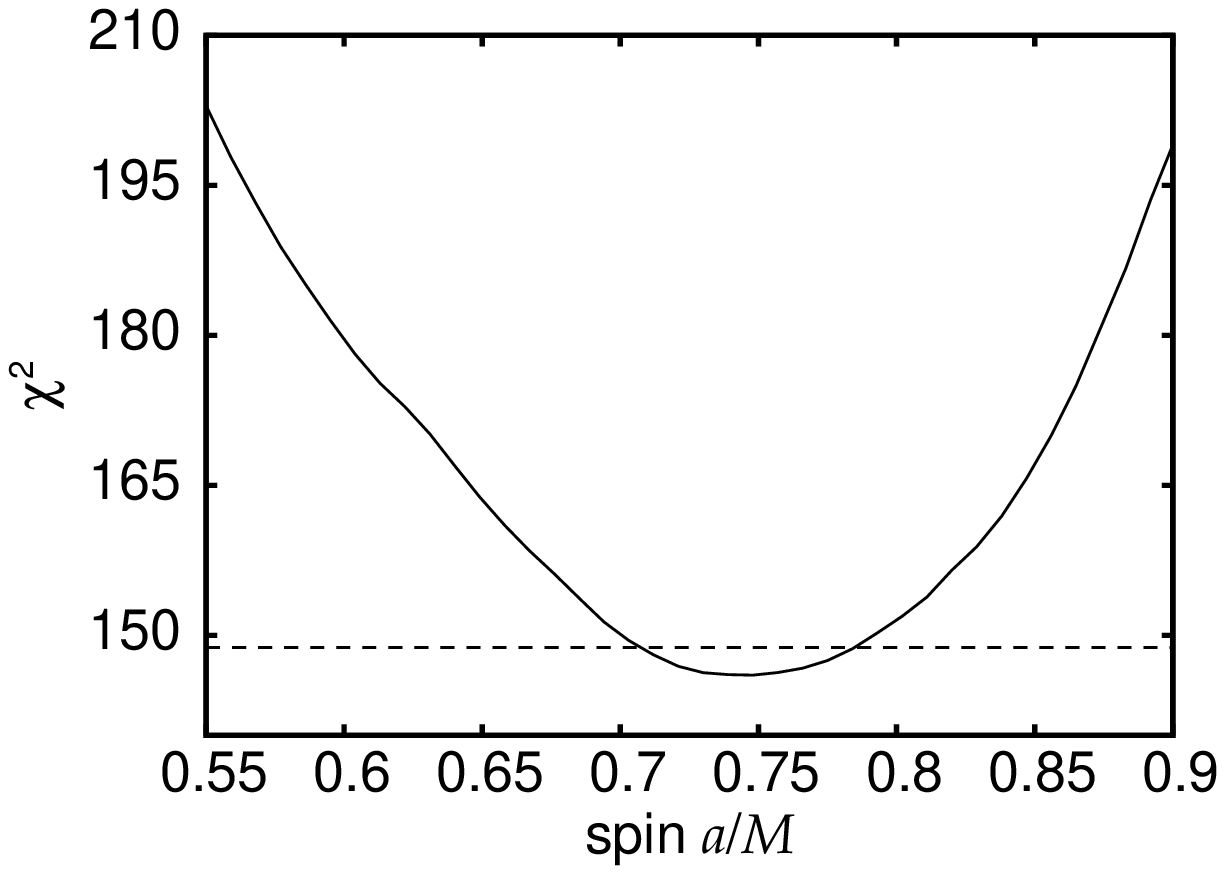}\\
\end{tabular}
\begin{tabular}{ccc}
  \includegraphics[width=0.31\textwidth]{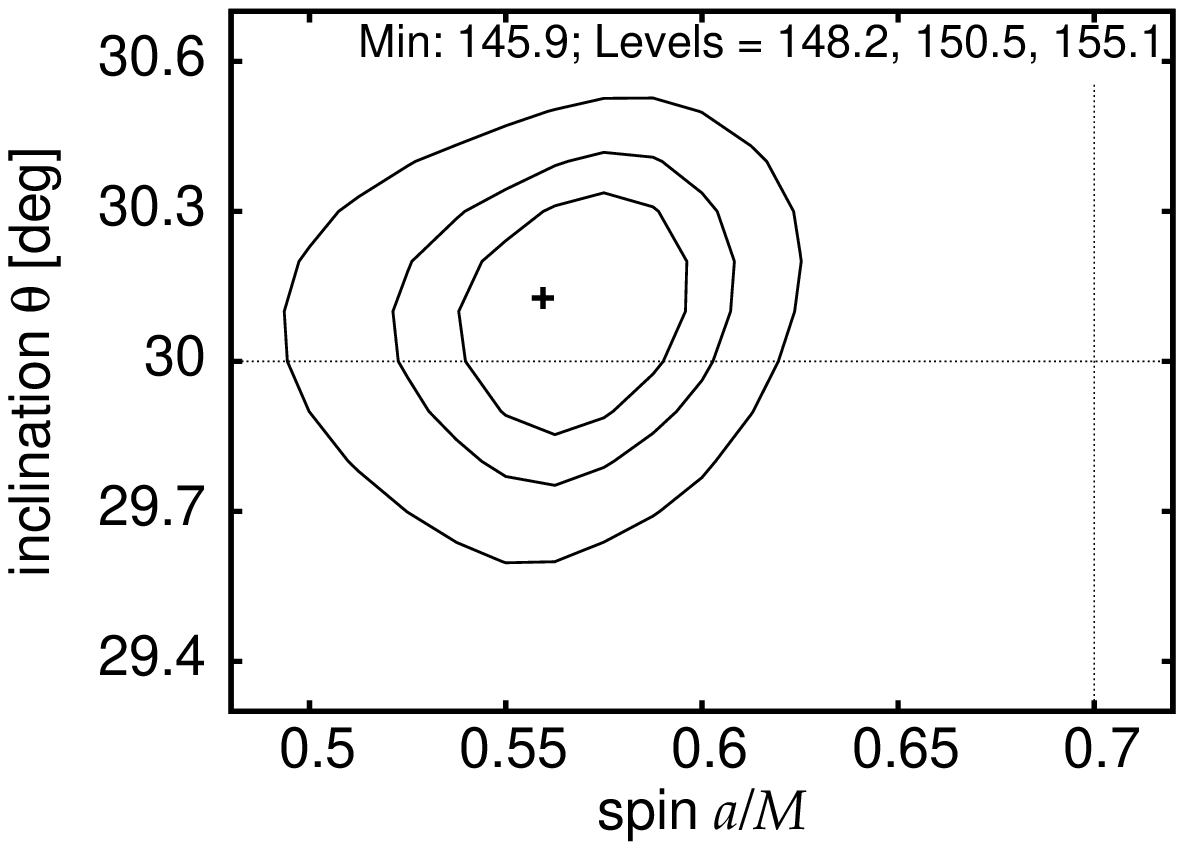} &
  \includegraphics[width=0.31\textwidth]{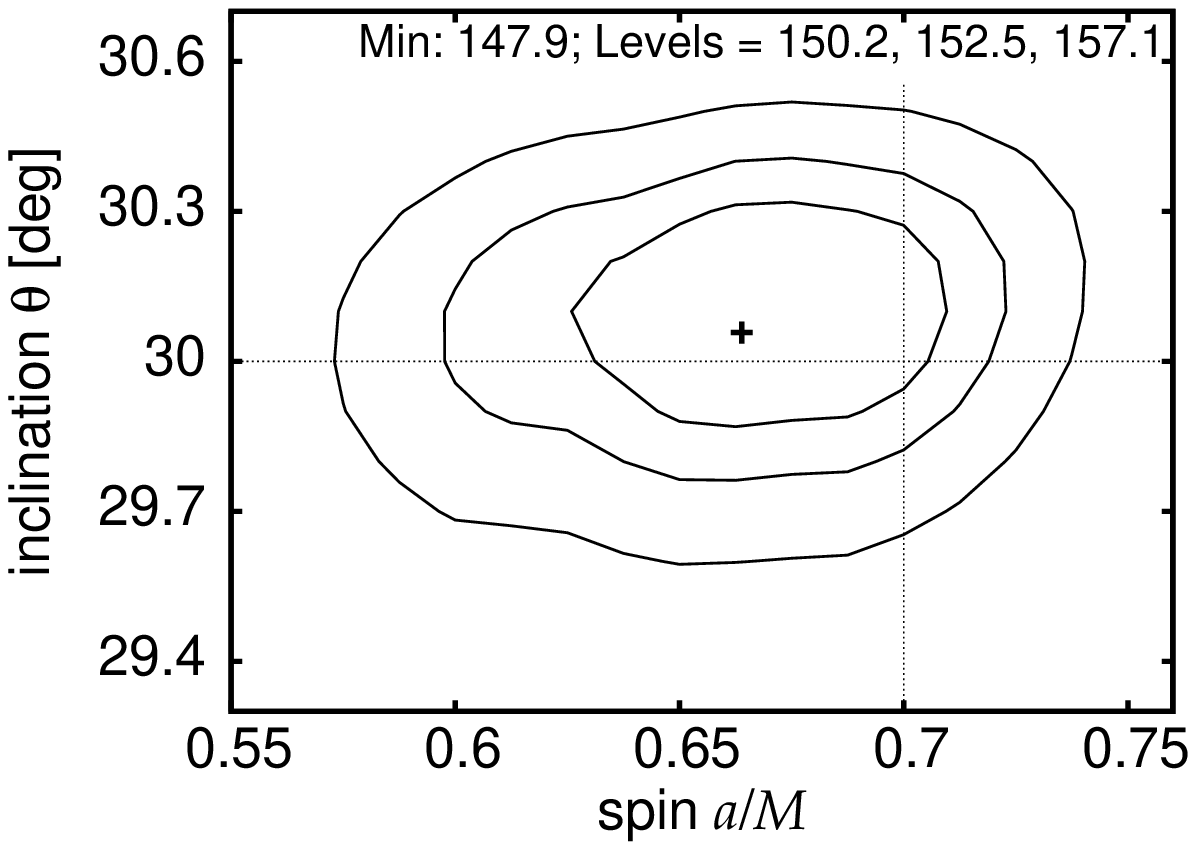} &
  \includegraphics[width=0.31\textwidth]{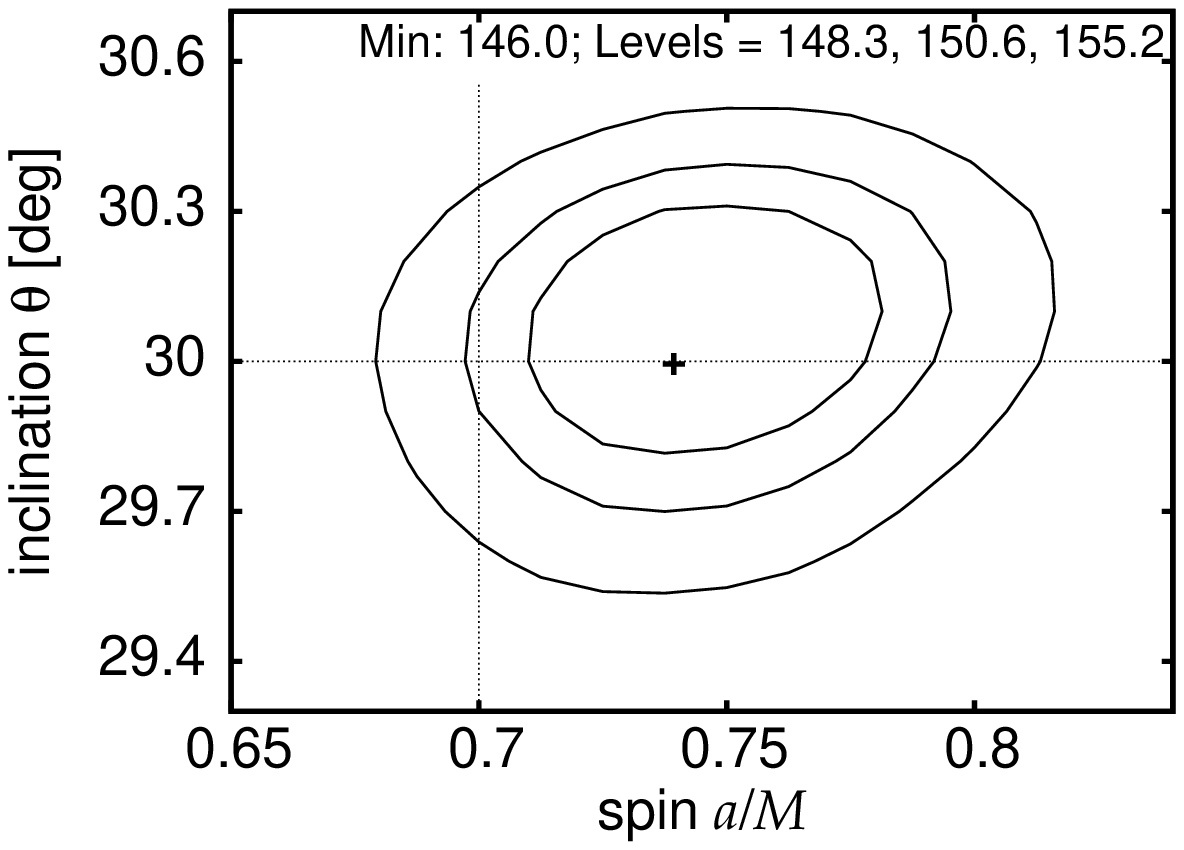}\\
\end{tabular}
\caption{Test results of the theoretical fits with 
$a_{\rm f}=0.7$, $\theta_{\rm f}=30\deg$ and three different profiles
of the emission directionality - left: limb brightening (Case~1 in eq.~(\ref{case123})), 
middle: isotropic (Case~2), right: limb darkening (Case~3). 
The simulated data were generated using the 
\textscown{powerlaw} + \textscown{kyrline} model
with isotropic directionality.
Top: dependence of the best fit $\chi^2$ values on the spin value.
The horizontal (dashed) line represents the 90\% confidence level.
Bottom: contour graphs of $a$ versus $\theta_{\rm o}$. The contour lines correspond to $1$, $2$,
and $3$ sigma. The position of the minimal value of $\chi^{2}$ is marked
with a small cross. The values of $\chi^{2}$ corresponding to the minimum
and to the contour levels are shown at the top of each contour graph.
The large cross indicates the position of the fiducial values of 
the angular momentum and the emission angle.
Besides the values of spin, inclination and normalisation constants of the model, 
the other parameters were kept fixed at their default values: 
$\Gamma=1.9$, $E_0=6.4$~keV, $r_{\rm in}=r_{\rm ms}$, $r_{\rm out}=400$,  $q=3$, 
(default values of normalisation constants: $K_\Gamma=10^{-2}$, $K_{\rm line}=10^{-4}$).}
\label{fig7}
\end{center}
\end{figure*}

\begin{figure*}[tbh!]
\begin{center}
\begin{tabular}{ccc}
  \includegraphics[width=0.31\textwidth]{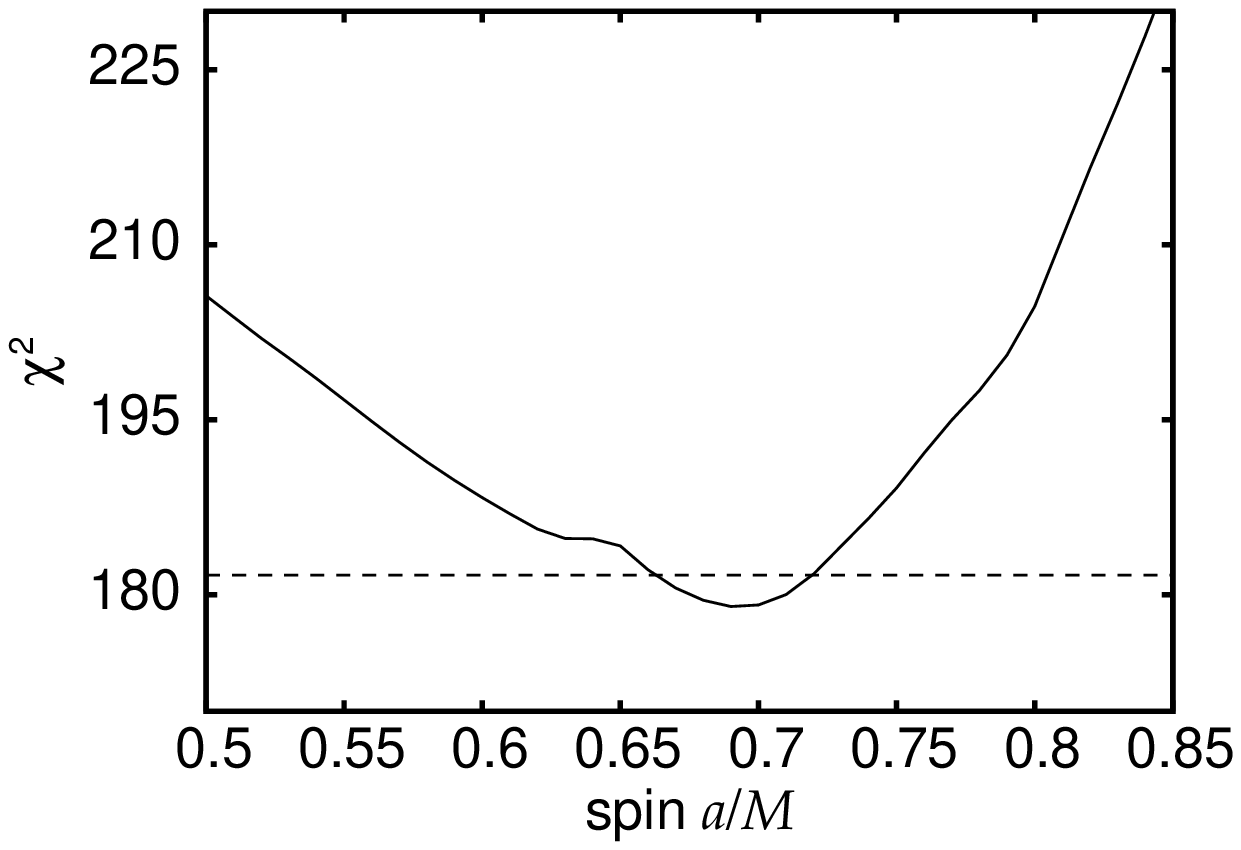} &
  \includegraphics[width=0.31\textwidth]{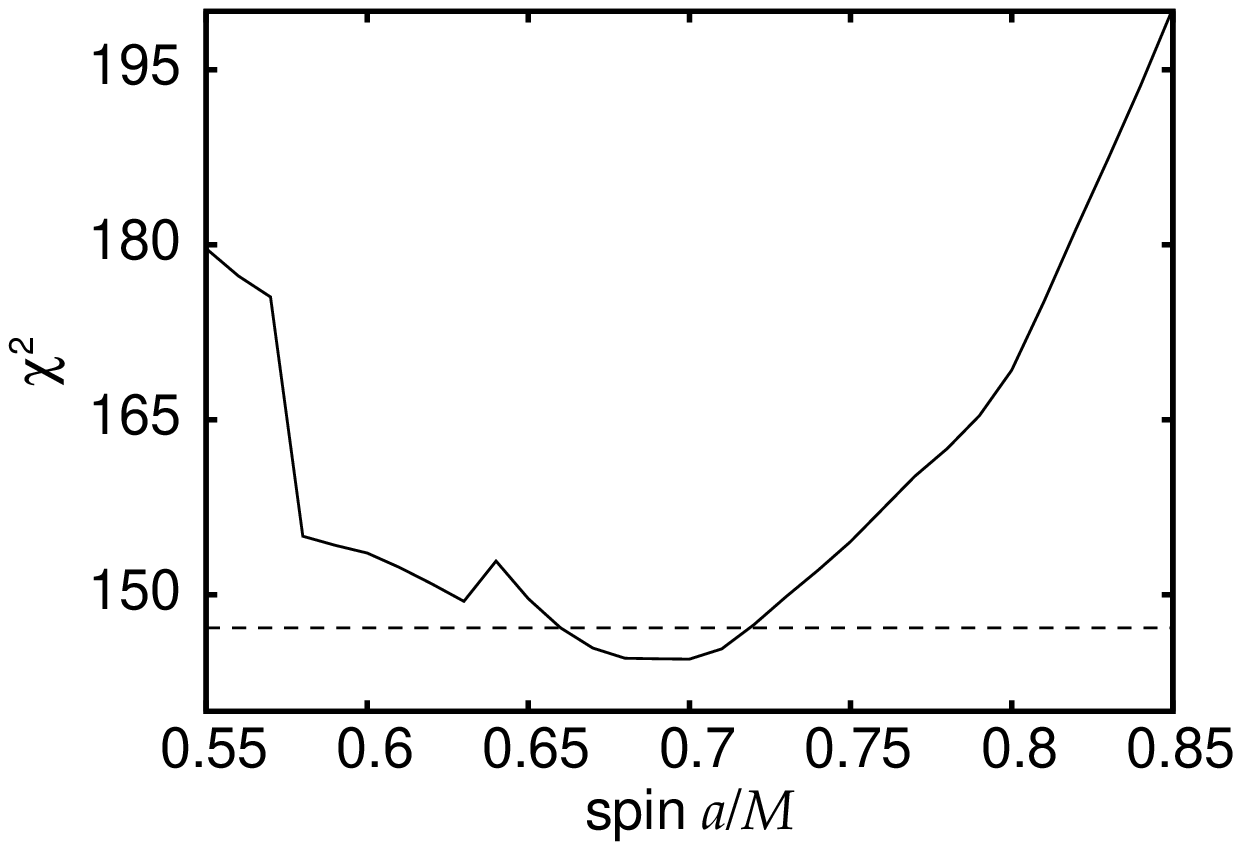} &
  \includegraphics[width=0.31\textwidth]{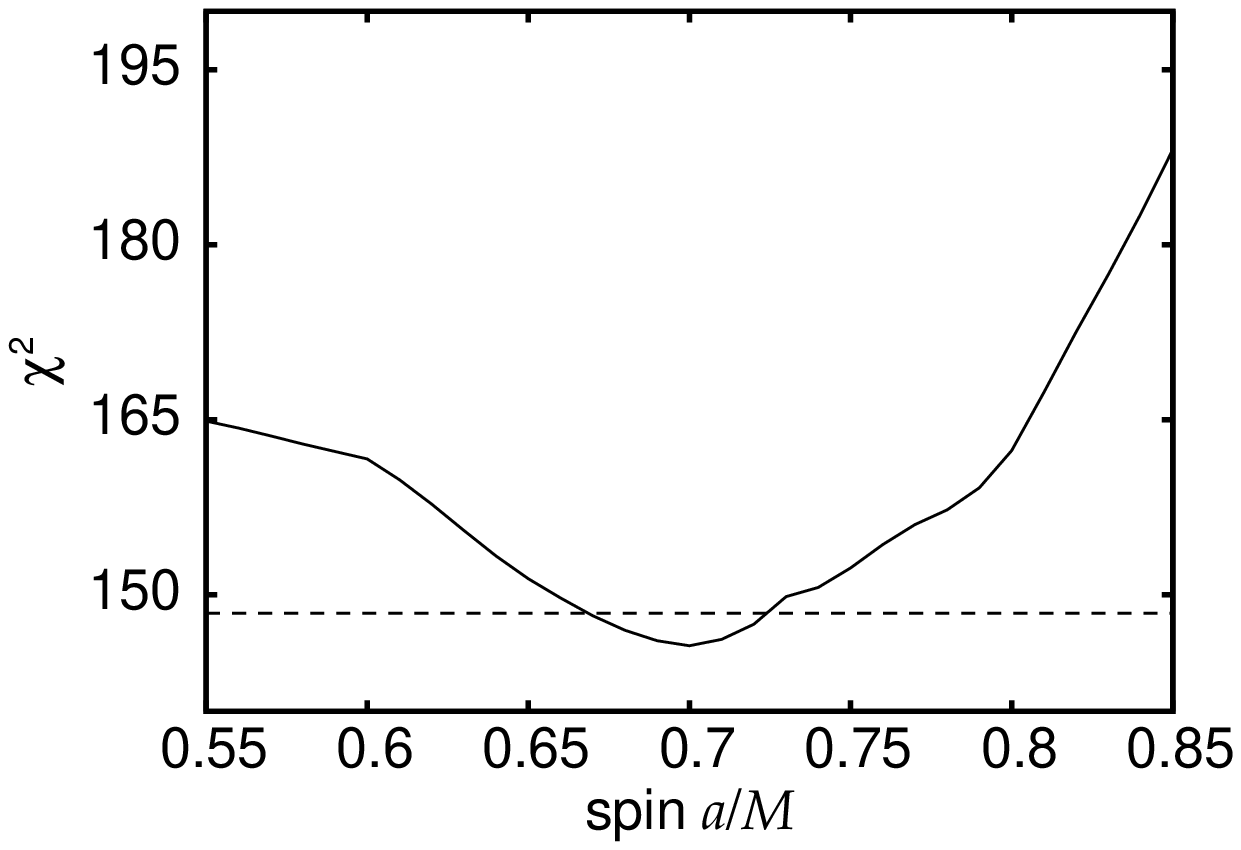}\\
\end{tabular}
\begin{tabular}{ccc}
  \includegraphics[width=0.31\textwidth]{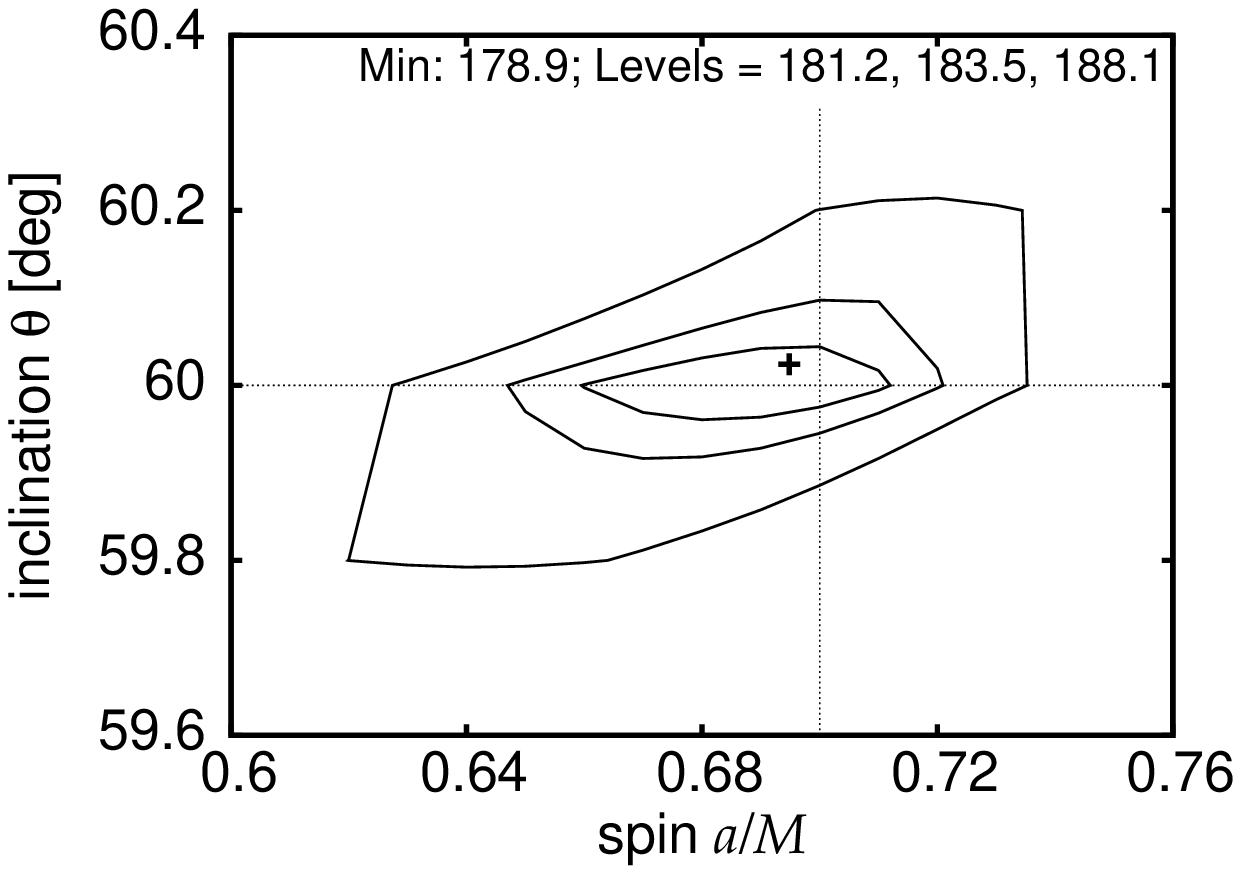} &
  \includegraphics[width=0.31\textwidth]{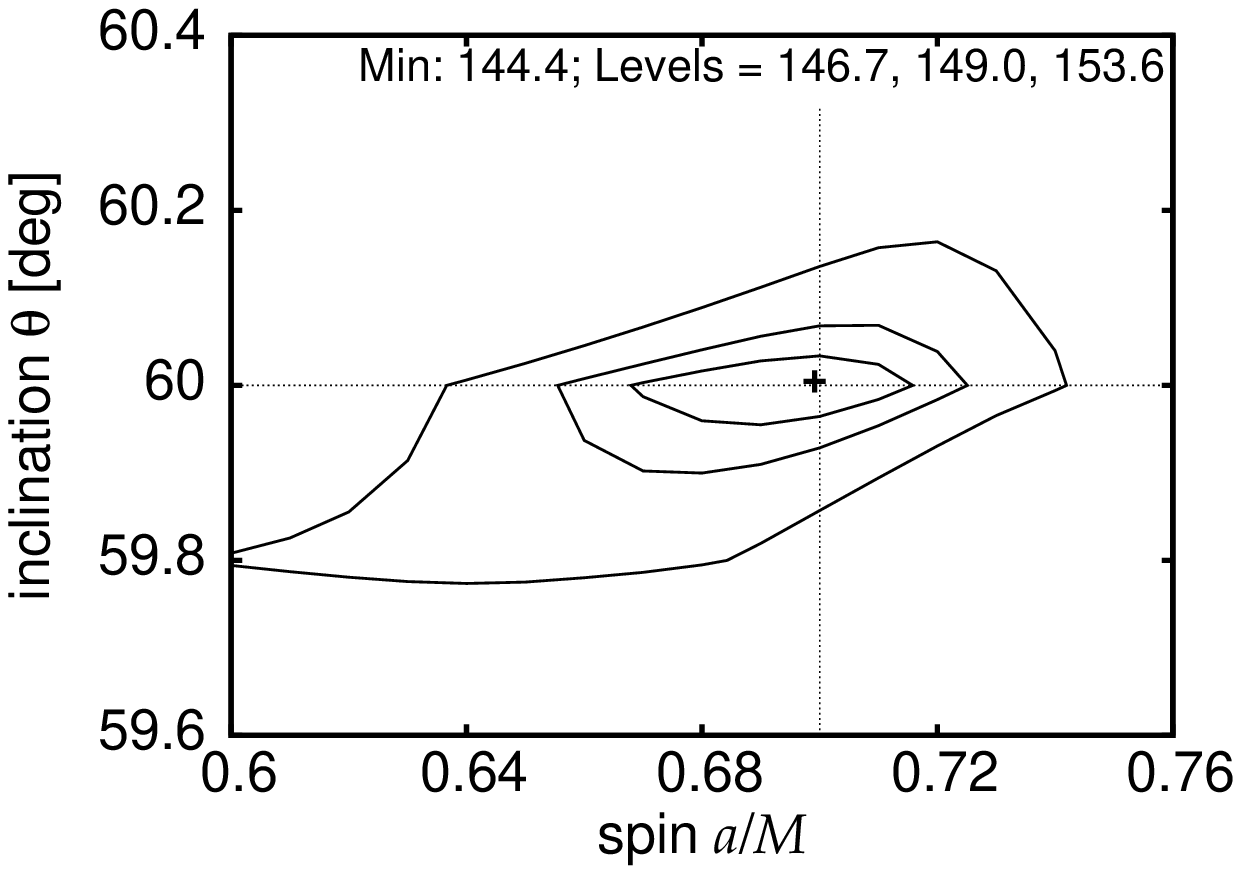} &
  \includegraphics[width=0.31\textwidth]{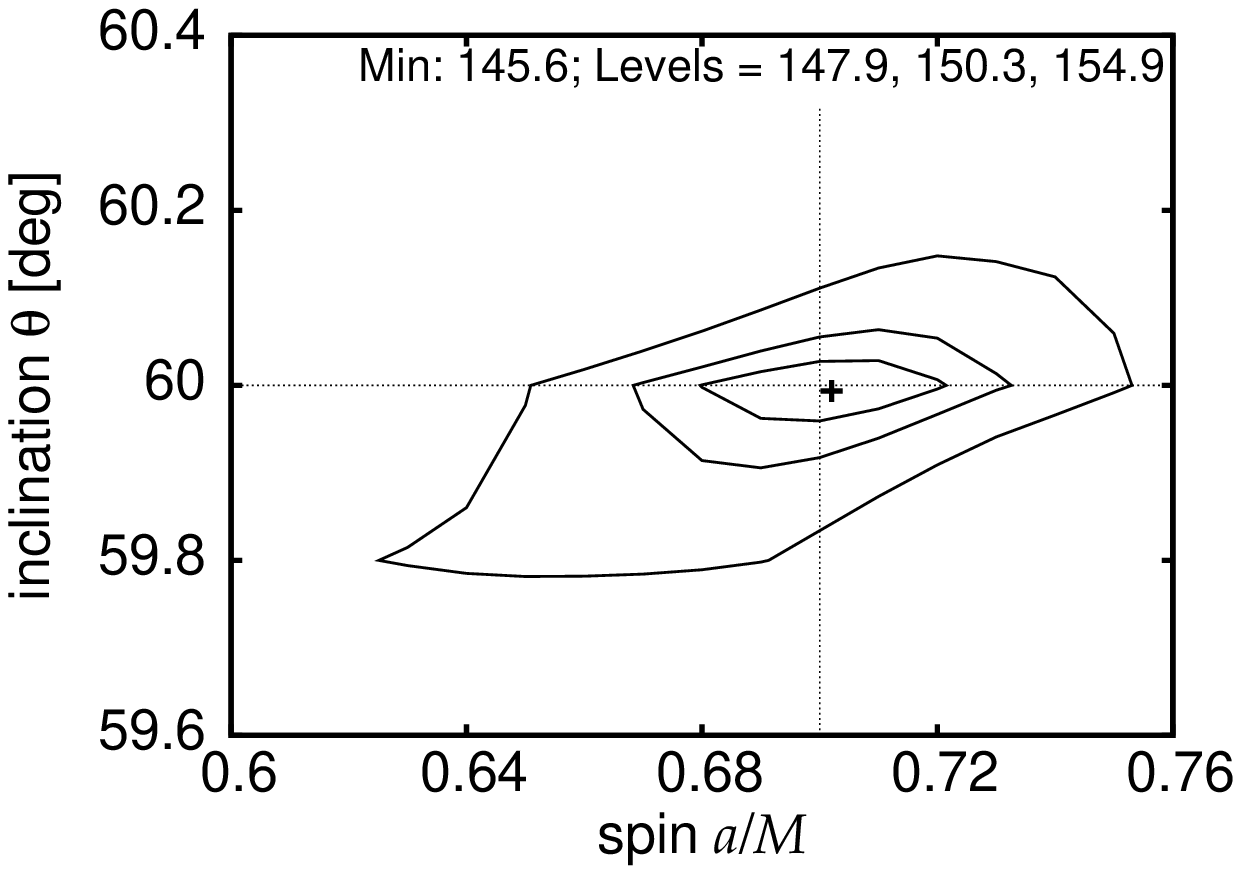}\\
\end{tabular}
\caption{The same as in Figure~\ref{fig7}, but for 
$a_{\rm f}=0.7$ and $\theta_{\rm f}=60\deg$.}
\label{fig8}
\end{center}
\end{figure*}

\begin{figure*}[tbh!]
\begin{center}
\begin{tabular}{ccc}
  \includegraphics[width=0.31\textwidth]{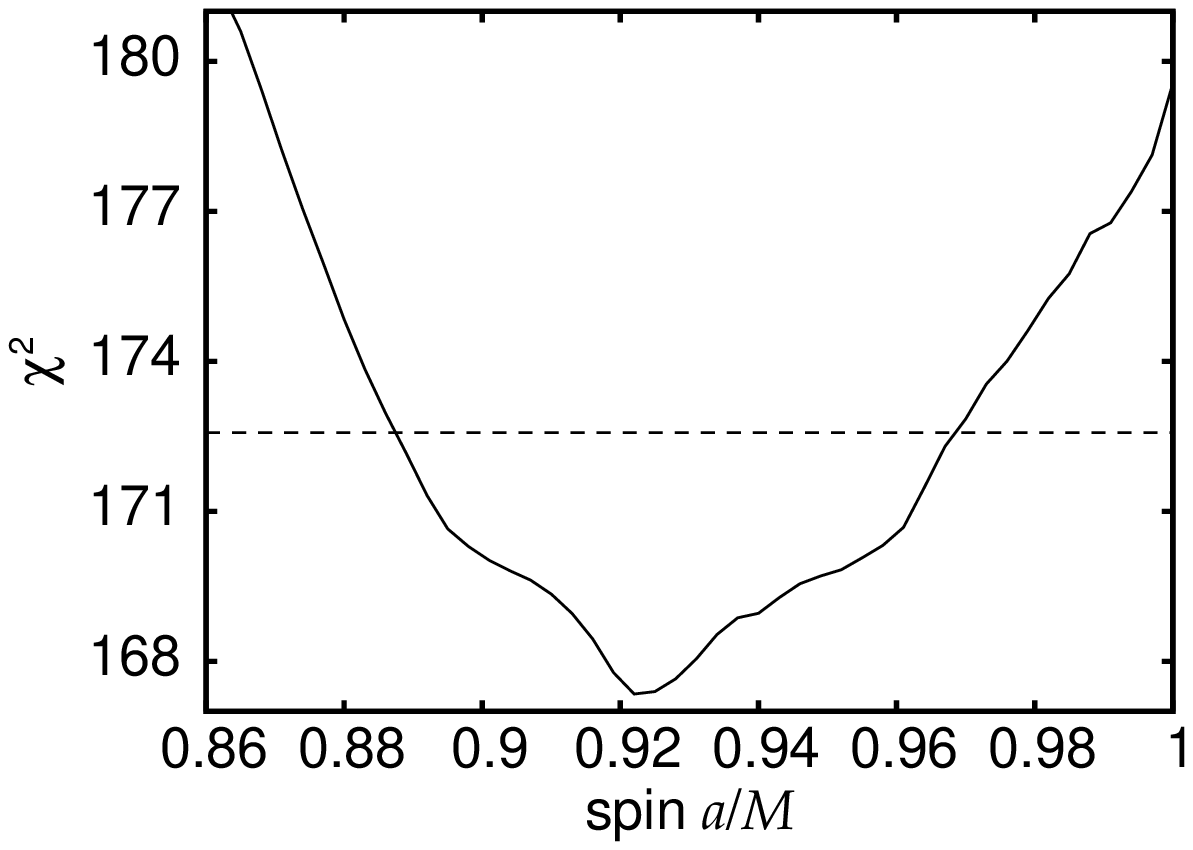} &
  \includegraphics[width=0.31\textwidth]{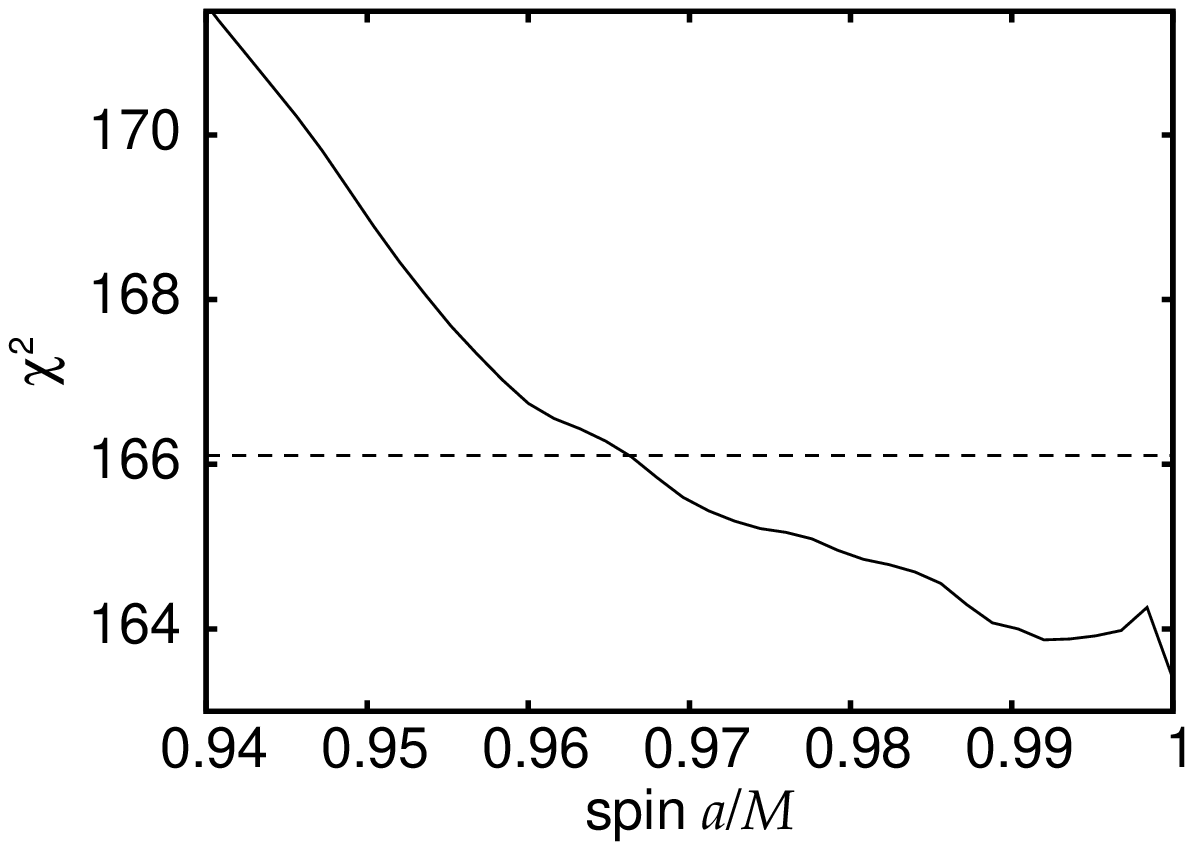} &
  \includegraphics[width=0.31\textwidth]{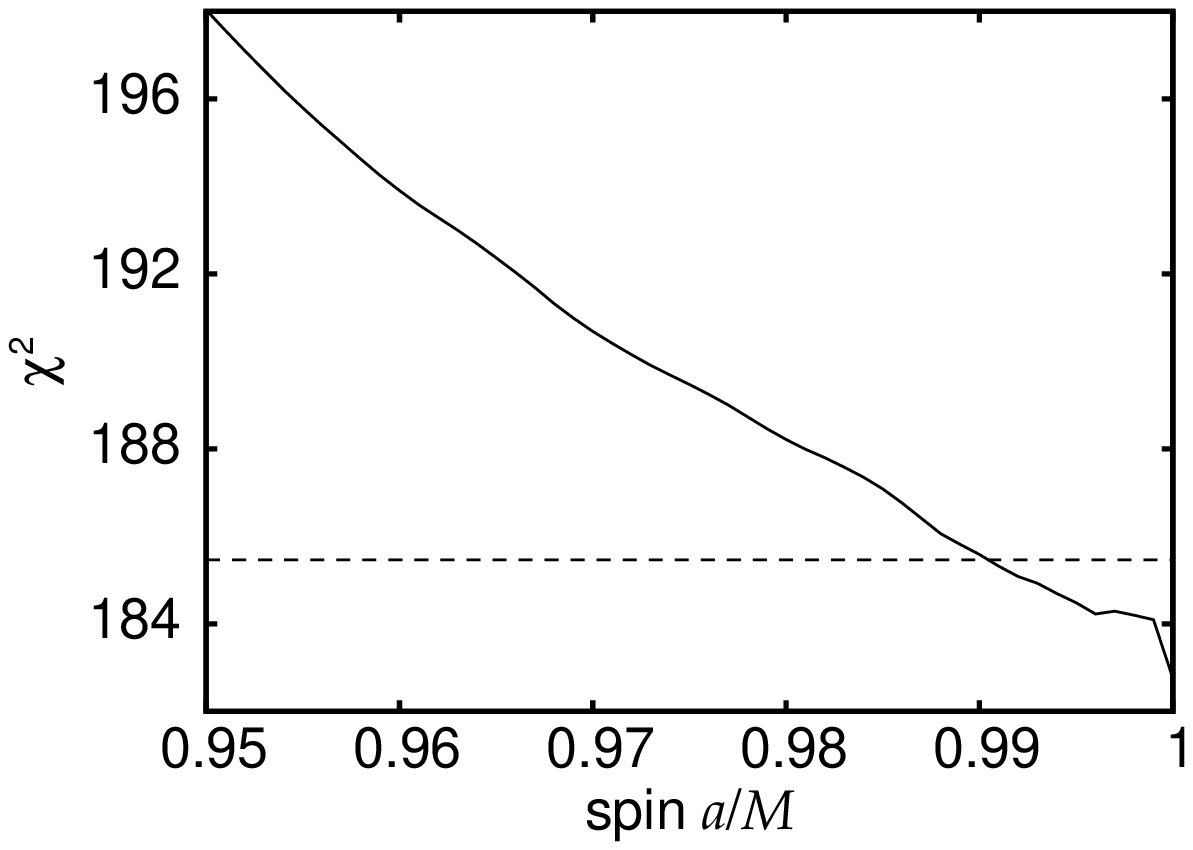}\\
\end{tabular}
\begin{tabular}{ccc}
  \includegraphics[width=0.31\textwidth]{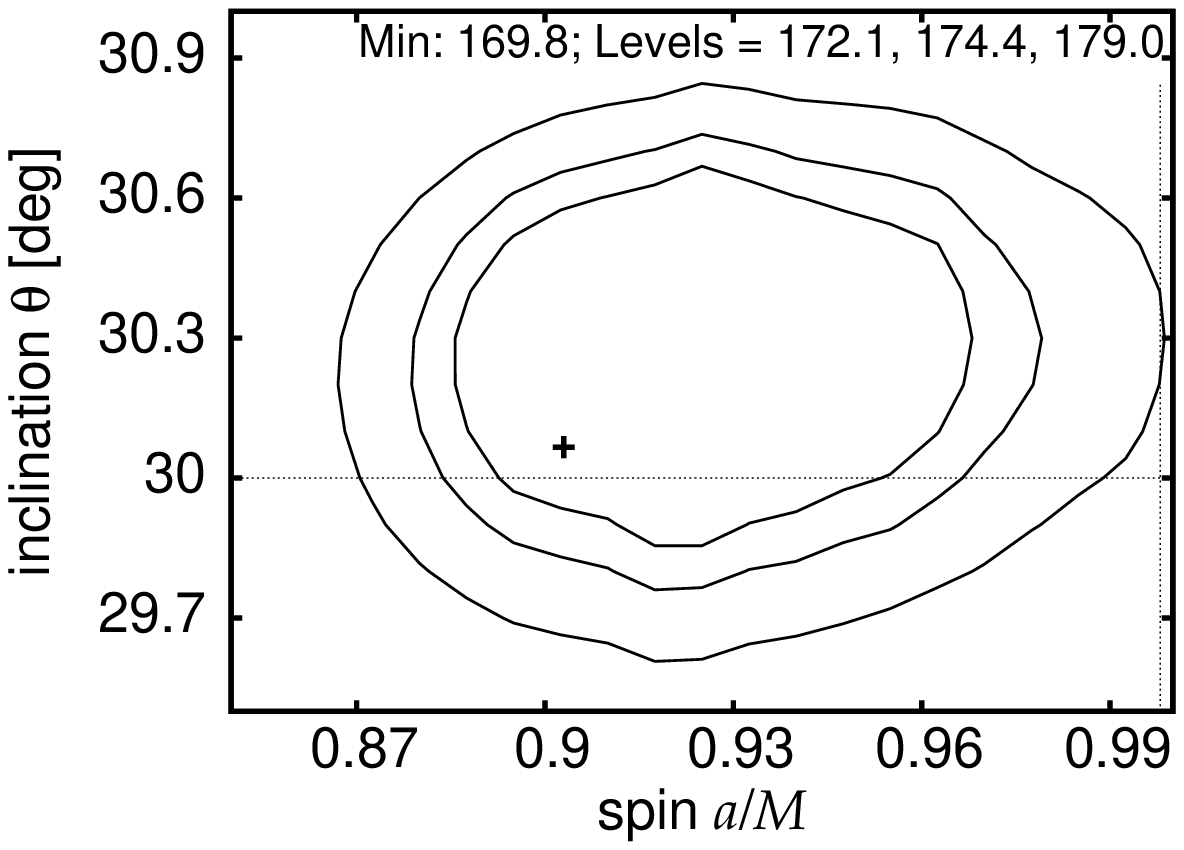} &
  \includegraphics[width=0.31\textwidth]{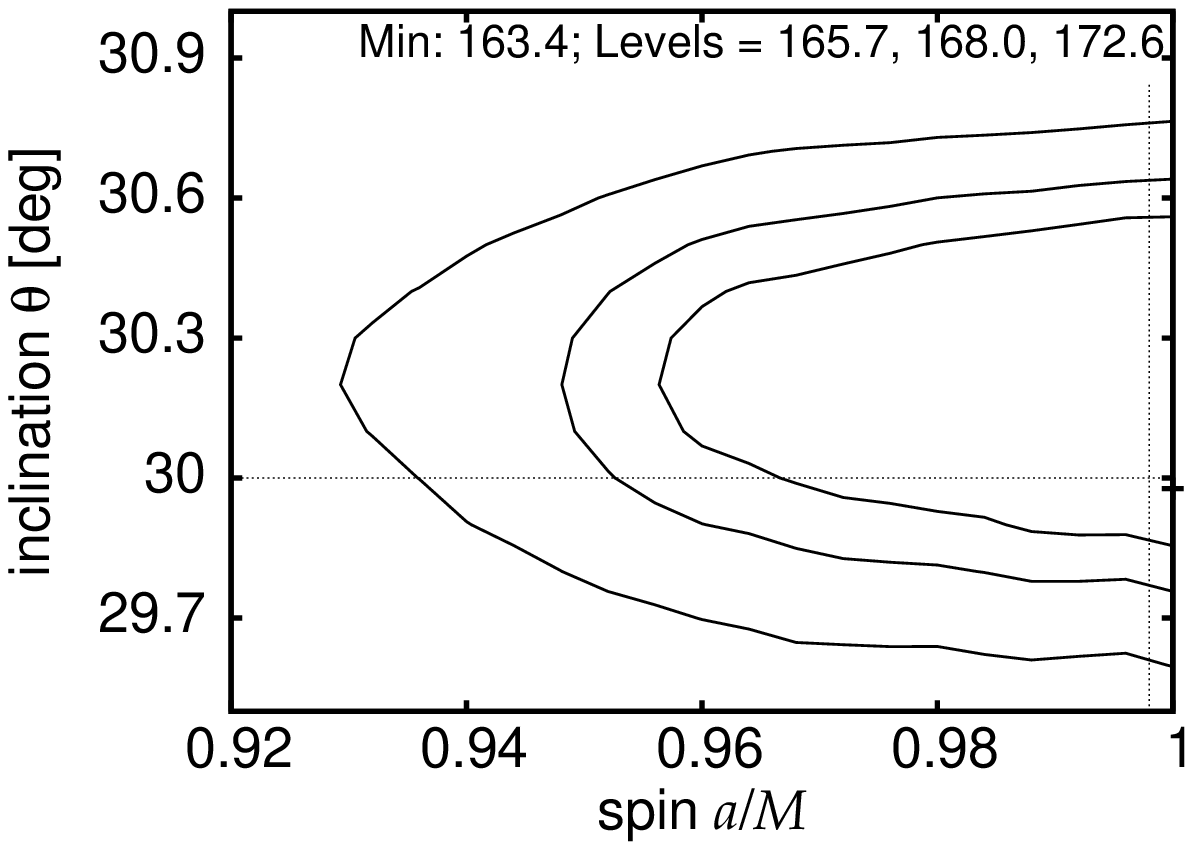} &
  \includegraphics[width=0.31\textwidth]{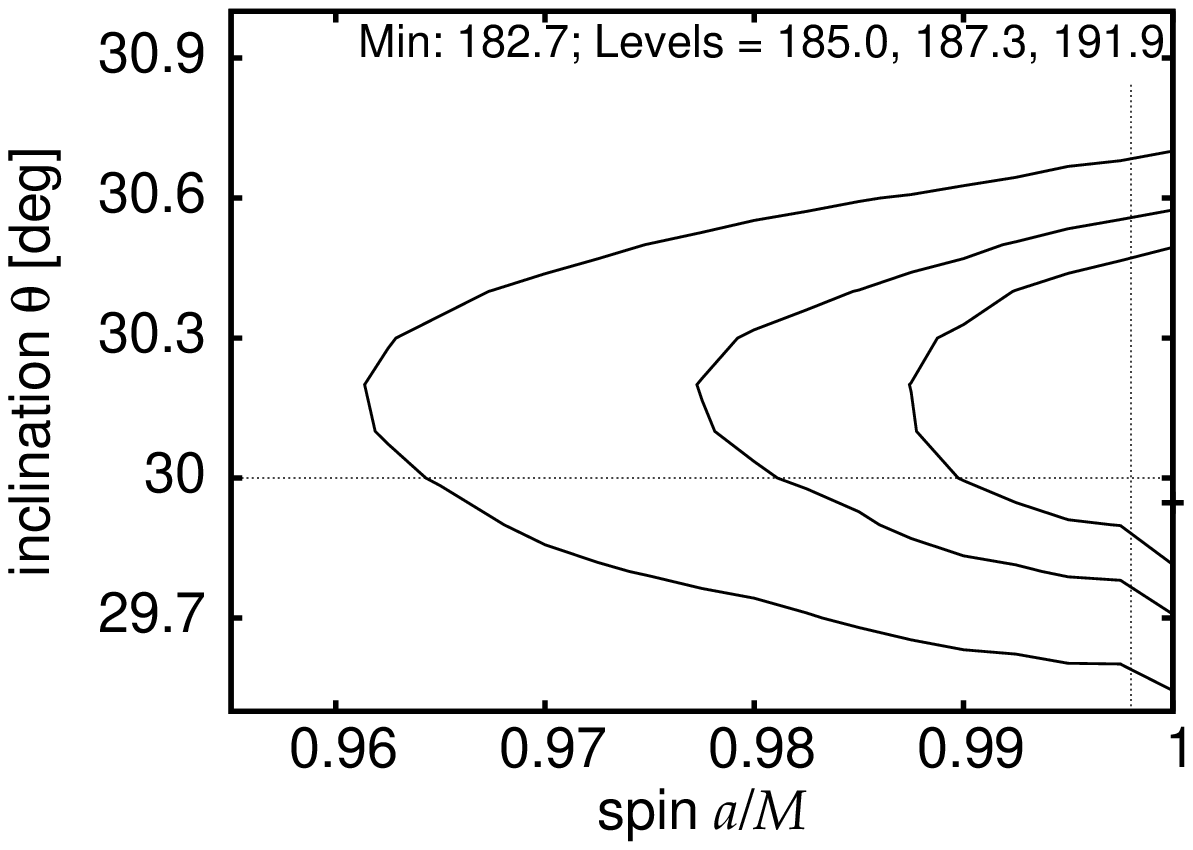}\\
\end{tabular}
\caption{The same as in Figure~\ref{fig7}, but for $a_{\rm f}=0.998$ and $\theta_{\rm f}=30\deg$.}
\label{fig9}
\end{center}
\end{figure*}

\begin{figure*}[tbh!]
\begin{center}
\begin{tabular}{ccc}
  \includegraphics[width=0.31\textwidth]{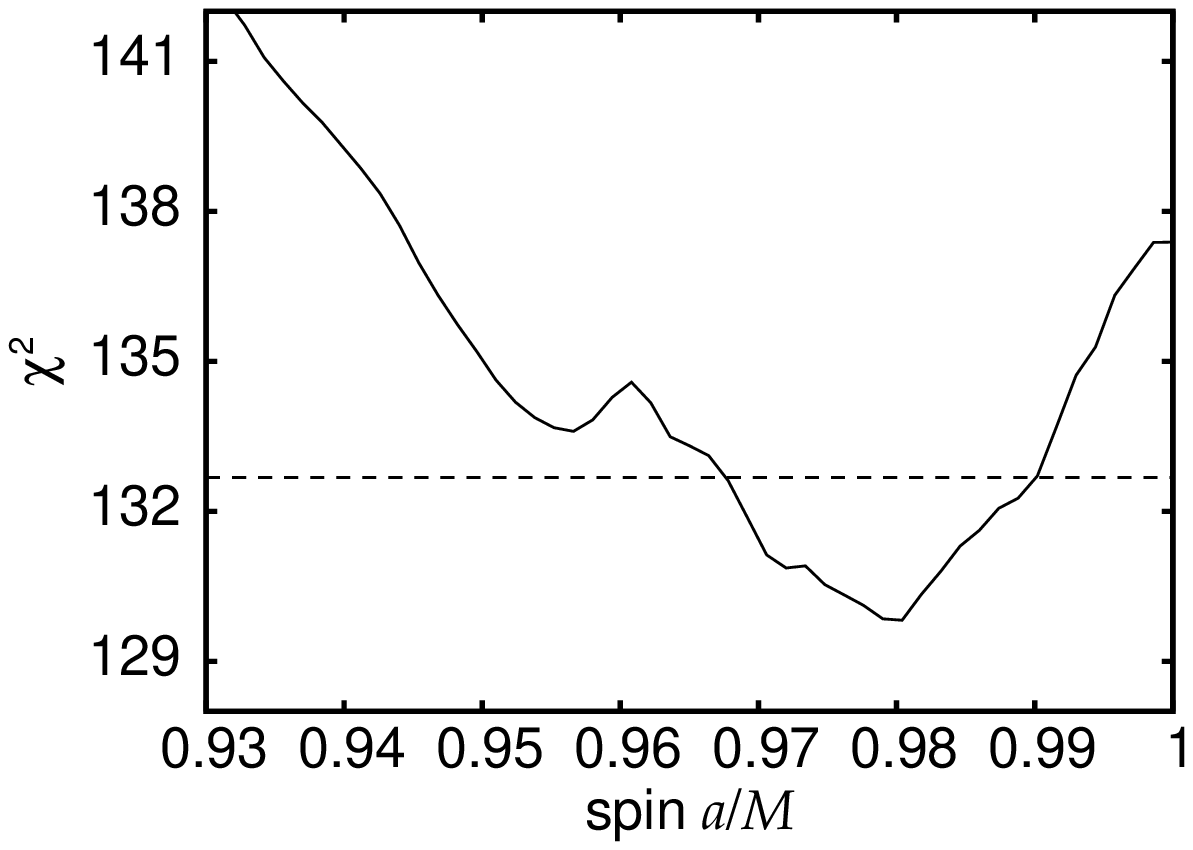} &
  \includegraphics[width=0.31\textwidth]{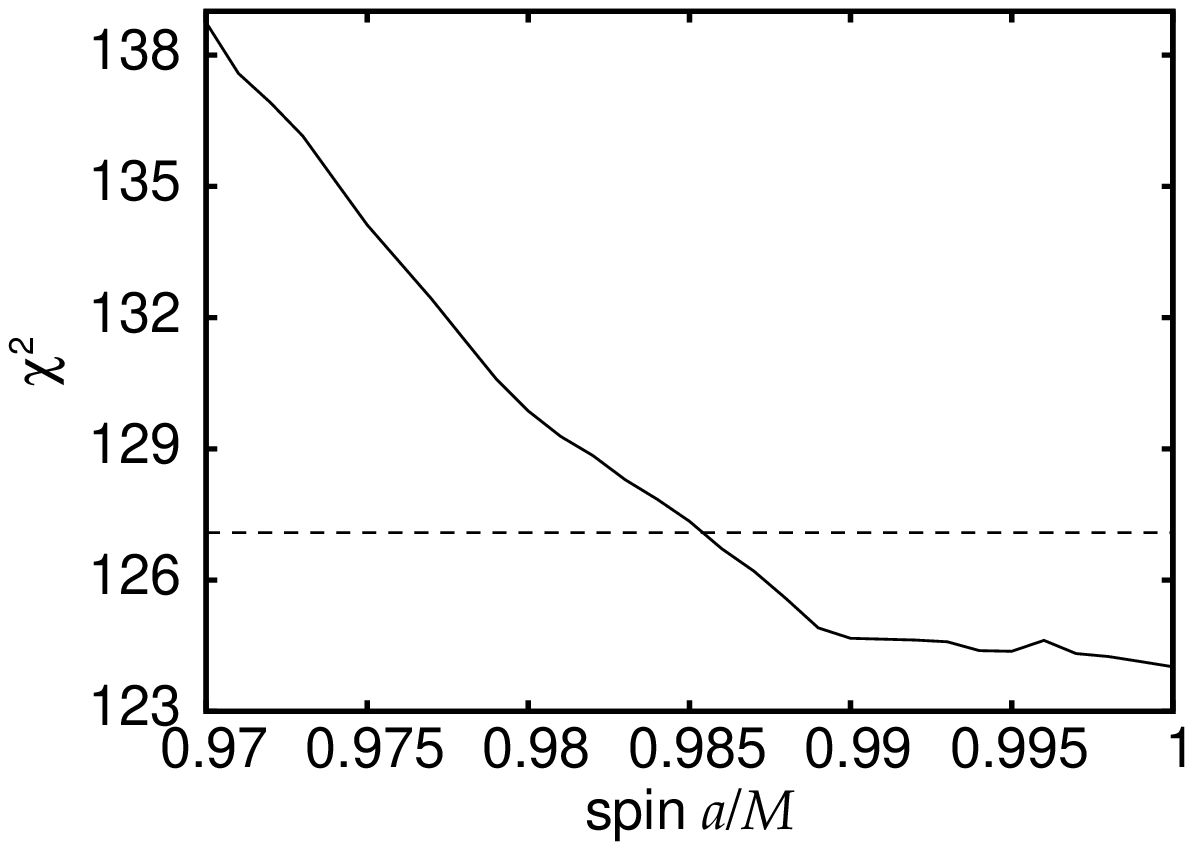} &
  \includegraphics[width=0.31\textwidth]{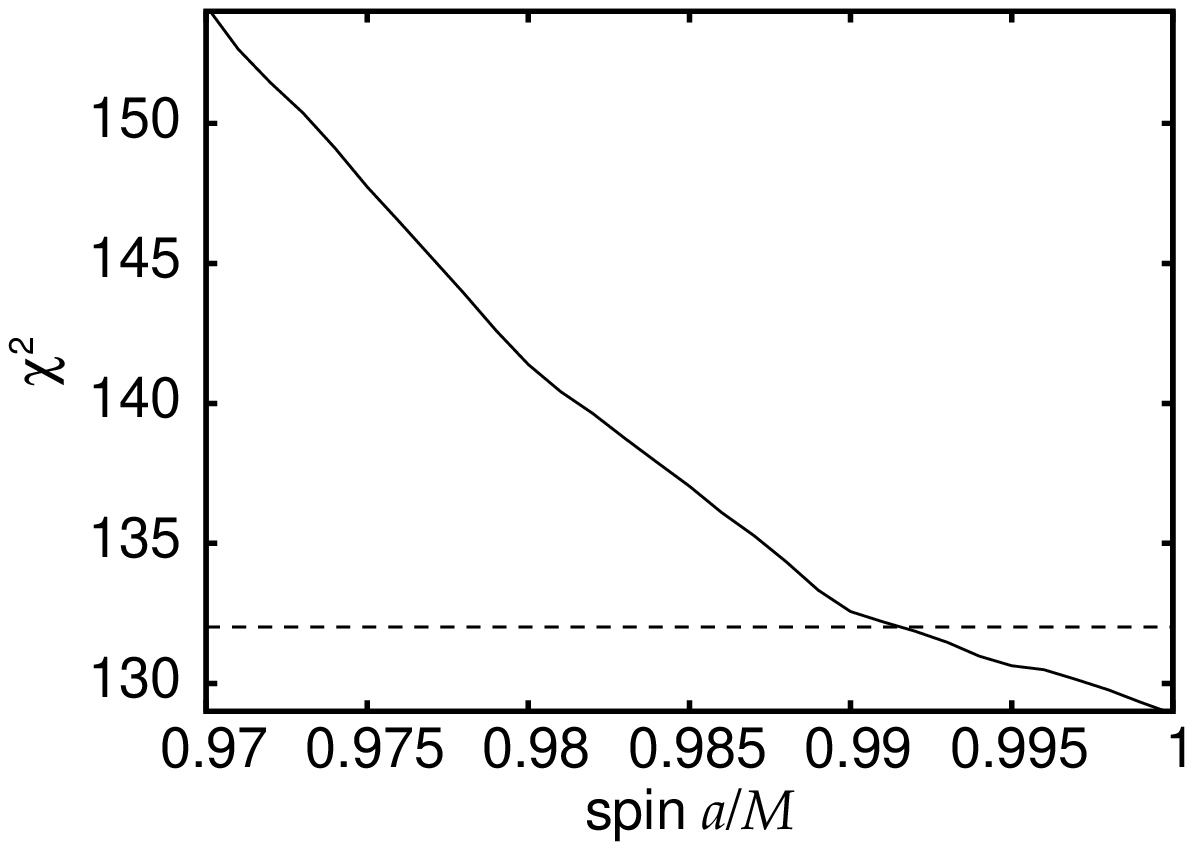}\\
\end{tabular}
\begin{tabular}{ccc}
  \includegraphics[width=0.31\textwidth]{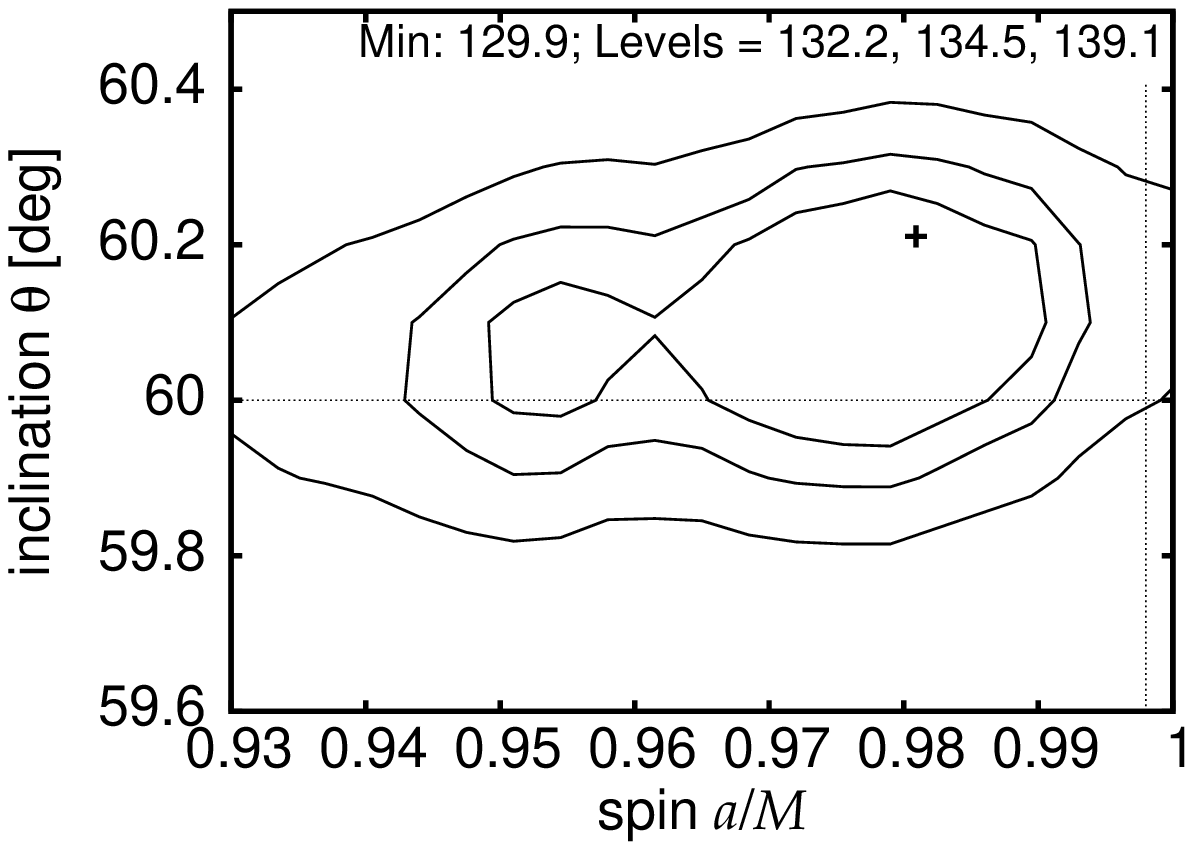} &
  \includegraphics[width=0.31\textwidth]{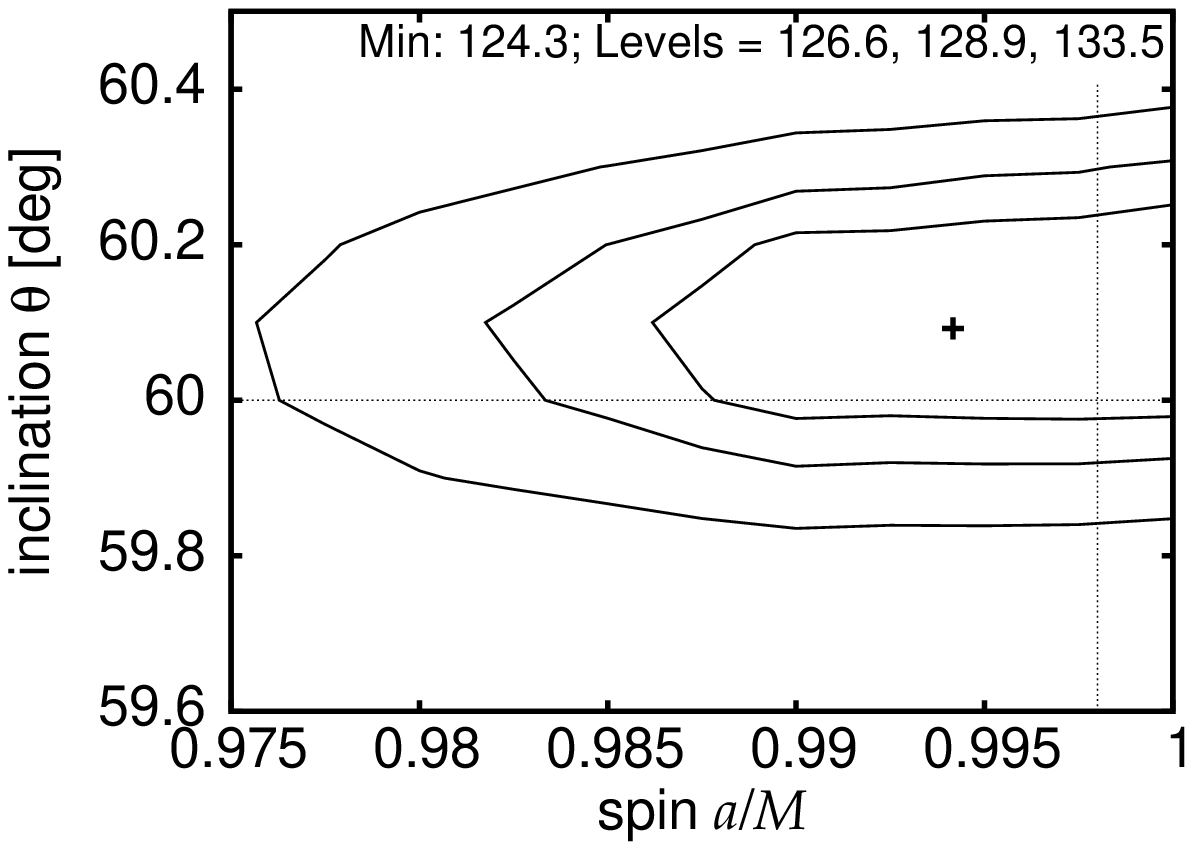} &
  \includegraphics[width=0.31\textwidth]{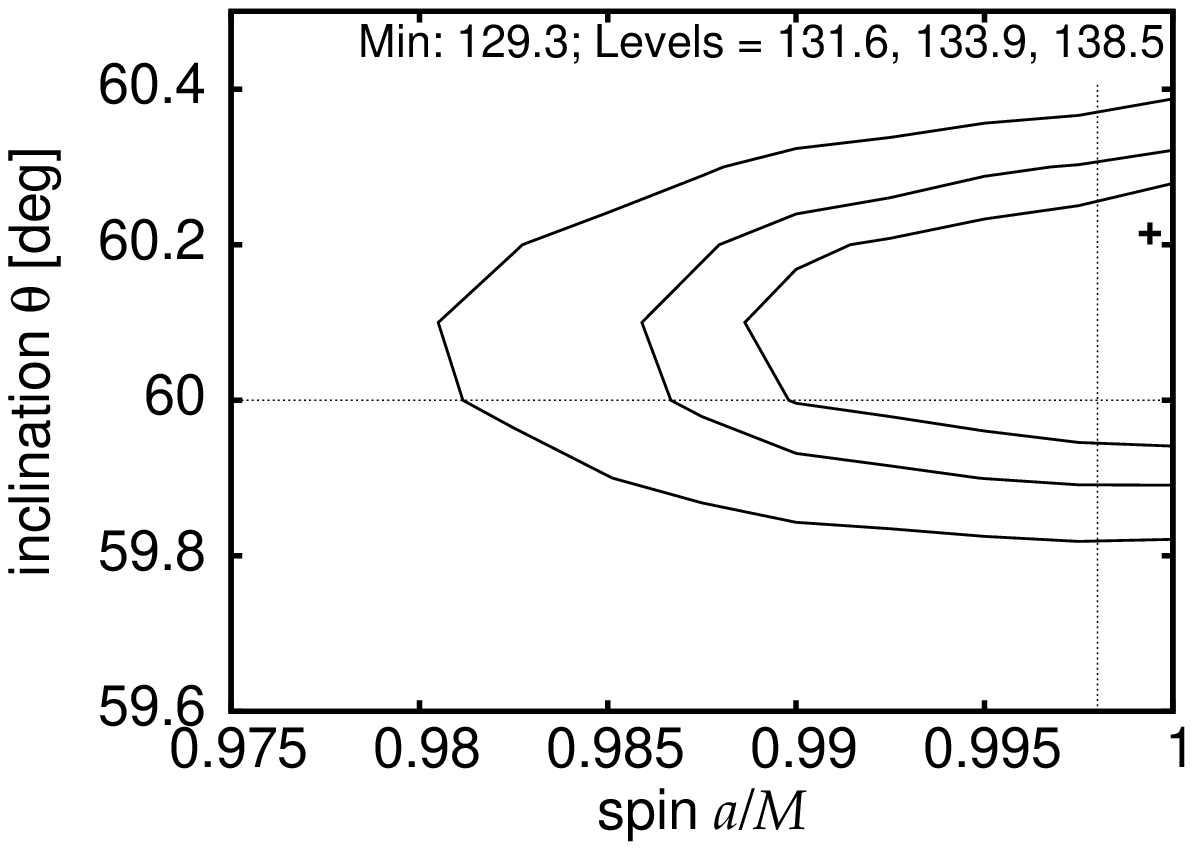}\\
\end{tabular}
\caption{The same as in Figure~\ref{fig7}, but for $a_{\rm f}=0.998$ and $\theta_{\rm f}=60\deg$.}
\label{fig10}
\end{center}
\end{figure*}

We generated a set of ``fake'' spectra (i.e., artificial spectra in the XSPEC 
terminology). These spectra were produced in a grid of angular 
momentum values while assuming isotropic directionality, Case~2 in eq.~({\ref{case123}).
We call the assumed angular moment of the black hole the fiducial spin
and we denote it as $a_{\rm f}$. We performed the fitting loop to these data points using 
each of the three angular emissivity profiles. Once the fit reached convergence, 
we recorded the {\em inferred spin $a$}. 
Figures \ref{fig7}--\ref{fig10} show the results in terms of best-fit 
$\chi^2$ profiles and the confidence contours
for two different {\em fiducial values of the spin}
(we assumed $a_{\rm f}=0.7$, and $0.998$).
We summarise the values for the inferred spin
for two inclination angles $i=30$\,deg and
$i=60$\,deg in Table~\ref{tab1}.

The fitting procedure was performed in two different ways -- having the rest
of the parameters free or keeping them frozen. Obviously the former approach
results in an extremely complicated $\chi^2$ space. Therefore, for simplicity of
the graphical representation, we plot only the results of the second approach which, 
however, gives broadly consistent results (though it misses some 
local minima of $\chi^2$). In other words, the plots have the parameters 
of the power law continuum, the energy of the line and the radial 
dependence parameter fixed at $\Gamma=1.9$, $E_0=6.4$~keV, and $q=3$. 

The conclusion from this analysis is that 
the determination of $a$ indeed seems to be sensitive within certain 
limits to the assumed directionality of the intrinsic emission. 
The suppression of the flux of
the reflection component at high values of $\theta_{\rm e}$ may
lead to overestimating the spin, and vice versa.
The middle panels of Figs.\ \ref{fig7}--\ref{fig10} show the fit results
for isotropic directionality, which was also the seed model
used to generate the test data, and so these contours illustrate
the magnitude of combined dispersion due to the simulated noise 
and the degeneracy between the spin and the inclination.. The fiducial values 
are well inside the $1\sigma$ confidence contour in all the graphs
in the middle panels.

However, systematically lower values of the angular momentum are
obtained for the limb brightening profile and, vice versa, higher values
are found for the limb darkening profile. The magnitude of the difference
is larger for higher values of angular momentum.

\subsection{The angular emission profile of the detailed reprocessing model}
\label{sec-titan-noar-modelling}

\begin{figure*}[tbh!]
\begin{center}
\includegraphics[width=0.48\textwidth,angle=0]{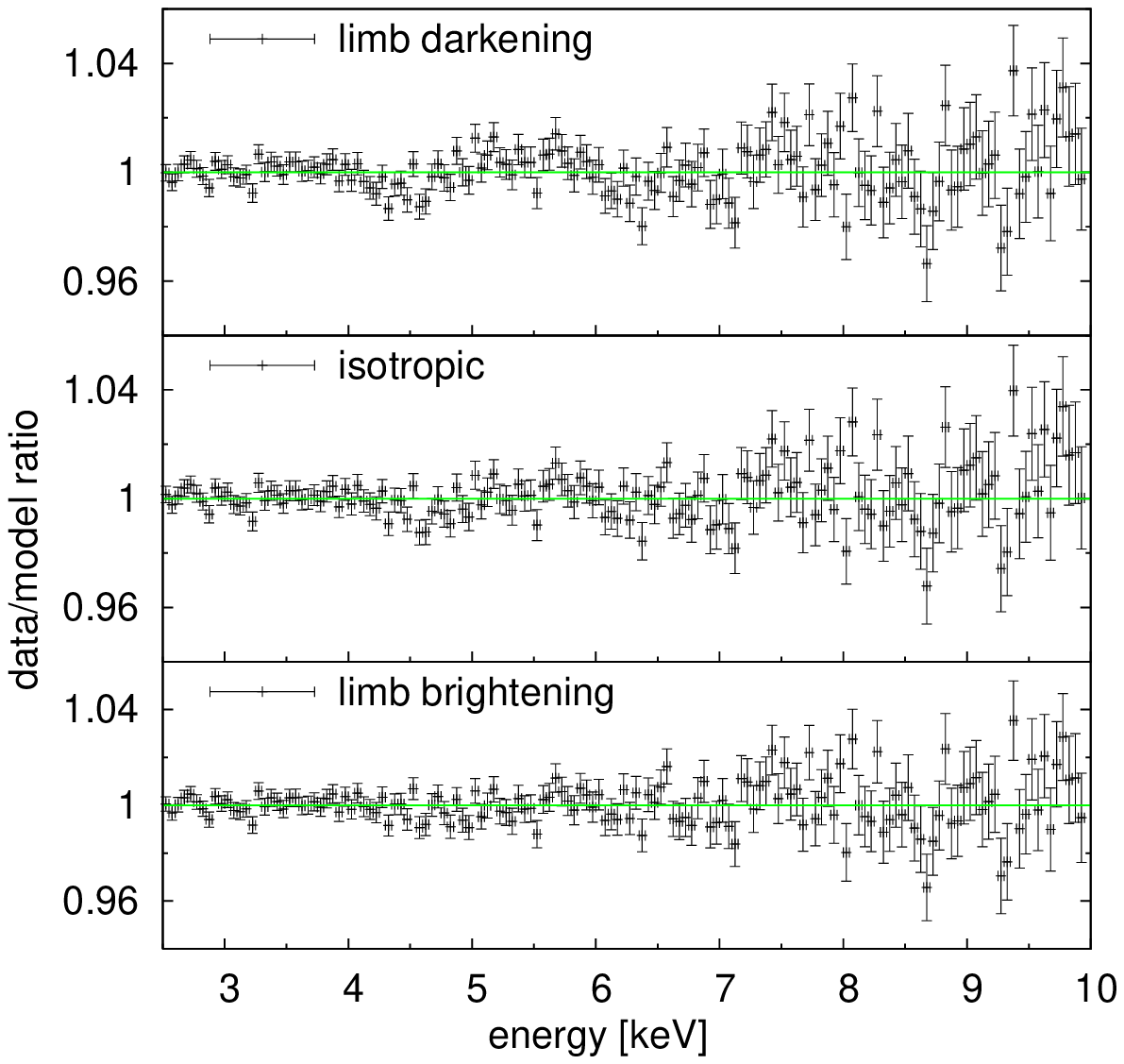}
\hfill
\includegraphics[width=0.48\textwidth,angle=0]{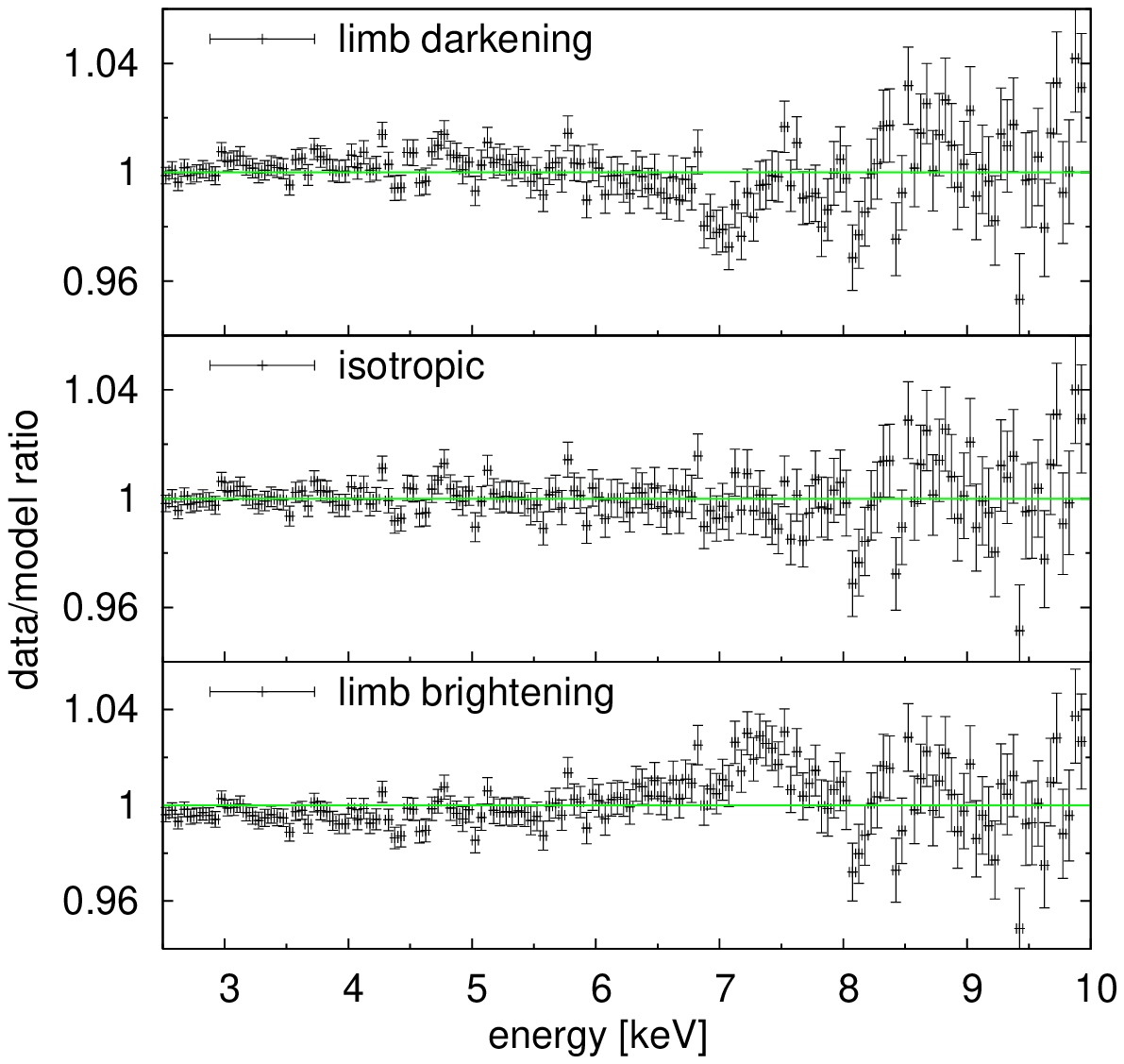}
\caption{Plots of data/model ratio, where the data are simulated as
\textscown{powerlaw} + \textscown{kyl2cr} and the model applied to the data
is \textscown{powerlaw} + \textscown{kyl3cr} with a particular analytical approach
of the directionality. The default angular momentum value is $a_{\rm f}=0.7$,
and the emission angle $\theta_{\rm f}=30^{\circ}$ (left), 
and $\theta_{\rm f}=60^{\circ}$ (right). These parameters 
and the normalisation of the reflection component were allowed 
to vary during the fitting procedure.
The plotted results correspond to the terminal values 
of the parameters obtained during the $\chi^{2}$ minimisation process.
Other parameters of the model were kept frozen at their default
values: $\Gamma=1.9$, $r_{\rm in}=r_{\rm ms}$, $r_{\rm out}=400$, 
$q=3$ and normalisation of the power law $K_{\Gamma} = 10^{-2}$.}
\label{fig_a07_ratio}
\end{center}
\end{figure*}

\begin{figure*}[tbh!]
\begin{center}
\includegraphics[width=0.48\textwidth,angle=0]{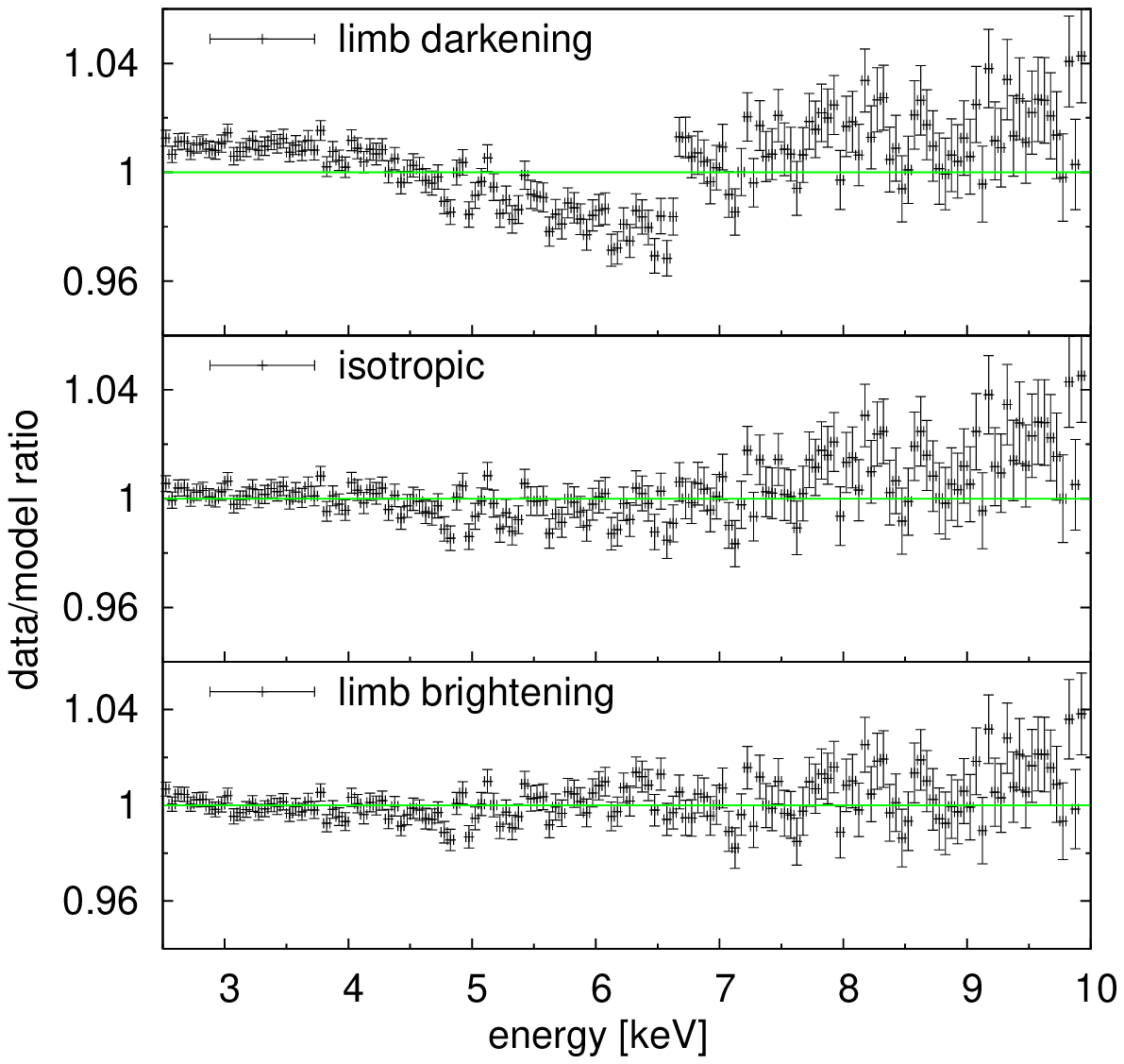}
\hfill
\includegraphics[width=0.48\textwidth,angle=0]{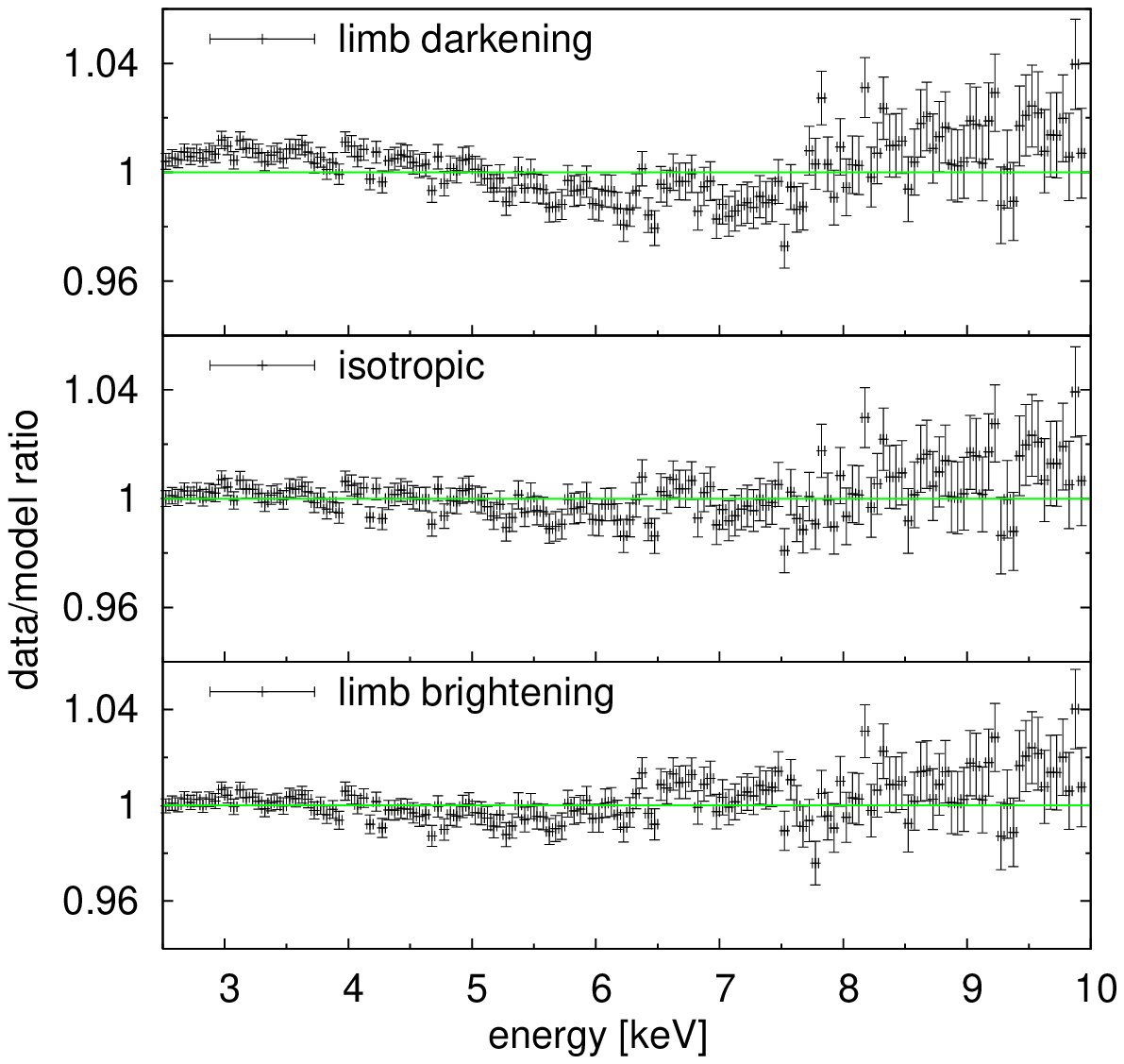}
\caption{The same plots of data/model ratio, but for $a=0.998$
(left: $\theta_{\rm f}=30^{\circ}$, right: $\theta_{\rm f}=60^{\circ}$). }
\label{fig_a0998_ratio}
\end{center}
\end{figure*}

\begin{figure*}[tbh!]
\begin{center}
\begin{tabular}{ccc}
\includegraphics[width=0.31\textwidth]{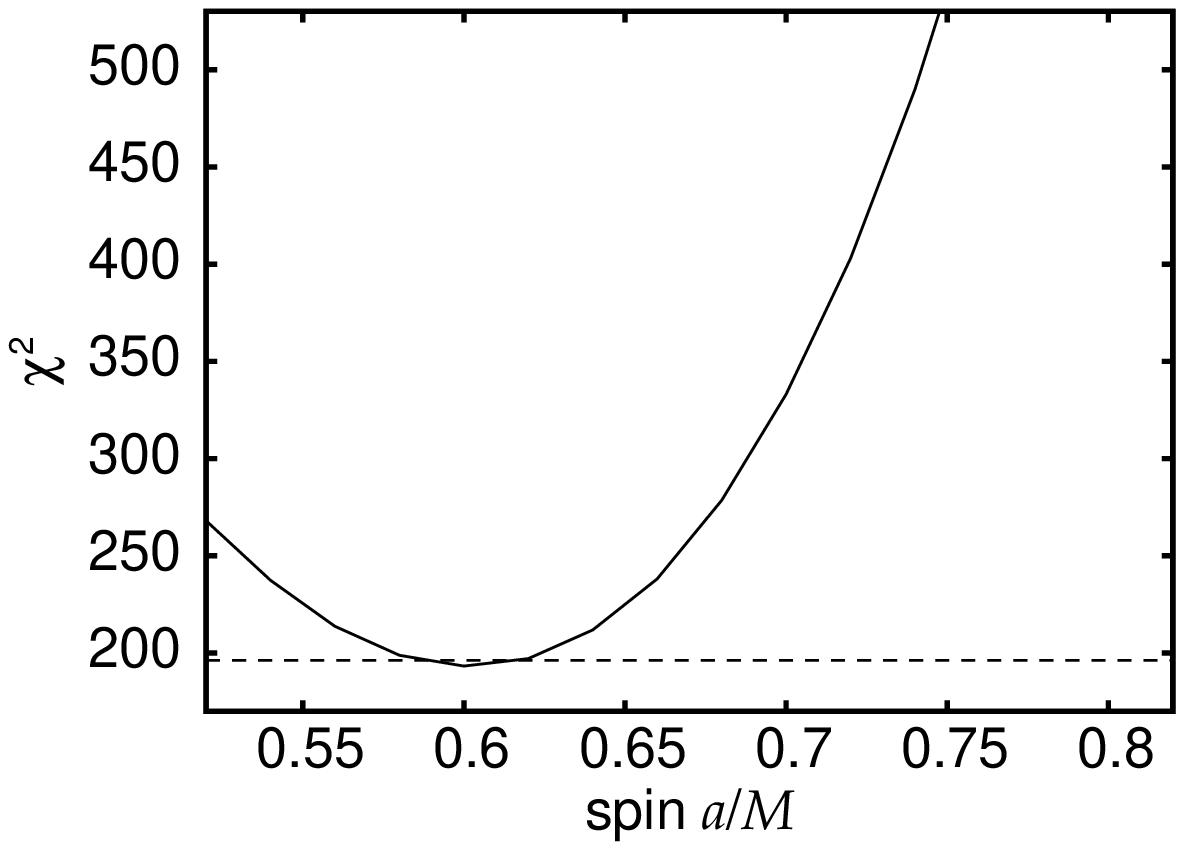} &
\includegraphics[width=0.31\textwidth]{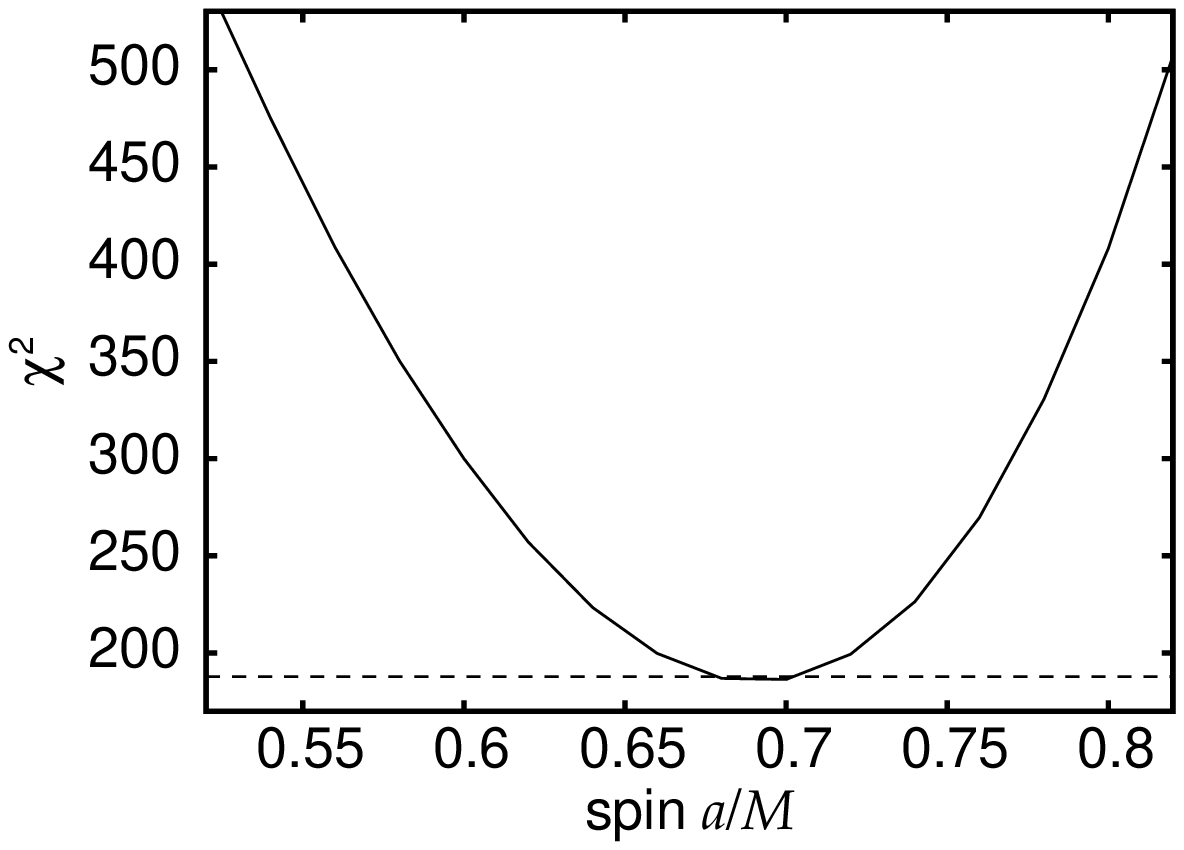} &
\includegraphics[width=0.31\textwidth]{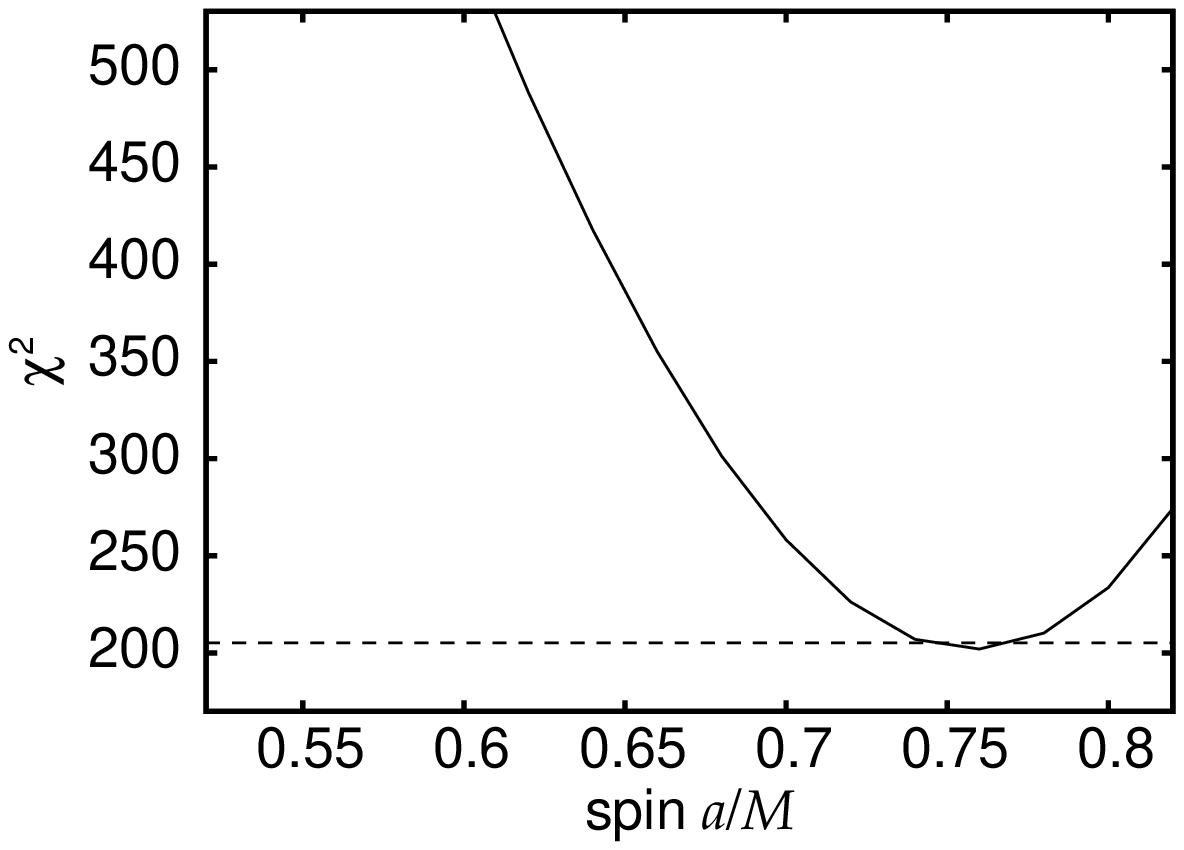} \\
\end{tabular}
\begin{tabular}{ccc}
\includegraphics[width=0.31\textwidth]{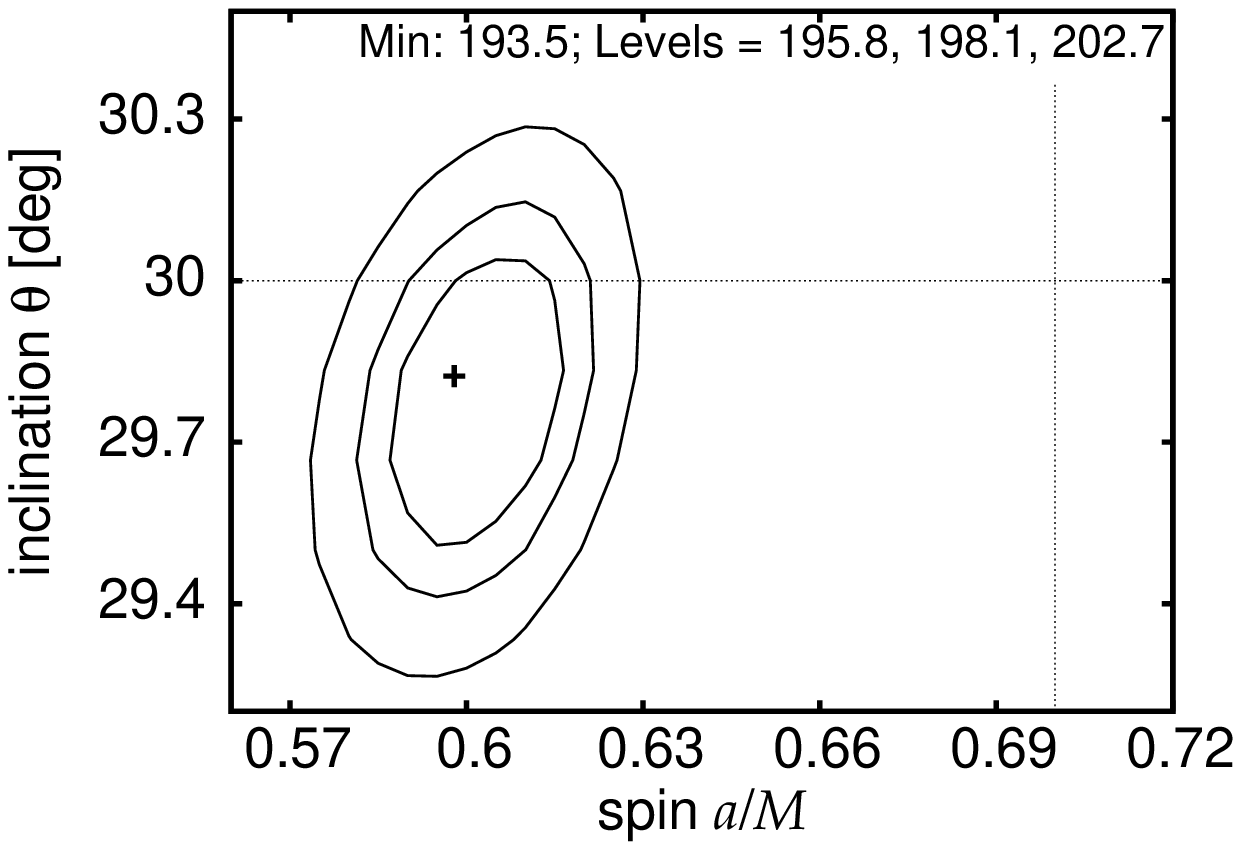} &
\includegraphics[width=0.31\textwidth]{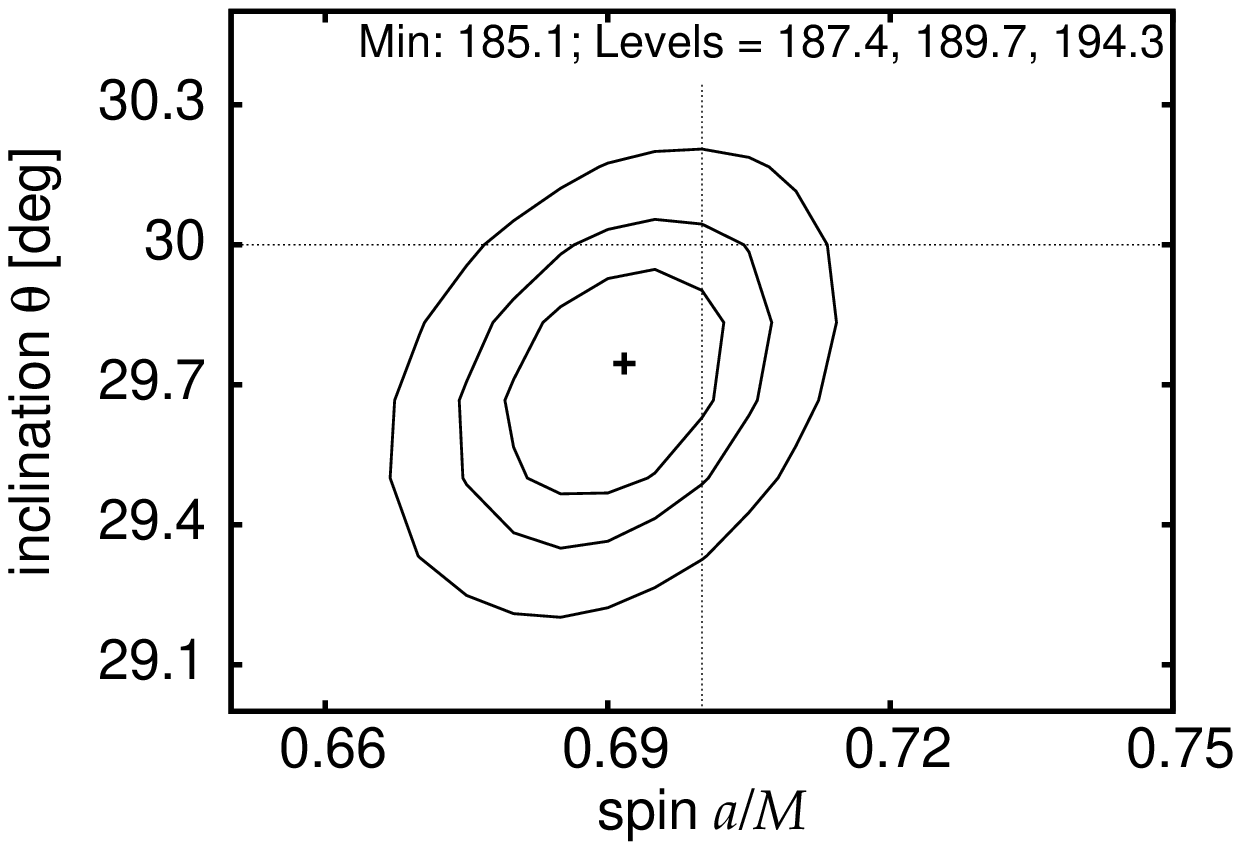} &
\includegraphics[width=0.31\textwidth]{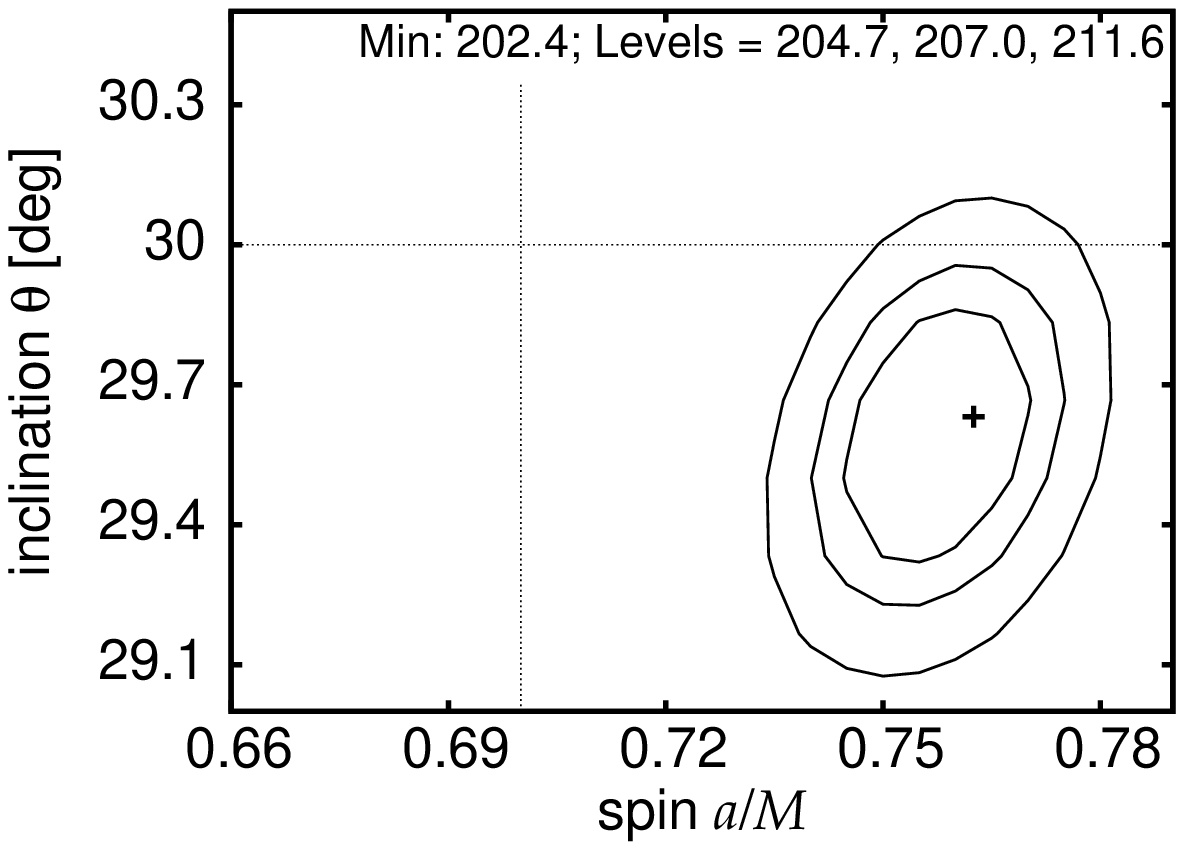}\\
\end{tabular}
\caption{Results from the test fits with $a_{\rm f}=0.7$ and $\theta_{\rm f}=30\deg$,
using the \textscown{powerlaw} + \textscown{kyl3cr} model applied to the data simulated
with \textscown{powerlaw} + \textscown{kyl2cr}. Three different profiles
of the emission directionality are shown in columns -- left:
limb brightening, middle: isotropic, right: limb darkening. 
Top: dependence of the best fit $\chi^2$ values on the fiducial spin value.
The horizontal (dashed) line represents the 90\% confidence level.
Bottom: contour graphs of $a$ versus $\theta_{\rm o}$. 
The contour lines refer to $1$, $2$, and $3$ sigma levels.
The position of the minimal value of $\chi^{2}$ is marked
with a small cross. The values of $\chi^{2}$ corresponding to the minimum
and to the contour levels are shown at the top of each contour graph.
The large cross indicates the position of the fiducial values of 
the angular momentum and the emission angle.
Other parameters of the model were kept fixed at default
values: $\Gamma=1.9$, $r_{\rm in}=r_{\rm ms}$, $r_{\rm
out}=400$,  $q=3$ and normalisation of the power law
$K_\Gamma=10^{-2}$.}
\label{fig_a07i30}
\end{center}
\end{figure*}

\begin{figure*}[tbh!]
\begin{center}
\begin{tabular}{ccc}
\includegraphics[width=0.31\textwidth]{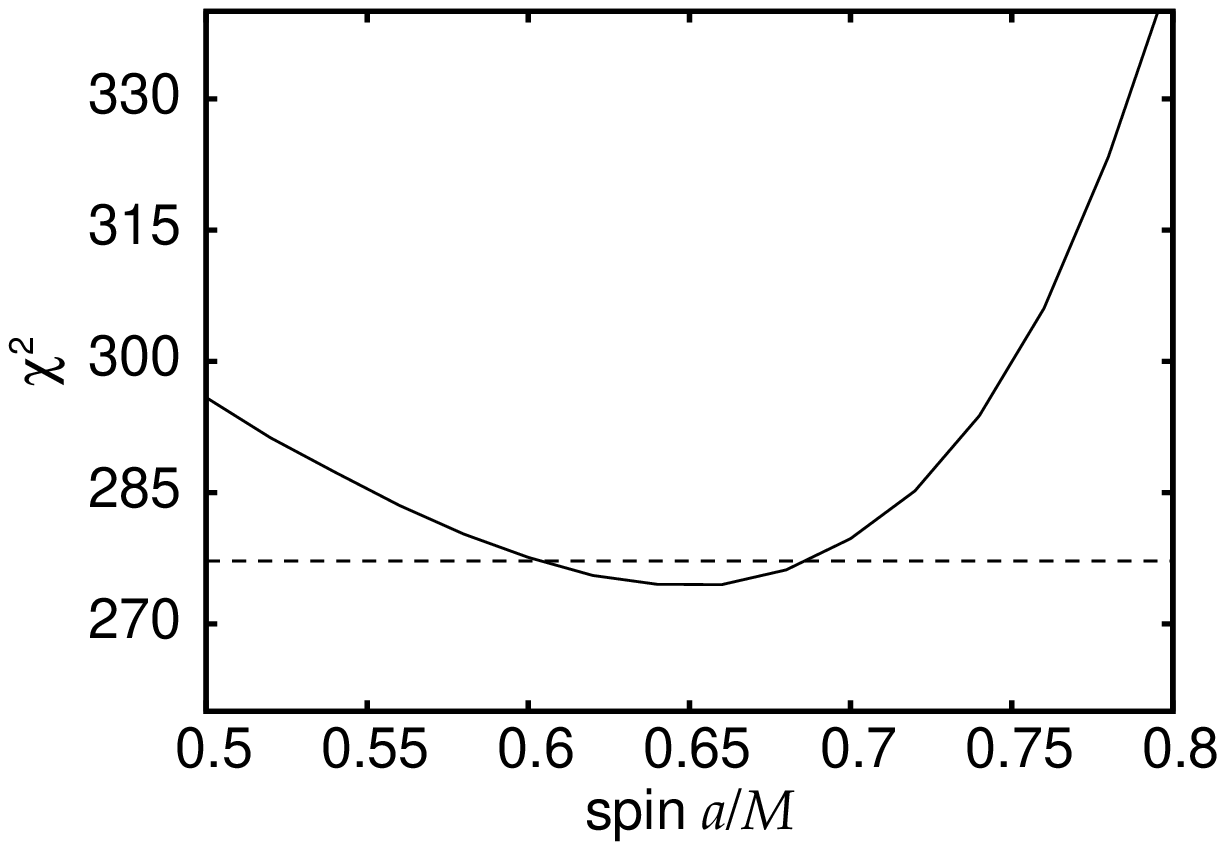} &
\includegraphics[width=0.31\textwidth]{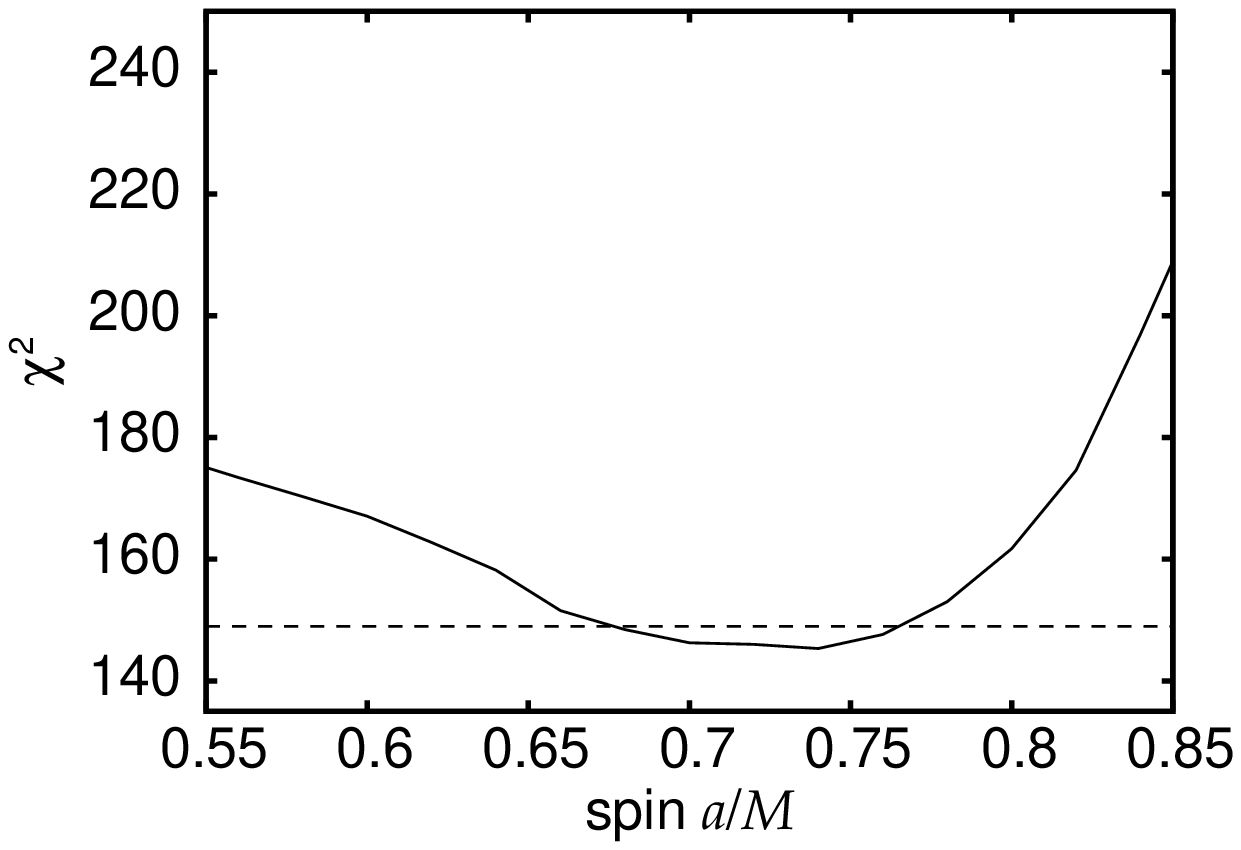} &
\includegraphics[width=0.31\textwidth]{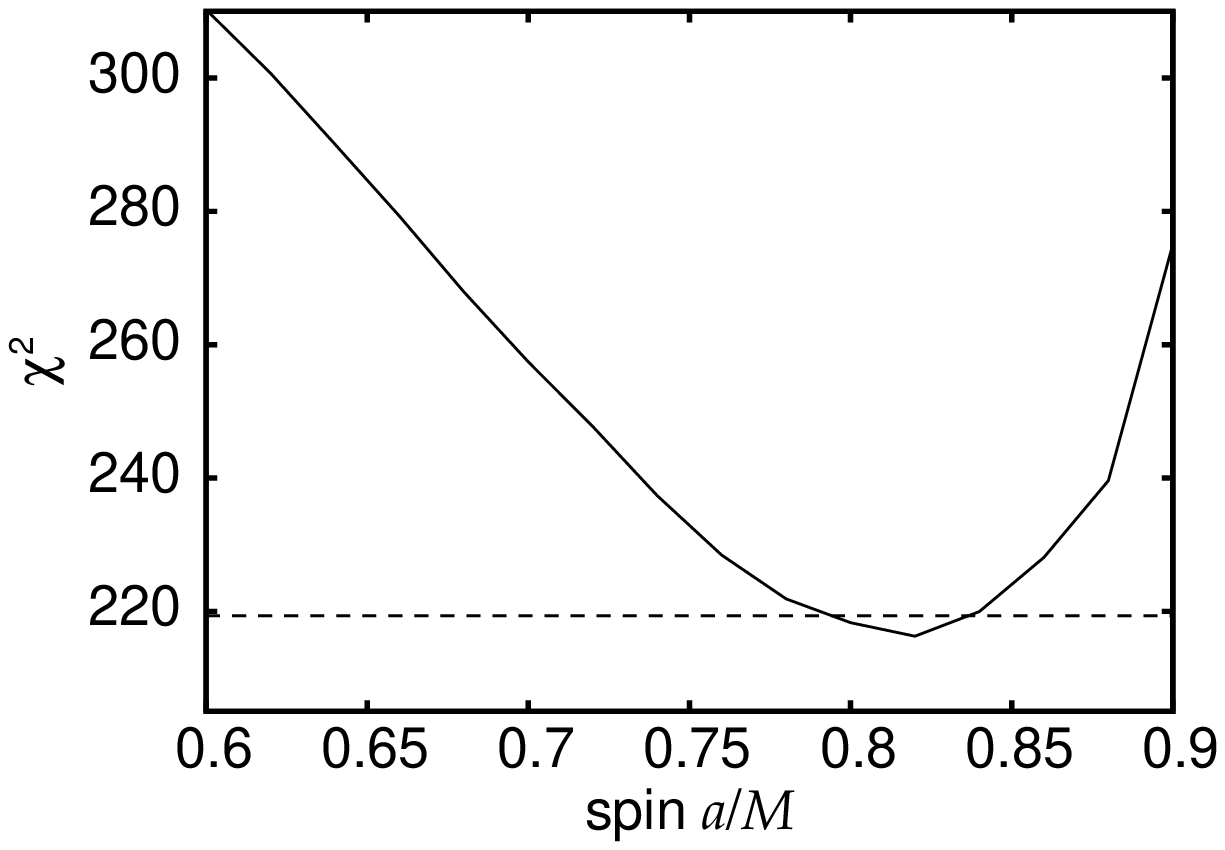} \\
\end{tabular}
\begin{tabular}{ccc}
  \includegraphics[width=0.31\textwidth]{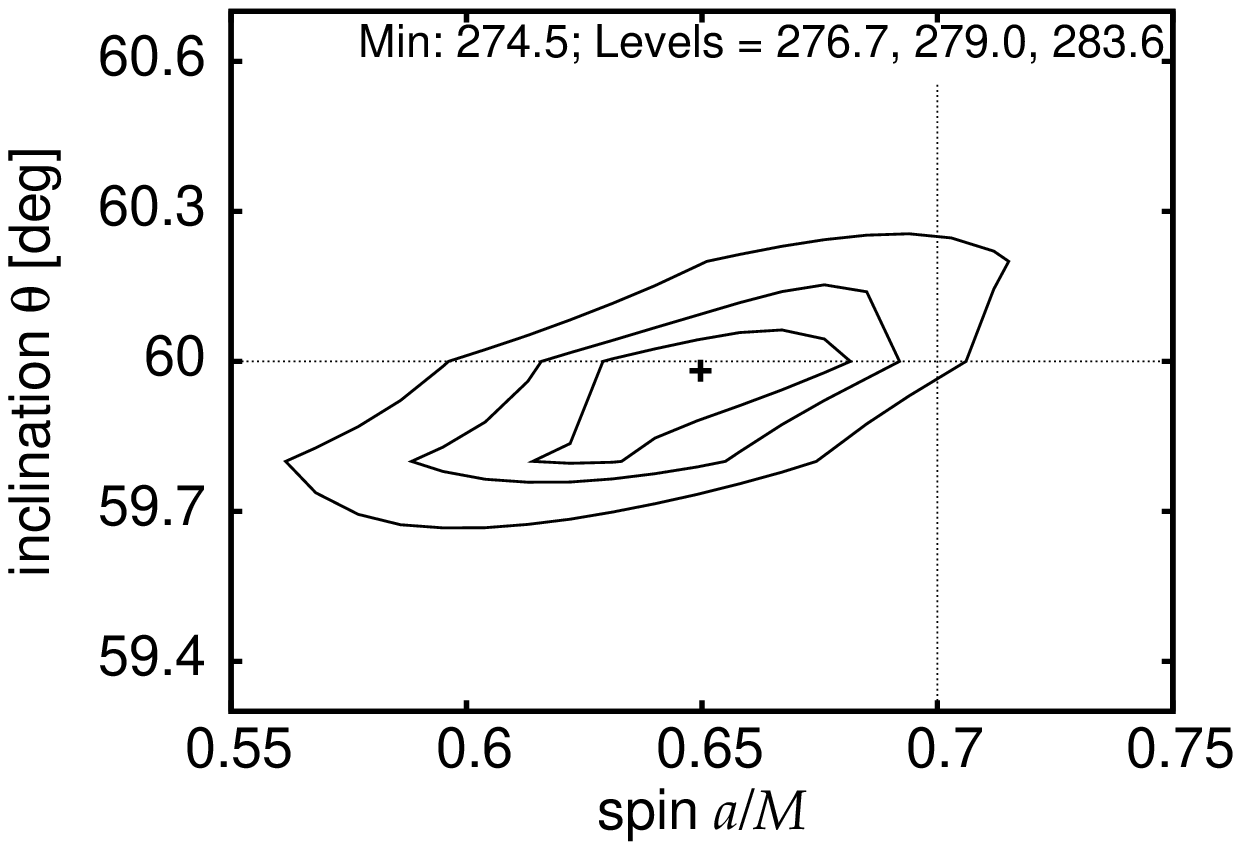} &
  \includegraphics[width=0.31\textwidth]{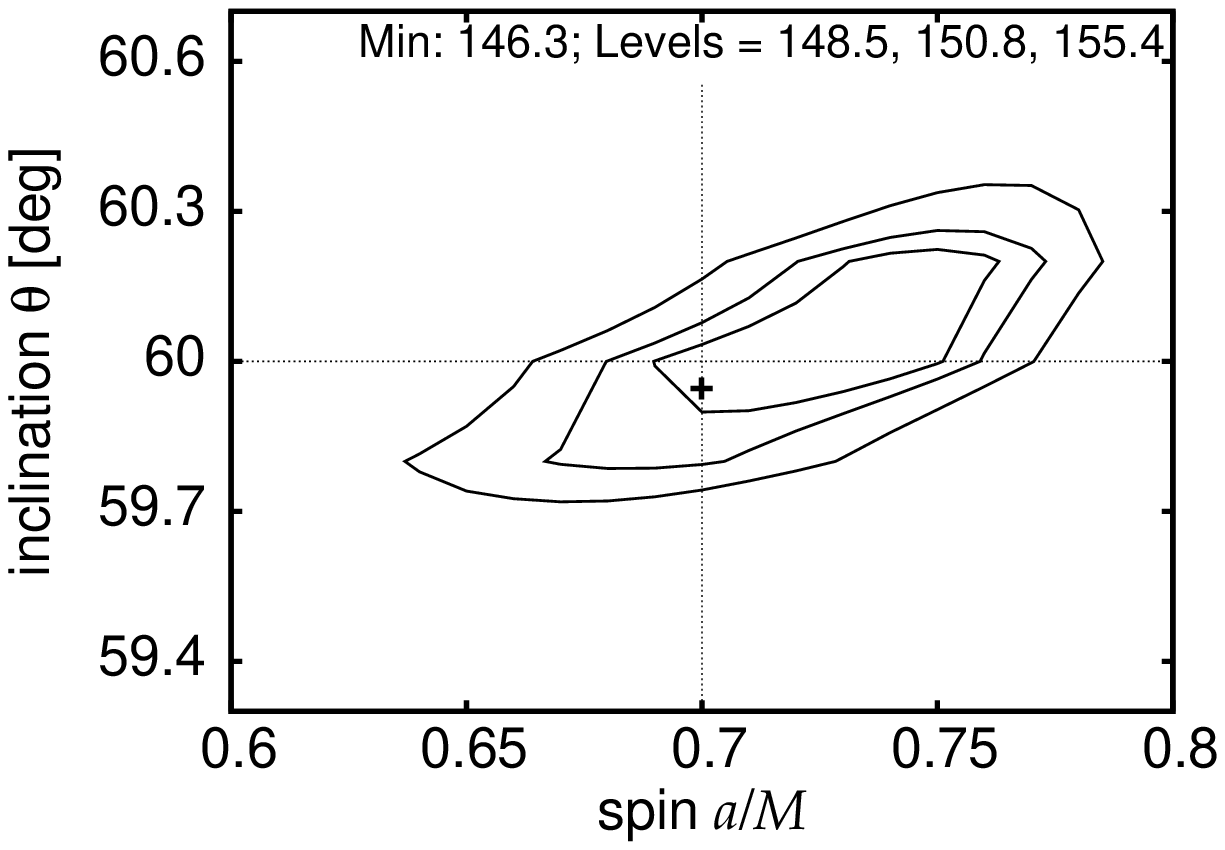} &
  \includegraphics[width=0.31\textwidth]{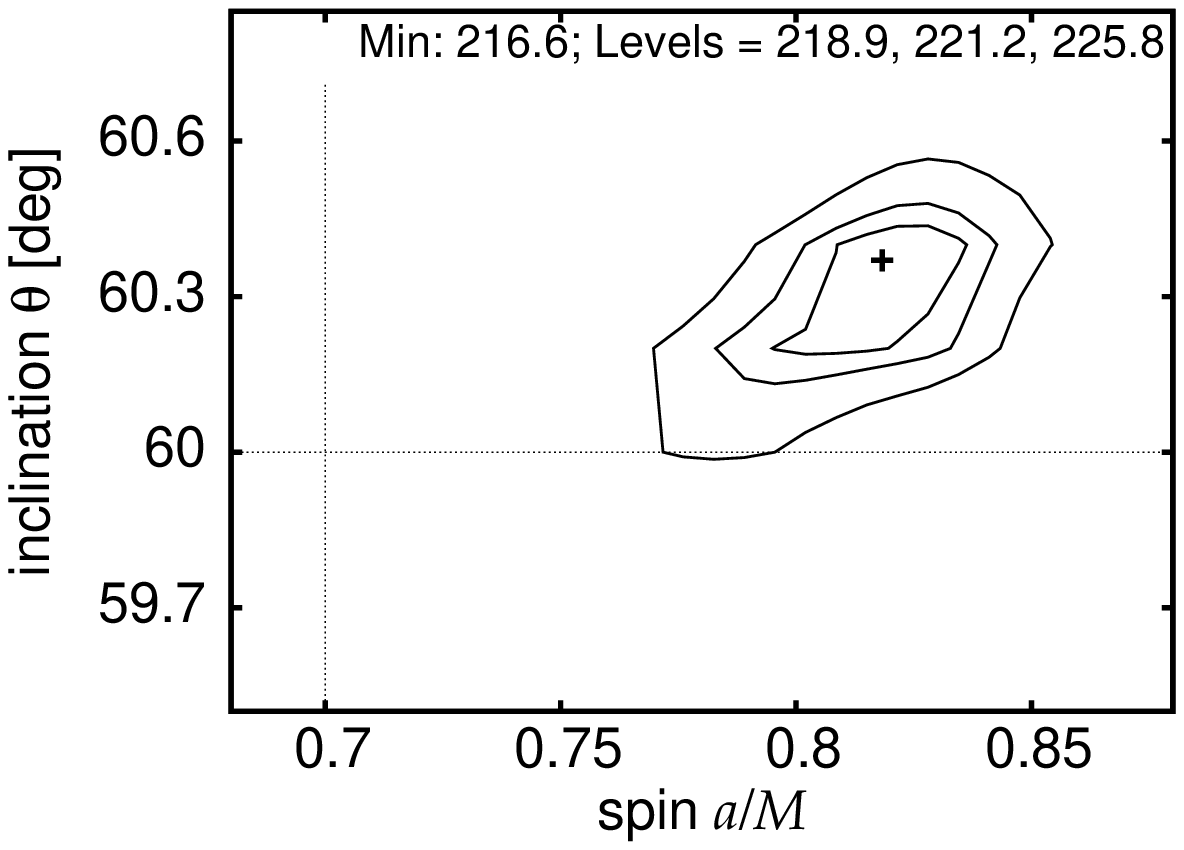}\\
\end{tabular}
\caption{The same as in Figure~\ref{fig_a07i30}, but for 
$a_{\rm f}=0.7$ and $\theta_{\rm f}=60\deg$.}
\label{fig_a07i60}
\end{center}
\end{figure*}

\begin{figure*}[tbh!]
\begin{center}
\begin{tabular}{ccc}
\includegraphics[width=0.31\textwidth]{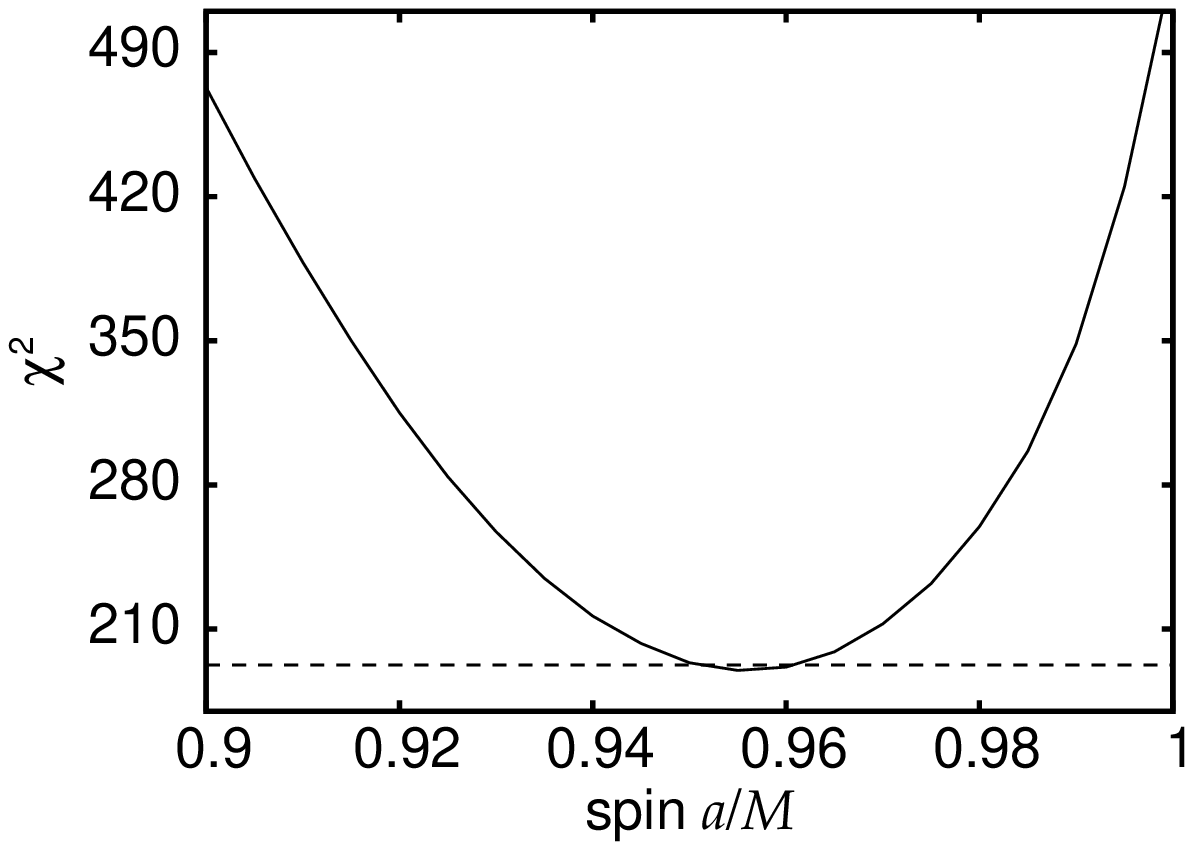} &
\includegraphics[width=0.31\textwidth]{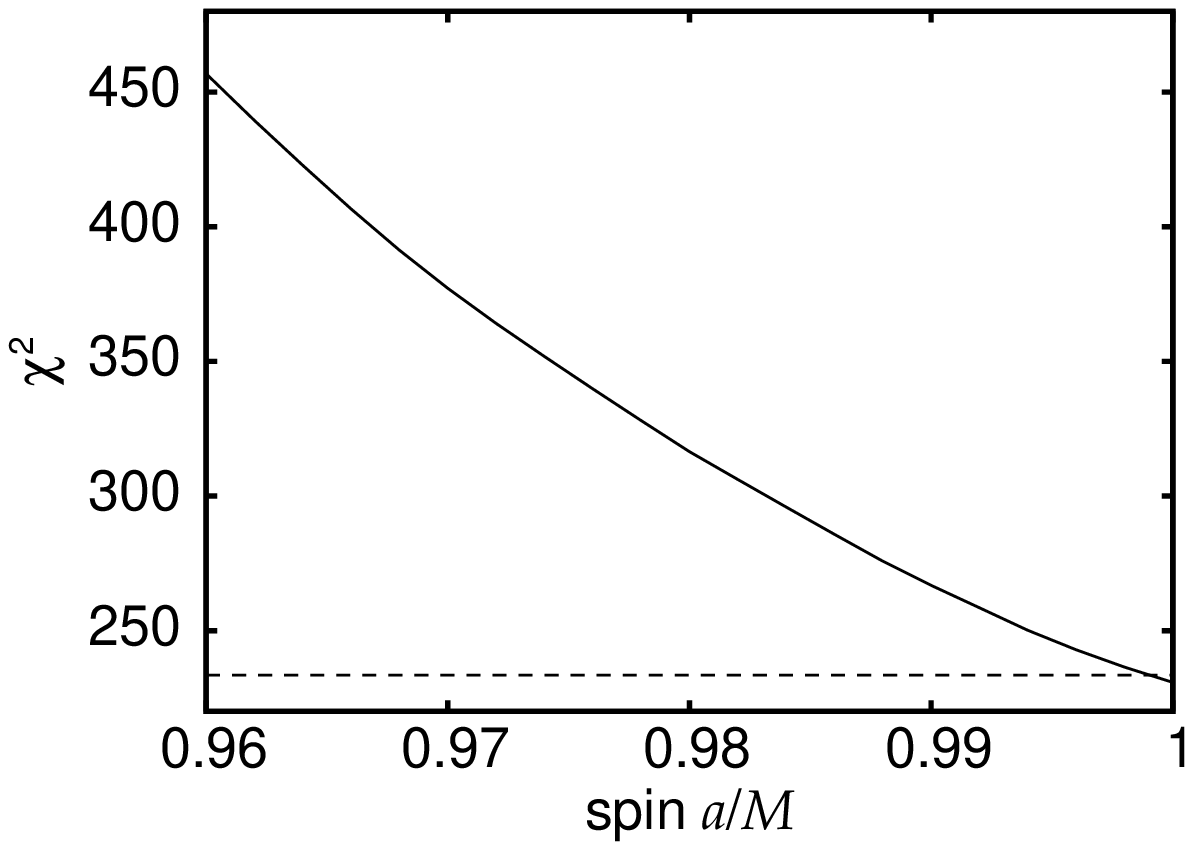} &
\includegraphics[width=0.31\textwidth]{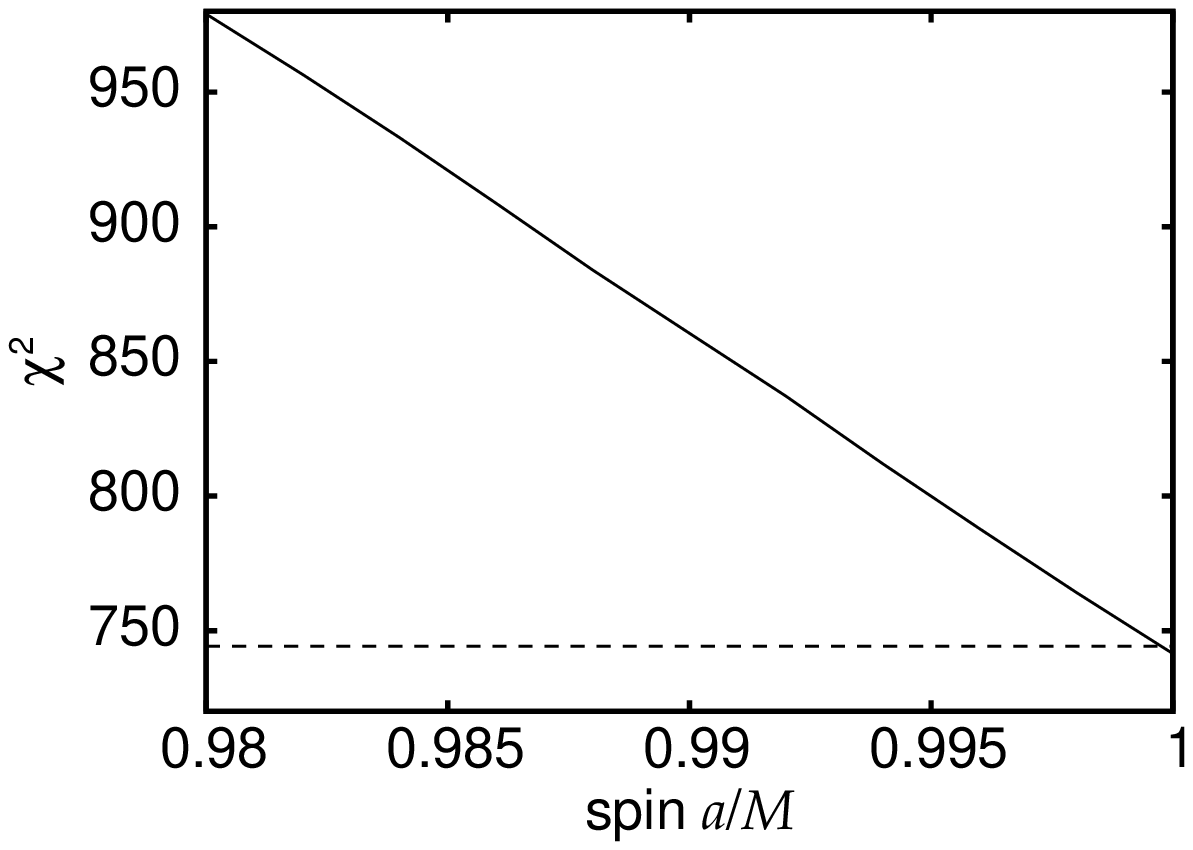} \\
\end{tabular}
\begin{tabular}{ccc}
  \includegraphics[width=0.31\textwidth]{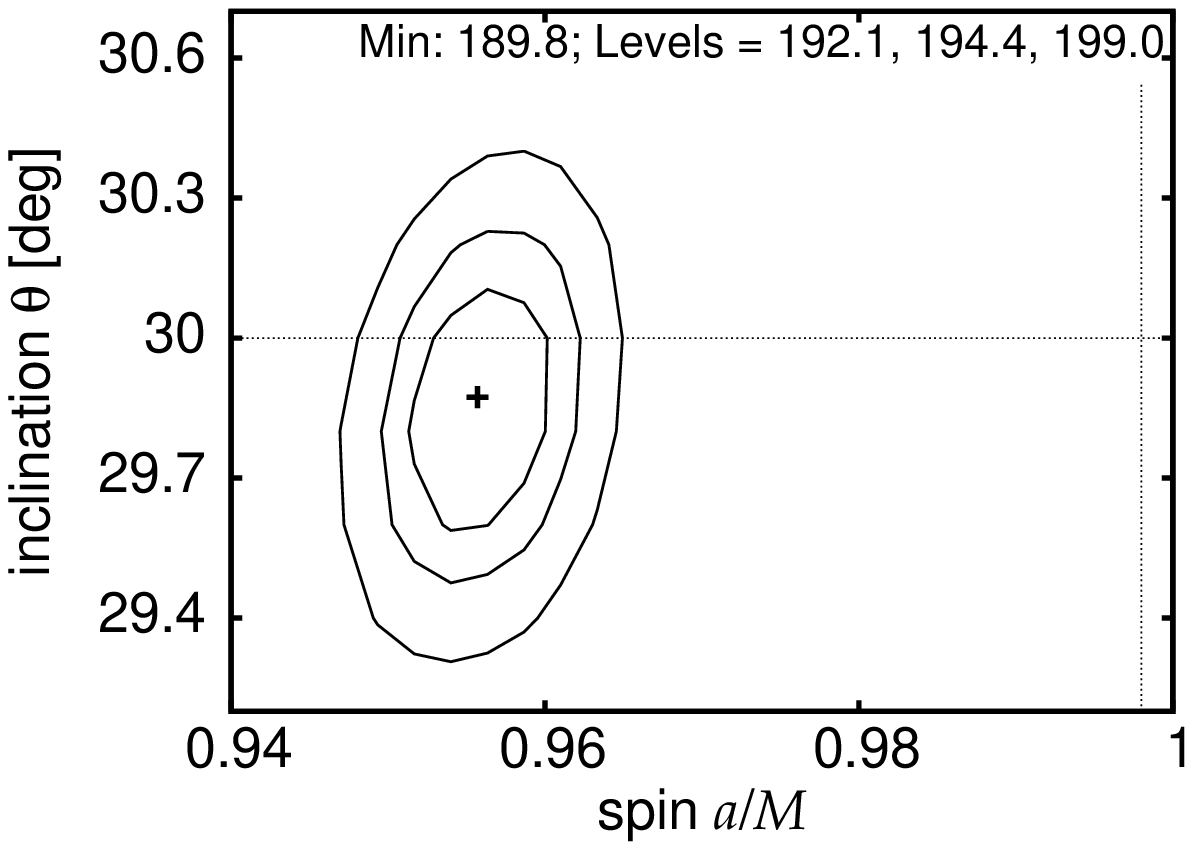} &
  \includegraphics[width=0.31\textwidth]{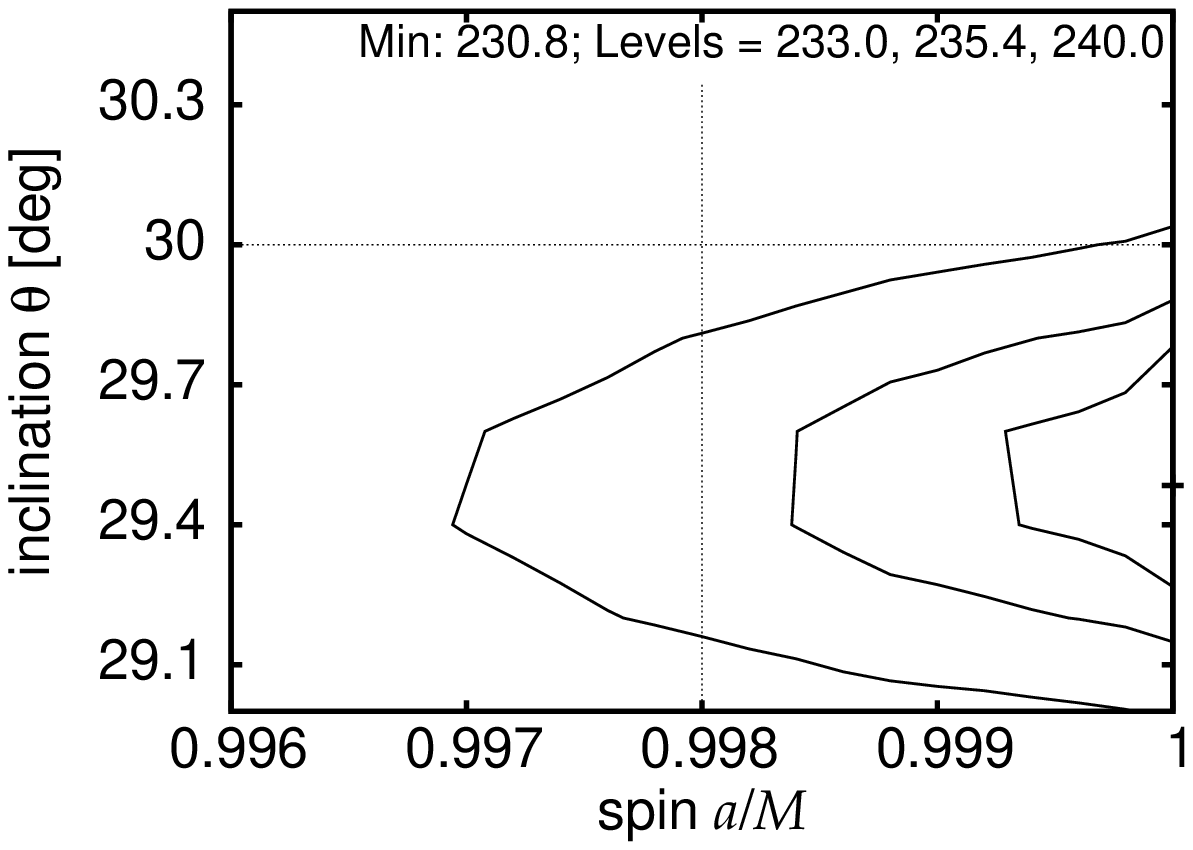} &
  \includegraphics[width=0.31\textwidth]{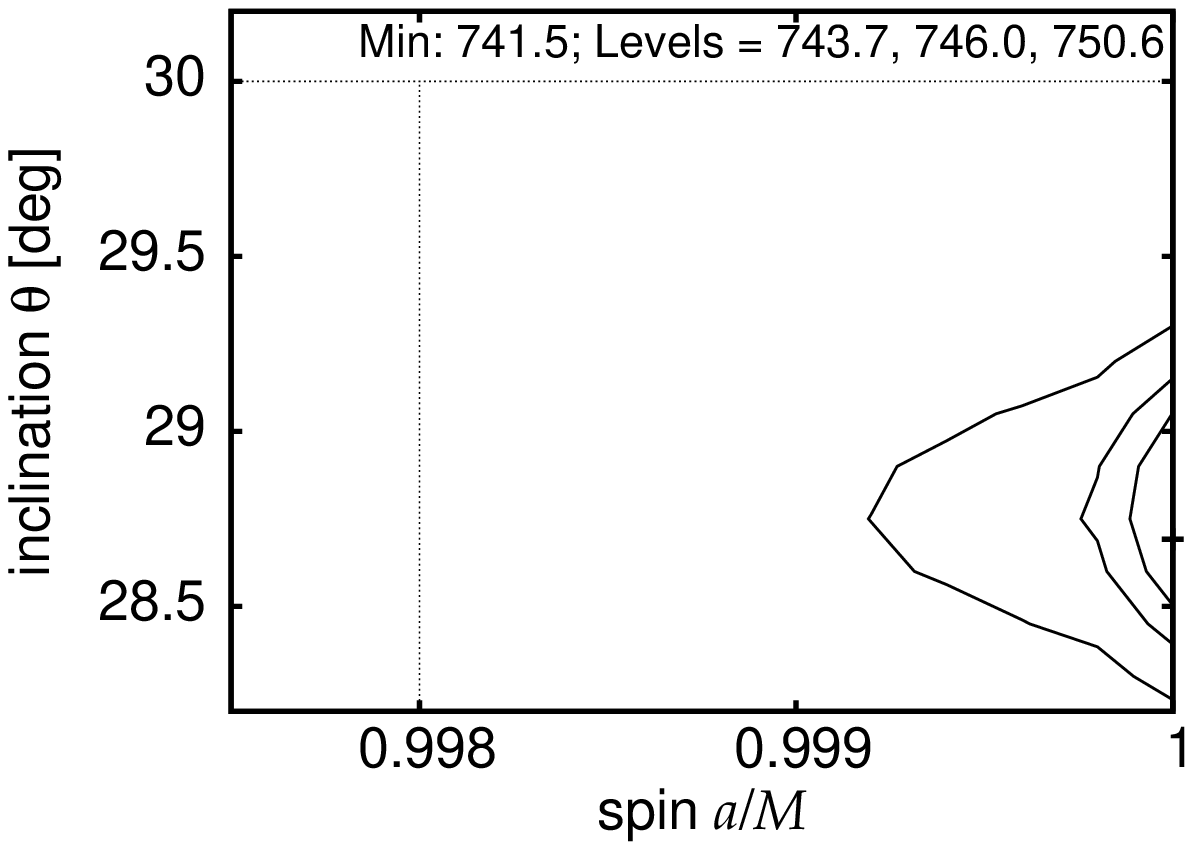}\\
\end{tabular}
\caption{The same as in Figure~\ref{fig_a07i30}, but for 
$a_{\rm f}=0.998$ and $\theta_{\rm f}=30\deg$.}
\label{fig_a0998i30}
\end{center}
\end{figure*}

\begin{figure*}[tbh!]
\begin{center}
\begin{tabular}{ccc}
\includegraphics[width=0.31\textwidth]{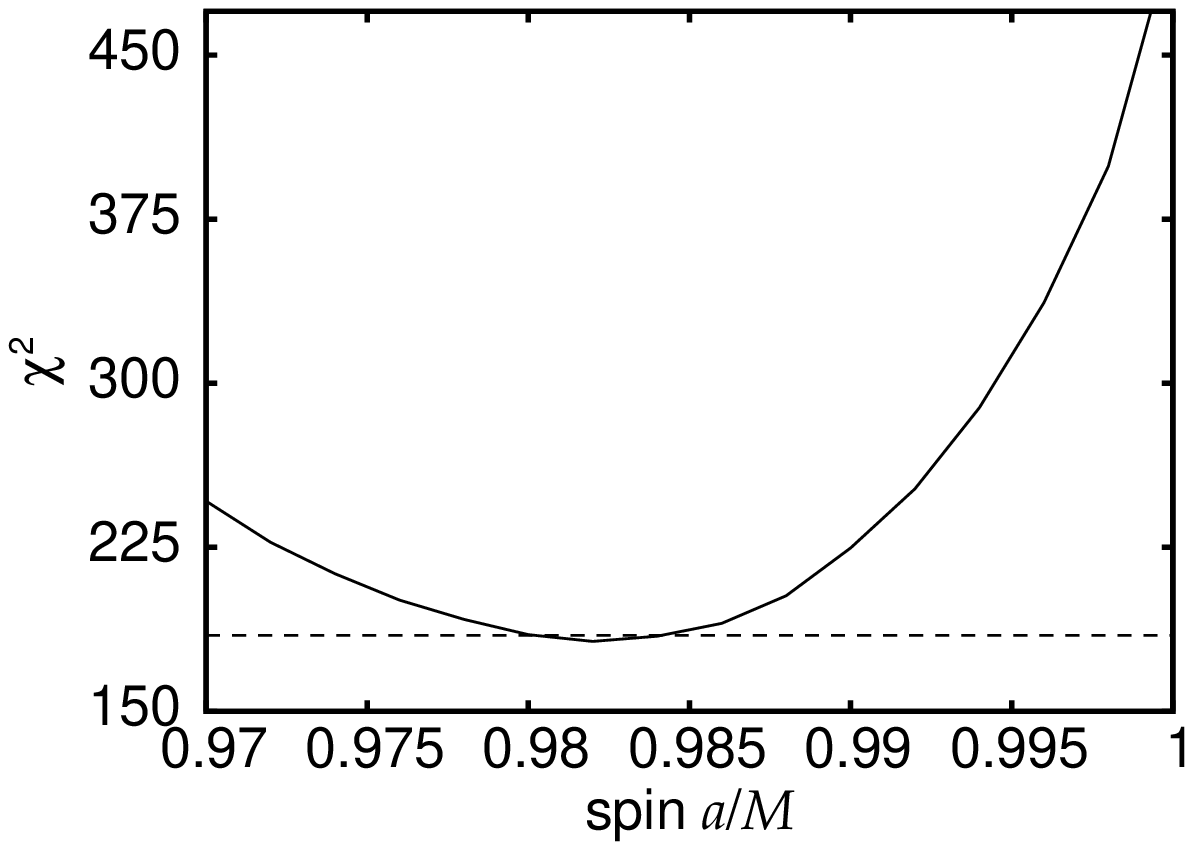} &
\includegraphics[width=0.31\textwidth]{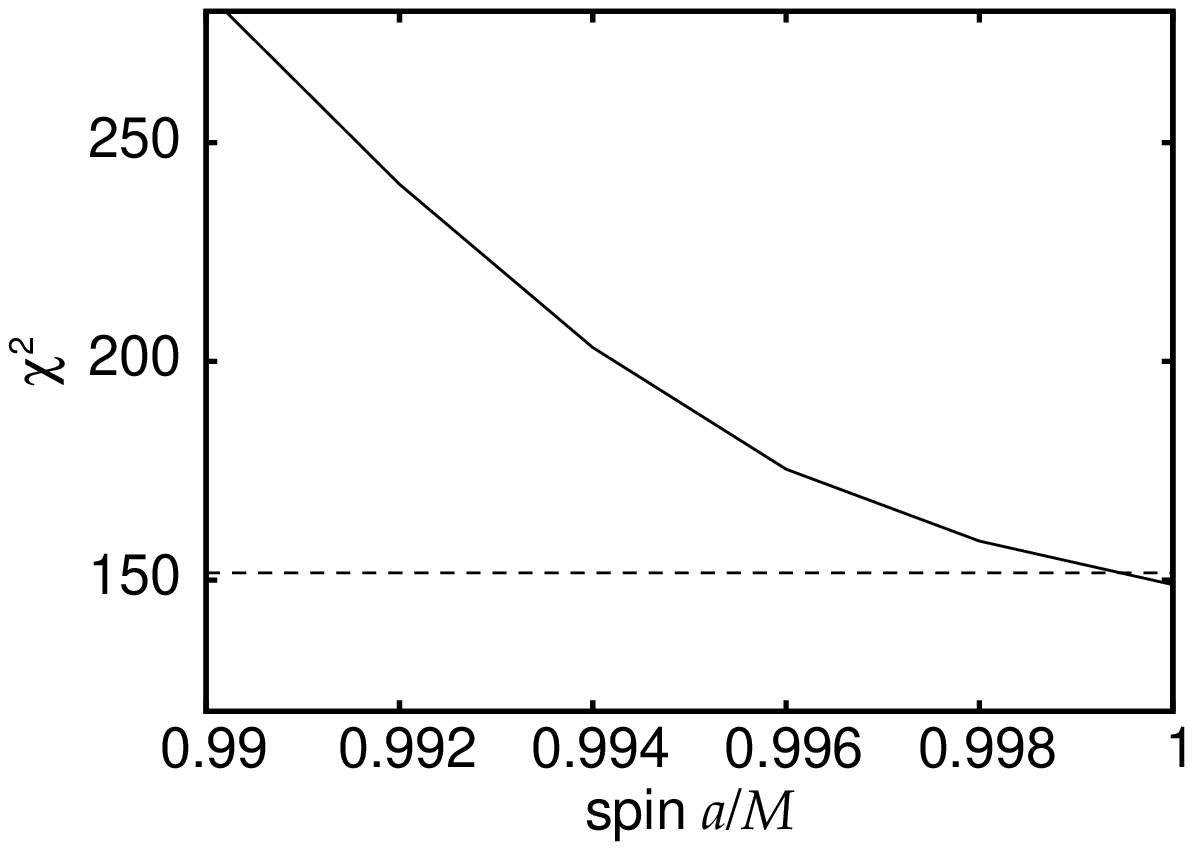} &
\includegraphics[width=0.31\textwidth]{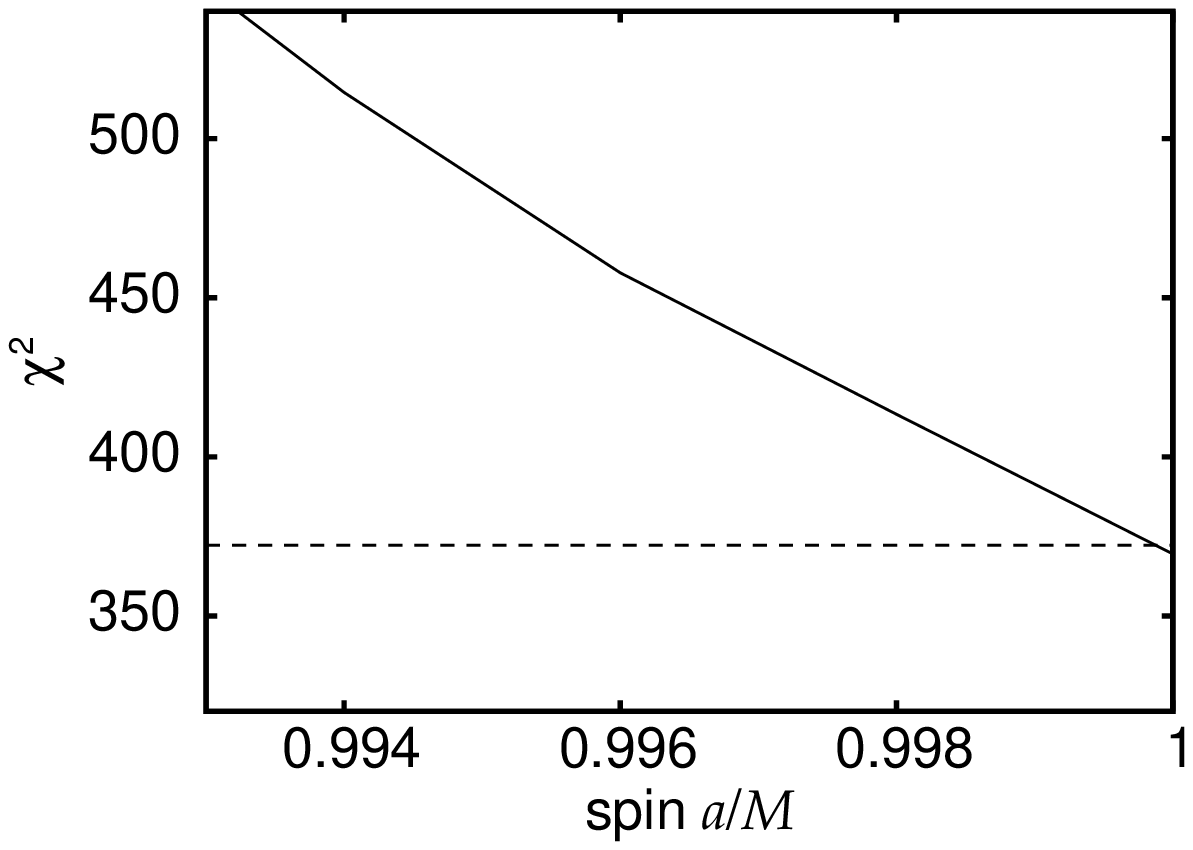} \\
\end{tabular}
\begin{tabular}{ccc}
  \includegraphics[width=0.31\textwidth]{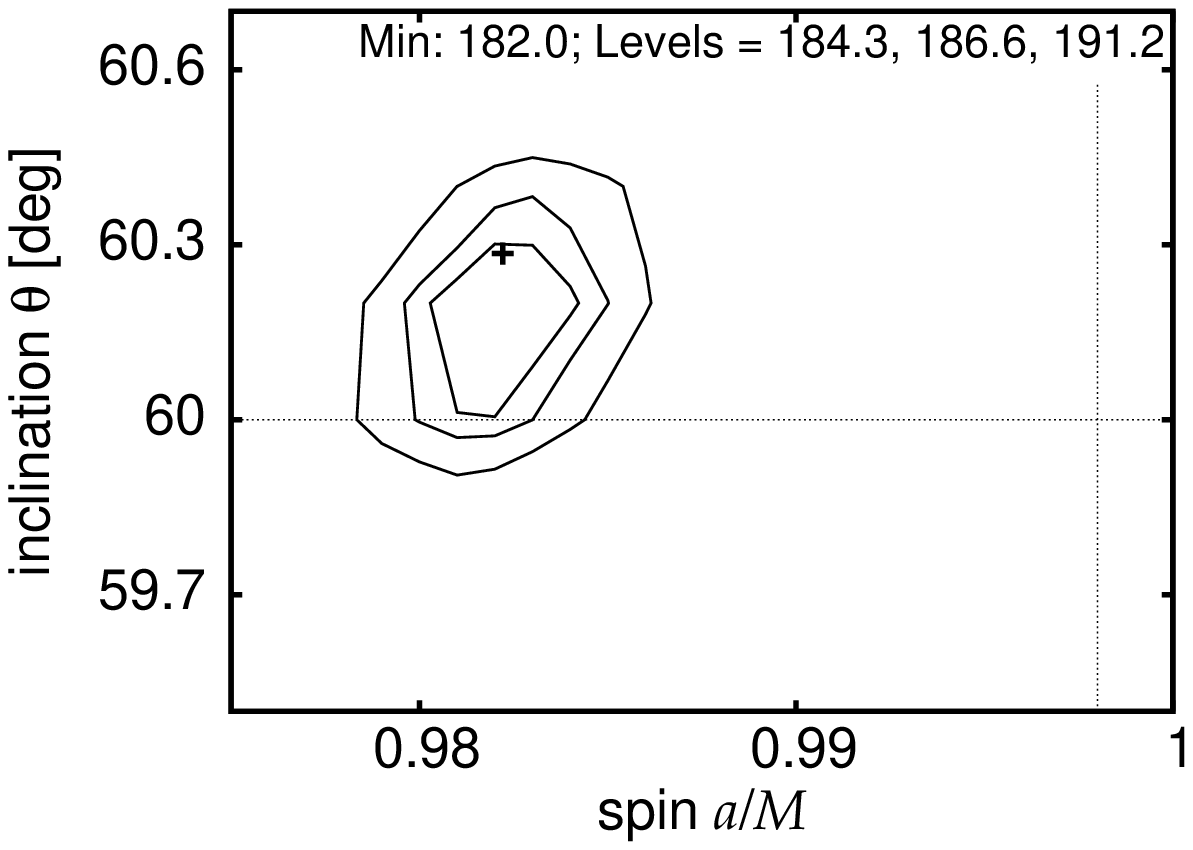} &
  \includegraphics[width=0.31\textwidth]{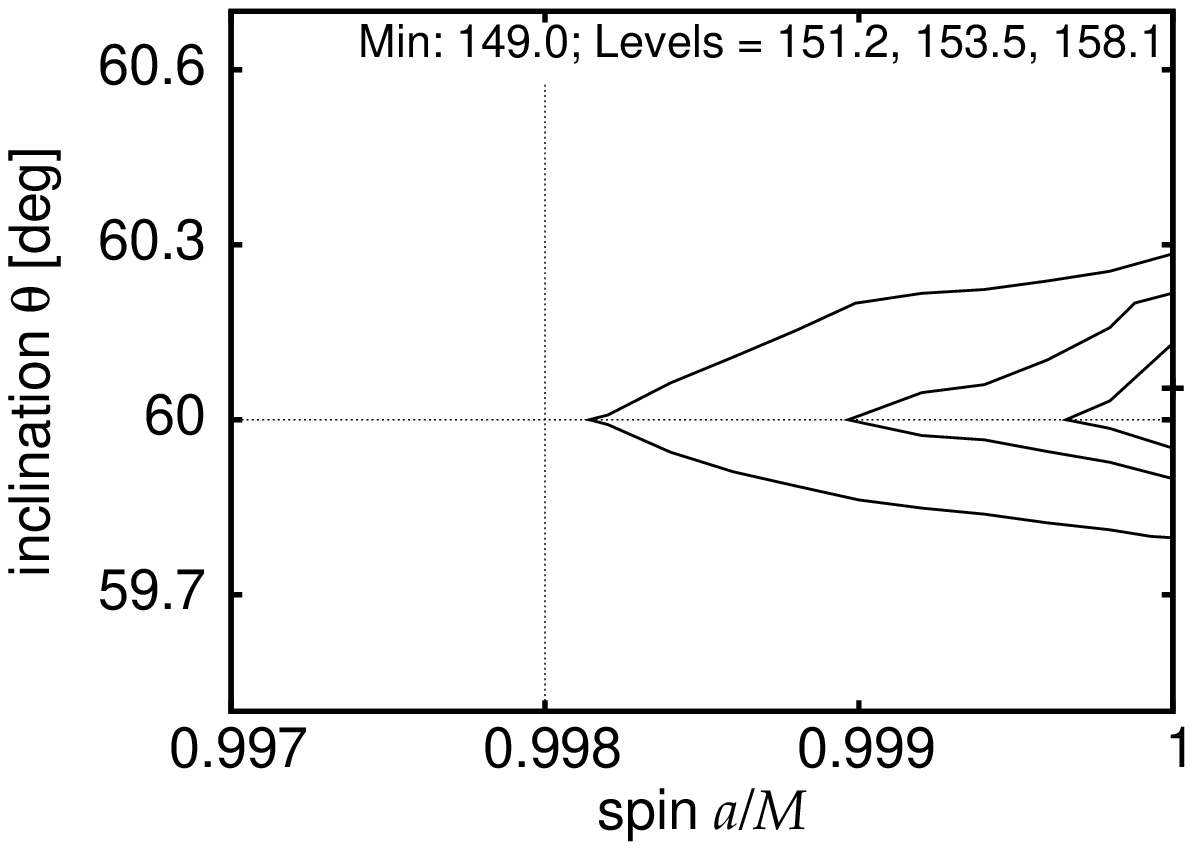} &
  \includegraphics[width=0.31\textwidth]{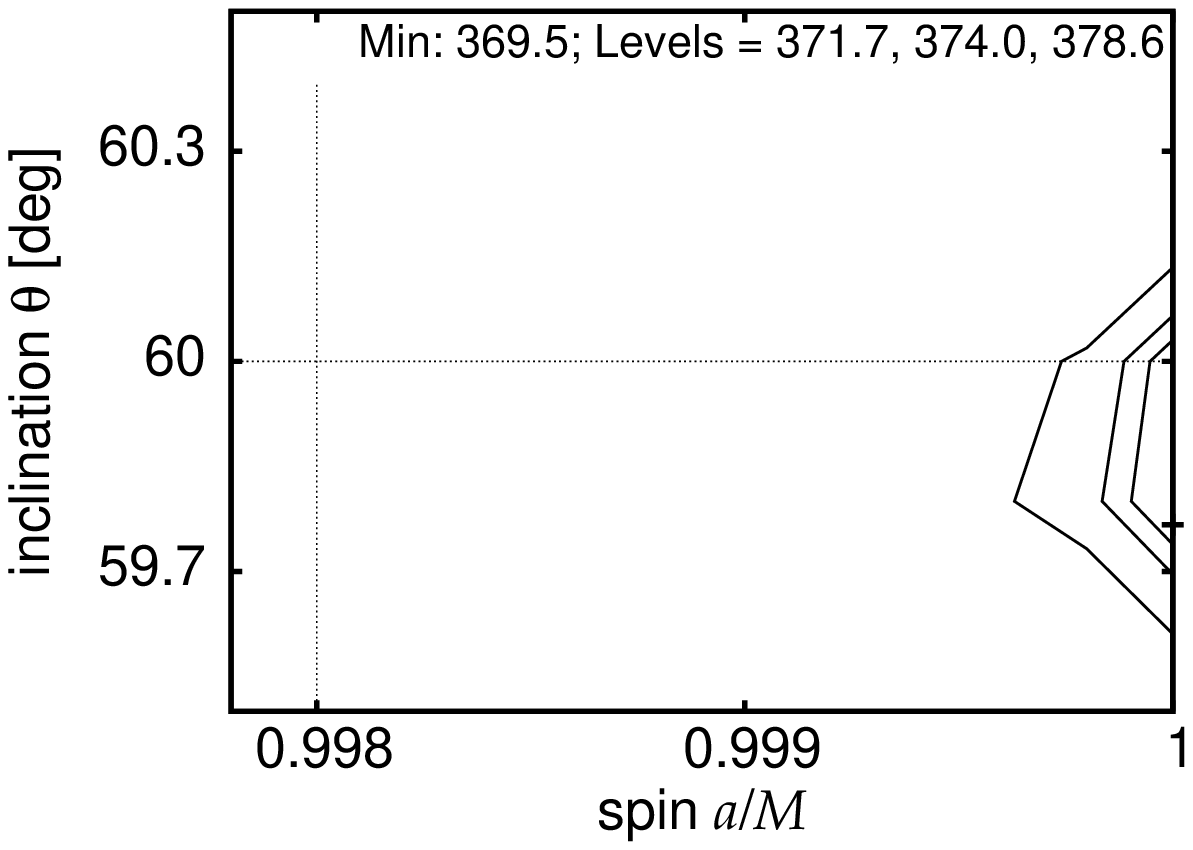}\\
\end{tabular}
\caption{The same as in Figure~\ref{fig_a07i30}, but for 
$a_{\rm f}=0.998$ and $\theta_{\rm f}=60\deg$.}
\label{fig_a0998i60}
\end{center}
\end{figure*}

\begin{table}
\begin{center}
\begin{tabular}{c|ccc} 
\rule[-0.7em]{0pt}{2em}  & Case 1 & Case 2 & Case 3 \\ \hline
\rule[-1.3em]{0pt}{3em} & \multicolumn{3}{c}{$a_{\rm f}=0.7$, $\theta_{\rm f}=30^{\circ}$}\\
\rule[-0.7em]{0pt}{2em} $a$ & $0.60^{+0.02}_{-0.01}$ & $0.69^{+0.01}_{-0.01}$ & $0.76^{+0.01}_{-0.02}$ \\ 
\rule[-0.7em]{0pt}{2em} $\theta_{\rm{}o}$ [deg] & $29.8^{+0.2}_{-0.3}$ & $29.7^{+0.3}_{-0.3}$ & $29.6^{+0.3}_{-0.3}$ \\
\rule[-0.7em]{0pt}{2em} $\chi^{2}/\nu$ & 	1.33	   & 1.27	    & 1.39 \\
\hline
\hline
\rule[-1.3em]{0pt}{3em}  & \multicolumn{3}{c}{$a_{\rm f}=0.7$, $\theta_{\rm f}=60^{\circ}$}\\
\rule[-0.7em]{0pt}{2em} $a$ & $0.65^{+0.03}_{-0.05}$ & $0.73^{+0.03}_{-0.04}$ & $0.82^{+0.02}_{-0.02}$ \\ 
\rule[-0.7em]{0pt}{2em} $\theta_{\rm{}o}$ [deg] & $60.0^{+0.1}_{-0.2}$ & $60.0^{+0.2}_{-0.1}$ & $60.3^{+0.1}_{-0.1}$ \\
\rule[-0.7em]{0pt}{2em} $\chi^{2}/\nu$ &	1.87	   & 1.00	    & 1.48  \\
\hline
\hline
\rule[-1.3em]{0pt}{3em} & \multicolumn{3}{c}{$a_{\rm f}=0.998$, $\theta_{\rm f}=30^{\circ}$}\\
\rule[-0.7em]{0pt}{2em} $a$ & $0.956^{+0.005}_{-0.005}$ & $1^{+0}_{-1{\rm E}-3}$ & $1^{+0}_{-8{\rm E}-5}$ \\ 
\rule[-0.7em]{0pt}{2em} $\theta_{\rm{}o}$ [deg] & $29.9^{+0.3}_{-0.3}$ & $29.5^{+0.3}_{-0.3}$ & $28.7^{+0.4}_{-0.2}$  \\
\rule[-0.7em]{0pt}{2em} $\chi^{2}/\nu$ & 	1.30	   & 1.58	    & 5.08 \\
\hline
\hline
\rule[-1.3em]{0pt}{3em} & \multicolumn{3}{c}{$a_{\rm f}=0.998$, $\theta_{\rm f}=60^{\circ}$}\\
\rule[-0.7em]{0pt}{2em} $a$ & $0.982^{+0.002}_{-0.002}$ & $1^{+0}_{-3{\rm E}-4}$ & $1^{+0}_{-6{\rm E}-5}$ \\ 
\rule[-0.7em]{0pt}{2em} $\theta_{\rm{}o}$ [deg] & $59.9^{+0.1}_{-0.2}$ & $60.1^{+0.1}_{-0.2}$ & $60.2^{+0.1}_{-0.2}$ \\
\rule[-0.7em]{0pt}{2em} $\chi^{2}/\nu$ & 	1.24	   & 1.02	    & 2.51 \\
\end{tabular}
\caption{The best-fit spin and inclination angle values inferred for the three cases of the
limb darkening/brightening law in the \textscown{kyl3cr} model. 
Data were generated using the \textscown{kyl2cr} model. See the main text for details. 
The quoted errors correspond to the 90$\%$ confidence level.}
\label{tab_angdirfp}
\end{center}
\end{table}

The results presented in the previous sections 
show that using different emission directionality approaches
leads to a different location of the $\chi^{2}$ minimum
in the parameter space. 
Doubt about the correct prescription for the emission directionality
thus brings some non-negligible inaccuracy into the evaluation of 
the model parameters. The magnitude of this error cannot be easily 
assessed as a unique number because other parameters are also involved.

A possible way to tackle the problem is to derive the intrinsic
spectrum from self-consistent numerical computations. This has the potential
of removing the uncertainty about the emission directionality (although,
to a certain degree this uncertainty is only moved to a different level 
of the underlying model assumptions).
In this section, we present such results from modelling the artificial
data generated by numerical simulations,
i.e.\ independently of an analytical approximation 
of the emission directionality presented in the previous sections.

Let us first remind the reader that the orbital speed within the inner
$\lesssim10r_{\rm g}$ reaches a considerable fraction of the speed of
light (Fig.~\ref{fig1}). Beaming, aberration, and the light-bending 
all affect the emitted photons very significantly in
this region. Less energetic photons come
from the outer parts where the motion slows down and the relativistic
effects are of diminished importance. This reasoning suggests that the
analysis of the previous section may be inaccurate because the adopted
analytical approximations (\ref{case123}) neglect any dependence on
energy and distance. 

We applied the Monte-Carlo radiative transfer code NOAR \citep[see
Section~5 in][]{2000A&A...357..823D} for the case of ``cold''
reflection, i.e.\ for neutral or weakly-ionised matter. 
The NOAR code computes absorption cross sections in each layer. 
Free-free absorption and the recombination continua of hydrogen-
and helium- like ions are taken into account, as well as the 
direct and inverse Compton scattering. The NOAR
code enables us to obtain the angle-dependent intensity for the
reprocessed emission. The cold reflection case serves as a reference
point that we will later, in a follow-up paper, compare against the
models involving  stronger irradiation and higher ionisation of the disc
medium.

The directional distribution of the intrinsic emissivity of the
reprocessing model is shown in Fig.~\ref{fig4} (right panel).
The continuum photon index $\Gamma=1.9$ is considered
and the energies are integrated over the $2$--$100$ keV range.
Although the results of the radiation transfer computations do 
show the limb-brightening effect, it is a rather mild one,
and not as strong as the Case~1.
In the same plot we also show the angular profile of the emissivity 
distribution in different energy ranges: (i)~2--6~keV (i.e.\ below the iron 
K$\alpha$ line rest energy); (ii)~6--15~keV (including the iron K$\alpha$ 
line); (iii)~15--100~keV (including the `Compton hump'). We notice that
the energy integrated profile is dominated by the contribution from
the Compton hump, where much of the emerging flux originates.
However, all of the energy sub-ranges indicate the limb-brightening
effect, albeit with slightly different prominence.

We implemented the numerical results of NOAR modelling of a reflected
radiation from a cold disc as \textscown{kyl2cr} in the \textscown{ky}
collection of models. Furthermore, we produced an averaged model,
\textscown{kyl3cr}, by integrating \textscown{kyl2cr} tables over all angles.
Therefore, \textscown{kyl3cr} lacks information about the detailed 
angular distribution of the intrinsic local emission from the disc
surface. On the other hand, it has the advantage of increased
computational speed and the results are adequate if the 
emission is locally isotropic. Furthermore, \textscown{kyl3cr} can be {\em
a posteriori} equipped with an analytical prescription for the angular
dependence (Cases 1--3 in eq.~\ref{case123}), which brings the angular 
resolution back into consideration. This approach allows us to switch
between the three prescriptions for comparison and rapid evaluation. 

In order to constrain the feasibility of the aforementioned approaches, we
generated the artificial data using the \textscown{powerlaw + kyl2cr} model.
The parameters of the model are:
photon index of the power law $\Gamma=1.9$ and its normalisation
$K_\Gamma=0.01$, spin of the black hole $a$, inclination angle
$\theta_{\rm o}$, the inner and outer radii of the disc $r_{\rm in}=r_{\rm
ms}$ and $r_{\rm out}=400$, the index of the radial dependence of the
emissivity $q=3$, and the normalisation of the reflection component
$K_{\rm kyl2cr}=0.1$. We simulated the data for two different values of
the spin, $a=0.7$ and $a=0.998$, and for inclination angles
$\theta_{\rm o}=30^{\circ}$ and  $\theta_{\rm o}=60^{\circ}$.
The simulated flux of the primary power law component is the same
as in the previous section. However, now an important fraction
of the primary radiation is reflected from the disc. The total
flux depends on the extension of the disc and its inclination. 
Its value is $4.9-5.6\times10^{-11}$\,erg\,cm$^{-2}$\,s$^{-1}$ for our
choice of the parameters.

As a next step, we replaced \textscown{kyl2cr} by \textscown{kyl3cr} and
searched back for the best-fit results using the latter model. In this
way, using \textscown{kyl3cr} we obtained the values of the spin and the
inclination angle for different directionalities. The fitting results are
summarised in Table~\ref{tab_angdirfp}. Besides the spin and the
inclination angle, only normalisation of the reflection component was
allowed to vary during the fitting procedure. The remaining parameters
of the model were kept frozen at their default values. 

The resulting data/model ratios are shown in
Figures~\ref{fig_a07_ratio} and \ref{fig_a0998_ratio}. For  $a=0.7$
and $\theta_{\rm o}=30^{\circ}$ the graphs look very similar in all
three cases. However, the inferred spin value differs from the fiducial
value with which the test data were originally created.
The dependence of the best-fit $\chi^{2}$ statistic on the spin
and the corresponding graphs of the confidence contours 
for spin versus inclination angle are shown
in Figures \ref{fig_a07i30}--\ref{fig_a0998i60}, again for the three cases of
angular directionality. These figures confirm that for the limb-brightening
profile the inferred spin value comes out somewhat lower than the
correct value, whereas it is higher if the limb-darkening profile is
assumed. 

In each of the three cases the error of the resulting $a$ determination
depends on the inclination angle and the spin itself. However,
we find that the isotropic directionality reproduces our data to the
best precision. The limb darkening profile is not accurate at higher values
of the spin, such as $a=0.998$, when the resulting  $\chi^{2}/\nu$ value
even exceeds $2$. The limb darkening profile is characterised by an 
enhanced blue peak of the line while the height of the red peak is
reduced (see Figs.~\ref{fig4a}--\ref{fig4b}). Consequently, the model
profile is too steep to fit the data. This is clearly visible in the
data/model ratio plots for $a=0.998$ shown in Figures~\ref{fig_a0998_ratio}.
The flux is underestimated by the model below a mean energy value $E_{\rm mean}$ of the
line (for $a=0.998$ and $i=30$\,deg $E_{\rm mean} \approx 5$\,keV) 
and overestimated above $E_{\rm mean}$. This fact leads to
a noticeable jump in the data/model ratio plot. Only the $q$
parameter can mimic this limitation and suit the data by increasing its value. 
It could be the case of the analysis of the MCG--6-30-15 data in the previous
section. 

A noteworthy result appears in comparing 
of the contours produced by the model with 
limb brightening and limb darkening for 
$a=0.7$ and $i=60$\,deg (Fig.~\ref{fig_a07i60}). 
Although the former model (limb brightening) gives 
a statistically worse fit with $\chi^{2}/\nu=1.87$ 
than the limb darkening case ($\chi^{2}/\nu=1.48$),
the inferred values of the spin and the inclination 
angle are consistent with the fiducial values within
the $3\sigma$ level. On the other hand, the spin value 
inferred from the limb darkening model
is far from the fiducial (i.e., the correct) value.

\section{Discussion}
\label{sec:discussion}

We investigated whether the spin measurements of accreting black holes
are affected by the uncertainty of the angular emissivity law, 
${\cal M}(\mu_{\rm e},r_{\rm e},E_{\rm e})$,
in the relativistically broadened iron K$\alpha$ line models.
We employed three different approximations of the angular
profiles, representing limb brightening, isotropic and limb darkening emission profiles.
For the radius-integrated line profile of the disc emission,
and especially for higher values of the spin, the
broadened line has a triangle-like profile. The
differences among the considered profiles concern
mainly the width of the line's red wing.
However, the height of the individual peaks,
also affected by the emission directionality,
is important for the case of an orbiting spot 
(or a narrow ring), which produce a characteristic 
double-horn profile.

We reanalysed an XMM-Newton observation of MCG--6-30-15 
to study the emission directionality effect on
the broad iron line, as measured by current X-ray instruments.
We showed the graphs of $\chi^{2}$ values as a function
of spin for different cases of directionality. 
We can conclude that the limb darkening law favors higher values of
spin and/or steeper radial dependence of the line emissivity;
vice versa for the limb brightening profile.
Both effects are comprehensible after examining the 
left panel of Fig.~\ref{fig4a}. The limb darkening
profile exhibits a deficit of flux in the red wing compared
with the limb brightening profile. Both higher spin value
and steeper radial profile of the intensity can compensate for this deficit.

The higher spin value has the effect of shifting the 
inferred position of the marginally 
stable orbit (ISCO) closer to the black hole, in accordance 
with eq.~(\ref{velocity1}). Consequently, the accretion disc
is extended closer to the black hole.
The radiation comes from shorter radii 
and it is affected by the extreme gravitational redshift.
Hence, the contribution to the red wing of the total 
disc line profile is enhanced. Naturally, these considerations
are based on the assumption that the inner edge
of the line emitting region coincides with the ISCO.
Likewise, the steeper radial dependence of the emissivity means
that more radiation comes from the inner parts of
the accretion disc than from the outer parts, and
this produces a similar effect to decreasing the 
inner edge radius. With the limb brightening law 
the above-given considerations work the other way around.

We further simulated the data with the \textscown{powerlaw} + \textscown{kyrline} 
model. The simple model allows us to keep better control 
over the parameters
and to evaluate the differences in the spin determination.
We used one of the preliminary response matrices for
the IXO mission
and we chose the flux at a level similar
to bright Seyfert 1 galaxies observable
by current X-ray satellites \citep{2007MNRAS.382..194N}.
The simulations with the \textscown{kyrline} model confirm that the
measurements would overestimate the spin for the limb 
darkening profile and, vice versa, they tend to the lower
spin values for the limb brightening profile. 

Although the interdependence of the model parameters is essential
and it is not possible to give the result of our analysis in terms of a 
single number, one can very roughly estimate that the uncertainties
in the angular distribution of the disc emission will produce an uncertainty
of the inferred inner disc radius of about $20$\% for the high quality 
data by IXO.\footnote{Here, we refer to the inner disc radius instead of the spin
because the latter is not as uniform a quantity as the corresponding
ISCO radius to express the uncertainty simply as a percentage value.
However, we still suppose that the inner edge of the line emitting region of the 
disc coincides with the marginally stable orbit. In realistic circumstances
this assumption cannot be satisfied precisely, so it introduces an additional 
source of error. This has been extensively discussed \citep{2008MNRAS.390...21B}. The 
magnitude of the resulting error was constrained most recently in work by 
\citet{2008ApJ...675.1048R} who applied physical arguments about
the emission properties of the inner flow. It is very likely that this discussion will have
to continue for some time until the emission properties of
the general relativistic MHD flows are fully understood.} 
We consider this value as realistic.

In the next step, we applied the NOAR radiation transfer 
code to achieve a self-consistent simulation of the outgoing
spectrum without imposing an ad hoc formula
for the emission angular distribution. In this paper we assumed 
a cold isotropically illuminated disc with a constant 
density atmosphere.
We created new models for the KY suite, \textscown{kyl2cr} and \textscown{kyl3cr}.
The results of NOAR computations of the cold disc are implemented
in the \textscown{kyl2cr} model, while the \textscown{kyl3cr} model
uses the angle-integrated tables over the entire range of emission
directions. This enables us to include, a posteriori, the analytical 
formulae for directionality and check how precisely they reproduce the
original angle-resolved calculations.
We simulated the data using the preliminary IXO response matrix
and we analysed them using the \textscown{powerlaw} + \textscown{kyl2cr} model.
Then we used the \textscown{kyl3cr} model to test
the analytical directionality approaches on these artificial data.

We found that, for $a=0.998$, none of the three assumed cases 
of the directionality profile covered
the fiducial values for the spin and the inclination angle
within the $3\sigma$ contour line 
(see Figs. \ref{fig_a0998i30}--\ref{fig_a0998i60}).
The suitability of the particular directionality prescription
depends on the fiducial values of the spin and the inclination
angle. The limb brightening profile successfully minimises the 
$\chi^{2}$ values for $a=0.998$ and $i=30$\,deg,
but for $a=0.7$ and $i=60$\,deg it gives the worst fit
of all studied cases of the directionality laws.

On the whole, we found that the isotropic angular 
dependence of the emission intensity fits best.
Especially for higher values
of the black hole spin, the model with the 
limb darkening profile was not able to reproduce the data:
the best fit $\chi^{2}/\nu$ value exceeds $5$ 
for $a=0.998$ and $i=30$\,deg, which means 
a more than three times worse fit than using 
isotropic or limb brightening directionality. 
The inclination angle was underestimated by more than $1$\,deg.

This is an important result because much
of the recent work on the iron lines, both in
AGN and black hole binaries, has revealed
a significant relativistic broadening near
rapidly rotating central black holes \citep{2007ARA&A..45..441M}. 
In some of these works, the limb darkening law was employed and different 
options were not tested. The modelled broad lines are typically
characterised by a steep power law in the
radial part of the intensity across the inner region
of the accretion disc, as in the mentioned MCG--6-30-15 
observation. This behaviour has been interpreted as a case of
a highly spinning compact source where the black hole rotational energy
is electromagnetically extracted \citep{2001MNRAS.328L..27W}.
We conclude that the significant steepness of the radial part of the intensity 
also persists in our analysis, however, the exact values depend partly 
on the assumed angular distribution of the emissivity 
of the reflected radiation.

It should be noted that, in reality, the angular distribution
of the disc emission is significantly influenced by the vertical
structure of the accretion disc. However, our comprehension 
of accretion disc physics is still evolving. In recent years, 
several detailed models have been developed for irradiated black hole 
accretion discs in hydrostatic equilibrium 
\citep[see e.g.][]{2000ApJ...540L..37N, 2001MNRAS.327...10B, 2002MNRAS.332..799R}. 
These models combine radiative transfer simulations with calculations of the
hydrostatic balance in the stratified disc medium. Aside from
reprocessing spectra, the models provide solutions for the vertical
disc profile of the density, temperature, and ionisation fractions. 
In \citet{2007A&A...475..155G} the effects of general relativity and
advection on the disc medium were added.

In \citet{2000ApJ...540L..37N}, \citet{2007A&A...475..155G}, and
\citet{2008MNRAS.386.1872R} the reprocessed spectra are evaluated at
different local emission angles. The shape and
normalisation of these spectra depend on various model
assumptions. Until we know in more detail how accretion discs
work, it is hard to choose which is the ``correct'' reprocessing
model. One could always argue that more physical processes and properties
should be included in the radiative transfer simulations, such as the impact
of magnetic fields, macroscopic turbulence, or different chemical
compositions of the medium. Using the above-mentioned models for an
accretion disc in hydrostatic equilibrium may then easily become
computationally intense.

For the practical purpose of data analysis, however, the computation
of large model grids is necessary, which requires sufficiently
fast methods. This is why simple constant density models are most often
used to analyse the observational data. In fact,
\citet{2001MNRAS.327...10B} have shown that their reprocessing
spectra for a stratified disc medium in hydrostatic equilibrium can be
satisfactorily represented by spectra that are computed for irradiated
constant density slabs. Therefore, we include in our present analysis
the angular emissivity obtained from the modelling of neutral
reprocessing in a constant density slab. For the purpose of this paper
we have not discussed in any further detail the dependence on the
ionisation parameter, which we expect to be rather important. It will
be addressed in a future article.

We emphasise that the main strategy of the present paper
is not intended to find the ``correct'' angular emissivity, as this is 
still beyond our computational abilities and understanding of all the physical 
processes shaping the accretion flow. Instead, we examined the three 
different prototypical dependencies which are mutually disparate 
(i.e., the limb-darkening, isotropic, and limb-brightening cases), 
applied in current data analysis, and which presumably reflect the range of possibilities. 
By including these different cases we mimic various uncertainties, 
such as those in the vertical stratification, and we estimate 
the expected error that these uncertainties can produce in the spin determinations. 
Further detailed computations of reprocessing models 
and the angular emissivity are needed in the future 
in order to understand its role in different spectral states of
accreting black holes.

\section{Conclusions}
\label{sec:conclusions}

Black hole spin measurements using X-ray spectroscopy
of relativistically broadened lines depend on the 
definition employed of the angular distribution of the disc emission.
We studied three cases of the directionality profile - limb brightening,
isotropic and limb darkening.

We found that isotropic directionality is consistent with our
radiation transfer computations of 
the reflection spectra of an irradiated cold disc. This
eliminates any need for ad hoc approximations to the limb-darkening profile.
Using an improper directionality profile could impact on
the other parameters inferred for the relativistic broad line model.
Especially for the often used case of limb darkening,
the radial steepness can interfere with the line parameters
of the best-fit model by enhancing the red wing of the line.

Uncertainties in the precise position of the inner edge
can further increase the error of the spin determination; however,
it appears that the expected magnitude of these errors does not
prevent us from setting interesting and realistic constraints on the
spin parameter. Our present treatment of the problem is incomplete
by neglecting the magnetohydrodynamical effects and their influence
on the ISCO location. Future improvements in our theoretical 
understanding of the inner edge location are highly desirable and 
will help to improve the confidence in the determination of the 
model parameters.

We verified our conclusions by comparing them with the results based on
an XMM-Newton long observation of MCG--6-30-15, and we found them 
to be broadly consistent. The expected improvement of sensitivity of
the future IXO mission will significantly advance the 
reliability of the spin determinations.

\begin{acknowledgements} 
The authors are grateful for useful comments and suggestions from the
participants of two `FERO' (Finding Extreme Relativistic Objects) workshops,
held at the European Space Astronomy Centre (ESAC, Spain) and Laboratoire 
d'Astrophysique de Grenoble (France).
JS acknowledges the doctoral student program of the Czech Science Foundation, 
ref.\ 205/09/H033, and the student research grant of the Charles University, ref.\ 33308.
VK and MD appreciate the continued support from research grants of the Academy
of Sciences (300030510), the Czech Science Foundation (205/07/0052), and the
Ministry of Education international collaboration programme (ME09036). RG acknowledges 
the support from the ESA Plan for European Cooperating States (project No.\ 98040).
The Center for Theoretical Astrophysics is operated under the Czech Ministry
of Education, Youth and Sports programme LC06014.
\end{acknowledgements}

\bibliographystyle{aa} 
\bibliography{0001} 

\end{document}